\documentstyle[epsf,floats%
,preprint%
,eqsecnum,prd,aps]{revtex}

\def\0{\over } \def\2{{1\over2}} \def\4{{1\over4}}
\def\5{\hat } \def\6{\partial }

\def\a{\alpha } \def\b{\beta }  \def\d{\delta }
\def\e{\epsilon }   
  \def\l{\lambda }
 
 \def\o{\omega }
   
\def\O{\Omega } \def\S{\Sigma }

\def\({\left(} \def\){\right)} \def\<{\langle } \def\>{\rangle }

\newcommand{\bea}{\begin{eqnarray}}
\newcommand{\eea}{\end{eqnarray}}
\newcommand{\be}{\begin{equation}}
\newcommand{\ee}{\end{equation}}
\newcommand{\nn}{\nonumber\\ }
\newcommand \beq{\begin{eqnarray}}
\newcommand \eeq{\end{eqnarray}}

\def\bfgamma{\mbox{\boldmath$\gamma$}}

\def\Tr{{\,\mathrm Tr\,}}
\def\Im{{\,\mathrm Im\,}}
\def\Re{{\,\mathrm Re\,}}
\def\tr{{\,\mathrm tr\,}}
\def\del{\partial }

\begin{document}
\preprint{CERN-TH/2000-121, SACLAY-T00/059, TUW-00/13 
}
\tighten
\draft

\title{Approximately self-consistent resummations for\\
the thermodynamics
of the quark-gluon plasma:\\
I. Entropy and density}
\author{J.-P. Blaizot}
\address{Service de Physique Th\'eorique, CE Saclay,
        F-91191 Gif-sur-Yvette, France}
\author{E. Iancu}
\address{Theory Division, CERN, CH-1211 Geneva 23, Switzerland}
\author{A. Rebhan}
\address{Institut f\"ur Theoretische Physik,
         Technische Universit\"at Wien,\\
         Wiedner Hauptstra\ss e 8-10/136,
         A-1040 Vienna, Austria}
\date{\today }
\maketitle
\begin{abstract}
We propose a
gauge-invariant and manifestly UV finite resummation of the physics of
hard thermal/dense loops (HTL/HDL)
in the thermodynamics of the quark-gluon plasma. The starting point is
a simple, effectively one-loop expression for the entropy
or the quark density which is derived from the fully self-consistent
two-loop skeleton approximation to the free energy, but
subject to further approximations, whose quality is tested
in a scalar toy model. In contrast to the direct HTL/HDL-resummation
of the one-loop free energy, 
in our approach both the leading-order (LO) and the 
next-to-leading order (NLO) effects of interactions are correctly
reproduced and arise from kinematical 
regimes where the HTL/HDL are justifiable approximations.
The LO effects are entirely due to the (asymptotic) thermal masses
of the hard particles. The NLO ones
receive contributions both from soft excitations, as described by
the HTL/HDL propagators, and from corrections to the dispersion relation
of the hard excitations, as given by HTL/HDL perturbation theory.
The numerical evaluations of our final expressions
show very good agreement with lattice data for zero-density QCD,
for temperatures above twice the transition temperature.
\end{abstract}

\tableofcontents
\newpage
\section{Introduction}

Besides its obvious relevance for cosmology, astrophysics or
ultra-relativistic heavy ion collisions,
the study of QCD at high temperature and/or large baryonic 
density \cite{Kapusta,MLB}  presents exciting theoretical challenges. 
It offers opportunity to explore the properties of matter in a regime where,
unlike in ordinary hadronic matter, the fundamental fields  of QCD---the
quarks and gluons--- are the dominant degrees of freedom and the 
fundamental symmetries are explicit.

Unfortunately, analytical tools available for such a study are not
many. However, because of  asymptotic freedom, the gauge coupling becomes
weak at high temperature, which invites us to try a perturbative treatment
of the interactions. 
But explicit perturbative calculations of the QCD free energy at
high temperature, which have been pushed in recent years up to 
the order $\alpha_s^{5/2}$ \cite{QCDP,BN}, show an extremely
poor convergence except for coupling constants as low as
$\alpha \lesssim 0.05$, which would correspond to 
 temperatures as high as $\gtrsim 10^5 T_c$.
Already the next-to-leading order perturbative correction,
the so-called plasmon effect which is of order 
$\a_s^{3/2}\propto g^3$, 
signals the inadequacy of the conventional
thermal perturbation theory except for very small coupling,
because in contrast to the leading-order terms it leads to
a free energy in excess of the ideal-gas value.

Lattice results on the other hand show a slow approach of the
ideal-gas result from below with deviations of not more than 
some 10-15\% for temperatures a few times the deconfinement
temperature. Besides, these results can be accounted 
for reasonably well by phenomenological fits involving massive 
``quasiparticles'' \cite{Peshier,LH} with masses of the order
of the perturbative leading-order thermal masses.
This suggests that the failure of ordinary 
perturbation theory may not be directly related to the 
non-perturbative phenomena expected at the scale $g^2T$ and  which
 cause a breakdown  of the loop expansion at order $g^6$ and
higher
\cite{Kapusta}. Rather, the quasiparticle fits support the idea that one
should be able to give an accurate description of the
thermodynamics of the QCD plasma in terms of its 
(relatively weakly interacting) quasiparticle excitations.

It is worth emphasizing at this stage that, among the relevant degrees of
freedom,  the soft collective ones, with momenta of order $gT$, are already
non-perturbative. Although their leading order  contribution $\propto g^3$
to the pressure can be easily isolated \cite{Kapusta}, it does not  make
much physical sense to regard this contribution as a genuine perturbative 
correction. 

Indeed, to leading order in $g$, the dynamics of the soft 
modes is described by an  effective theory which includes the one-loop
thermal fluctuations of the ``hard'' modes with momenta $\sim T$. 
The relevant generalization of
the Yang-Mills equation reads \cite{QCD,BIO} :
\be\label{ava}
D_\nu F^{\nu\mu}=\5 m_D^2\int\frac{{\rm d}\Omega}{4\pi}\,
\frac{v^\mu v^i}{v\cdot D}\,E^i
\equiv\hat\Pi_{\mu\nu}^{ab}A_b^\nu
+\frac{1}{2}\, \hat\Gamma_{\mu\nu\rho}^{abc} A_b^\nu A_c^\rho+\,...
\ee
where the induced current in the right hand side describes
the polarization of the hard particles by the soft colour fields
$A^\mu_a$ in an eikonal approximation.
[In this equation, $\5 m_D\sim gT$ is the Debye mass, 
$E_a^i$ is the soft electric field, $v^\mu\equiv (1,\,{\bf v})$, and
the angular integral $\int {\rm d}\Omega$ runs over the 
orientations of the unit vector ${\bf v}$.] 
This current is non-local and  gauge symmetry, which forces the presence of
the covariant derivative $D^\mu=\del^\mu+igA^\mu$
in the denominator of Eq.~(\ref{ava}), makes it  also non-linear. When
expanded in powers of $A^\mu_a$, 
it generates an infinite series of non-local  self-energy and vertex
corrections, known as ``hard thermal loops'' (HTL) \cite{BP,QCD}. The
latter encompass important physical phenomena,  like screening effects and
non-trivial dispersion relations for  the soft excitations \cite{MLB,BIO}
(and references therein). Similar phenomena exist also in the
case of soft fermions, which, to leading order in $g$, obey the following
generalized Dirac equation \cite{QCD} (with $\5 M\sim gT$ and
$\not\!{v}=\gamma_\mu v^\mu$) :
\beq\label{avpsi}
i\!\not\!\!{D} \psi\,=\,\5 M^2\int\frac{{\rm d}\Omega}{4\pi}
\,\frac {\not\!{v}} {i(v\cdot D)}\,\psi\,
\equiv\, \hat\Sigma\psi + \hat\Gamma_\mu^a A^\mu_a \psi +\,...\eeq
At soft momenta $k\lesssim gT$, all HTL's are leading order effects,
as obvious in Eqs.~(\ref{ava}) and (\ref{avpsi}), and
must be consistently resummed. Analogs of HTL's exist at finite chemical
potential
$\mu$. In the regime
$\mu \gg T$ these are often referred to as ``hard dense loops'' (HDL).

In traditional perturbative calculations of the thermodynamics 
performed in imaginary time \cite{MLB}, the HTL's play almost no 
role: only the Debye mass  $m_D^2$ needs to be resummed  in the
static electric gluon propagator \cite{AE}. This
resummation is responsible for the occurrence of odd powers of $g$ in the
perturbative expansion. 

Such a simple  resummation however may become insufficient whenever a more
complete information on the quasiparticles needs to be taken into account. 
Quite generally, this  physical information
is contained in the spectral weight $\rho(k_0, k)$
related to the corresponding propagator by:
\beq\label{Dspec0}
D(\omega, k)&=&\int_{-\infty}^{\infty}\frac{{\rm d}k_0}{2\pi}
\,\frac{\rho(k_0, k)}{k_0-\omega}\,.\eeq
In the imaginary time formalism, and for bosonic fields,
$\omega=i\omega_n\equiv i2\pi nT$ with integer $n$.
Clearly, the restriction to the Matsubara mode with $n=0$
retains in the propagator only one moment of the spectral weight. In the
HTL approximation, we know that the spectral density is divided into a pole
at time-like momenta and  a continuum at space-like momenta. While there
exist physical observables which can be accurately described in
perturbation theory by a single moment of the spectral weight,
this does not appear to be the case in the calculations that we shall
present and in which the various pieces of the spectral functions 
contribute in different ways. 

In fact, since the thermodynamical functions are dominated by hard
degrees of freedom, an important effect of the soft modes will be to induce
corrections on the 
hard quasiparticle dispersion relations. As we shall find, the spectral
functions for large momenta will take the approximate form 
$\rho(\o, k)\approx
\delta(\o^2-k^2-m^2_\infty)$, where $m^2_\infty\sim g^2T^2$ is the 
leading-order thermal mass (or {\it asymptotic} mass) of the hard excitation.
Clearly, such an effect does not naturally emerge in a scheme where one 
resums just the $n=0$ Matsubara mode. 

In order to overcome all these limitations, it has been recently proposed
to perform full resummations of the HTL self-energies
$\Pi_{\mu\nu}$ and $\Sigma$ in calculations of the thermodynamical
functions. In Refs. \cite{ABS,BR}, this has been done
by merely replacing the free propagators by the corresponding
HTL-resummed ones in the expression of the free-energy of the ideal gas;
e.g. (in simplified notations) :
\beq\label{ABS0}
\Tr \log D_0^{-1} \longrightarrow \Tr \log (D^{-1}_0+\Pi)\,.\eeq
In principle, this is just the first step in a systematic
procedure which consists in resumming the HTL's by adding and
subtracting them to the tree-level QCD Lagrangian.
This would be the extension to QCD of the so-called
``screened perturbation theory''\cite{KPP,CH}, a method which, for
scalar field theories, has shown an improved convergence 
(in one- and two-loop calculations) as compared 
to the straightforward perturbative expansion.
But in its zeroth order approximation in Eq.~(\ref{ABS0}), 
this method over-includes the leading-order interaction term
$\propto g^2$ (while correctly reproducing the order-$g^3$ 
contribution), and gives rise to new, ultimately
temperature-dependent UV divergences
and associated additional renormalization scheme dependences.

Another drawback of such a direct HTL resummation appears
to be that the HTL's are kept in the hard momentum regime where they are
no longer describing actual physics, while hard momenta are
providing the dominant contributions to the thermodynamic potential.

Our approach on the other hand \cite{PRL,PLB} will be based on self-consistent
approximations using the skeleton representation of the 
thermodynamic potential \cite{LW}
which takes care of overcounting problems
automatically, without the need for thermal
counterterms. We shall mainly consider the so-called
2-loop-$\Phi$-derivable \cite{Baym} approximation, for which it turns out
that the first derivatives of the thermodynamic potential, the
entropy and the quark densities, take a rather simple,
effectively one-loop form\cite{Riedel,VB},
but in terms of fully dressed propagators.

In gauge theories, the generalized gap equations that determine
these dressed propagators are too complicated {to be solved}
exactly (even numerically). But an exact solution would anyhow
be unsatisfactory because 
$\Phi$-derivable approximations in general do not respect gauge invariance.
We therefore propose gauge independent but only
approximately self-consistent dressed propagators as obtained
from (HTL) perturbation theory. Using these in the entropy\footnote{For
brevity we refer only to the entropy explicitly, but all of the
following remarks apply to the density as well.}
expression obtained from the 2-loop-$\Phi$-derivable approximation
gives a gauge-independent and UV finite 
approximation for the entropy, which, while being
nonperturbative in the coupling, contains the correct leading-order (LO)
and the next-to-leading order (NLO) effects of interactions
in accordance with thermal perturbation theory. Both turn
out to arise from kinematical regimes where the HTL's are
justifiable approximations.

While also being effectively a resummed one-loop expression,
the approximately self-consistent entropy  
differs from the direct HTL-resummation of the free
energy in Eq.~(\ref{ABS0})
in that it includes correctly also the LO interaction effects.
Remarkably, in our approach
the latter are entirely 
determined by the (asymptotic) thermal masses of the hard excitations.
This agrees with and justifies the simple quasiparticle models of 
Ref.~\cite{Peshier,LH}, which assume constant masses equal to the
respective asymptotic thermal masses for quarks and as many (scalar) bosons
as there are transverse gluons. Whereas these models do not include
the correct NLO (plasmon) effect, our approach does, but
in a rather unconventional manner which demonstrates the nontriviality
of the resummation that has been achieved: only part of the
plasmon effect is coming directly from soft excitations; a larger
part arises from corrections to the dispersion relation of the
(dominant) hard excitations by soft modes, 
as determined by standard HTL perturbation theory \cite{BP}.

Because of the approximations that we have made, it does matter
whether the entropy  or the thermodynamic
potential is considered. Our approach however attempts to take
advantage of the fact that entropy is generally the
simpler quantity. Indeed, the way by which the LO and NLO 
 interaction contributions
can be traced to spectral properties of free quasiparticles
within our entropy  expressions indicates a posteriori
the adequateness of this particular resummation scheme {to}
the physics contained in the HTL propagators.

The present paper is organized as follows: In Sect.\ II, the general
formalism of $\Phi$-derivable self-consistent approximations is
reviewed and the central, effectively one-loop formula for the
entropy in a two-loop skeleton approximation to the thermodynamic
potential is derived in a scalar theory with cubic and quartic
interactions.
In the simple solvable model of large-$N$ scalar O($N$) theory
\cite{DJ,DHLR2},
where the two-loop $\Phi$-derivable approximation becomes exact,
the further approximations that will be considered in the QCD case are
compared with the exact solution and their
renormalization scale dependence is exhibited.

In Sect.\ III, the approximately self-consistent resummations are
introduced for purely gluonic QCD first, and equivalence with
conventional perturbation theory up to and including order $g^3$
is proved and analyzed in detail. Sect.\ IV generalizes this to
QCD with quarks and to the quark density as an additional
thermodynamic quantity. Some of the more technical details
of how the plasmon effect arises in our approach are relegated
to the Appendix.

In Sect.\ V, the various approximations are evaluated numerically.
We find that the plasmon effect, which is largely responsible for the
poor convergence properties of conventional thermal
perturbation theory, in our approach leads only to moderate
contributions when compared with the leading-order effects.
When combined with a two-loop renormalization group improvement,
our results are found to compare remarkably well with available
lattice data for temperatures above twice the deconfinement
temperature. Moreover, we also present numerical results
for the quark density at zero temperature and large chemical
potential.

\section{General formalism. The scalar field}

In this section we develop the formalism of propagator renormalization  using
techniques that allow systematic rearrangements of the perturbative expansion 
avoiding double-countings.  We shall recall in particular how 
self-consistent approximations can be used to obtain 
a simple expression for the entropy which isolates the
contribution of the elementary excitations as a leading contribution. To get
familiarity with the formalism, we demonstrate some of its important features with
the example of the scalar field. This provides,
in particular, a test of the validity
of approximations which will be used in dealing with 
QCD in the rest of the paper.

\subsection{Skeleton expansion for thermodynamical potential and entropy}

The thermodynamic potential $\O=-PV$ of the  scalar field can be written as the
following  functional of the full propagator $D$ \cite{LW,Baym}:
\be
\label{LW}
\b \O[D]=-\log Z=\2 \Tr \log D^{-1}
-\2 \Tr \Pi D+\Phi[D]\,,
\ee
where $\Tr$ denotes the trace in configuration space,
$\b=1/T$, $\Pi$ is the self-energy related to $D$ by
Dyson's equation ($D_0$ denotes the bare propagator):
\beq\label{Dyson}
D^{-1}=D^{-1}_0+\Pi, 
\eeq
and $\Phi[D]$ is the sum of the 2-particle-irreducible ``skeleton''
diagrams
\be\label{skeleton}
-\Phi[D]=
\epsfxsize=7cm
\epsfbox[50 390 550 430]{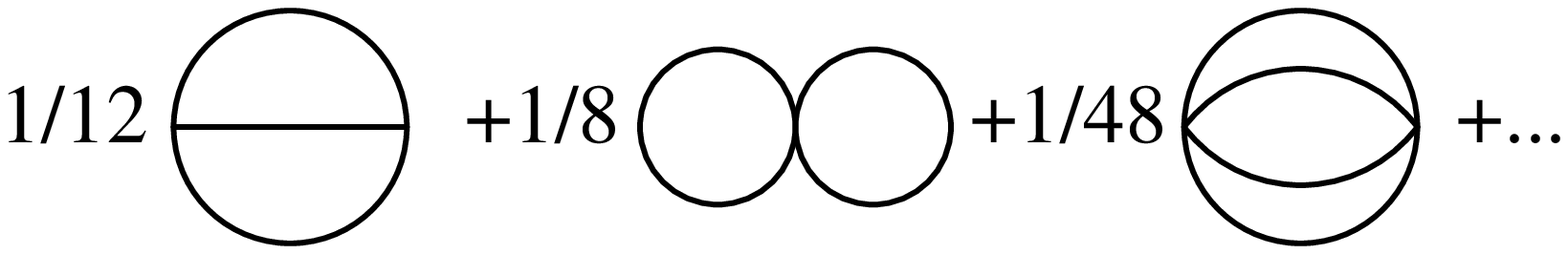}
\ee


The essential property of the functional $\O[D]$ is to be
stationary under variations of $D$ (at fixed $D_0$) around the
physical propagator. The physical pressure is then obtained
as the value of $\O[D]$ at its extremum. The stationarity condition,
\be\label{selfcons}
{\d\O[D] / \d D}=0,
\ee
implies the following relation 
\be\label{PhiPi}
\d\Phi[D]/\d D=\2\Pi,
\ee
which, together with Eq.~(\ref{Dyson}), defines the 
physical propagator and self-energy in a self-consistent way.
Eq.~(\ref{PhiPi}) expresses the fact that the skeleton 
diagrams contributing to $\Pi$ are
obtained by opening up one line of a two-particle-irreducible skeleton. 
Note that while the diagrams of the bare perturbation theory, 
i.e., those involving
bare propagators, are counted once and only once in the expression of
$\Pi$ given above, the
diagrams of bare perturbation theory 
contributing to the thermodynamic potential 
are counted several times in
$\Phi$. The extra terms in Eq.~(\ref{LW}) precisely correct for
this double-counting.

Self-consistent (or variational)
approximations, i.e., approximations which preserve the
stationarity property (\ref{selfcons}),
are  obtained by selecting a class of skeletons in
$\Phi[D]$ and calculating $\Pi$ from Eq.~(\ref{PhiPi}).
Such approximations are commonly called ``$\Phi$-derivable''
\cite{Baym}.

The traces over configuration space in Eq.~(\ref{LW}) involve integration over
imaginary time and over spatial coordinates. Alternatively, these can be
turned into summations over Matsubara frequencies and integrations over
spatial momenta:
\beq
\int_0^\beta {\rm d}\tau
\int {\rm d}^3x \rightarrow \beta V
 \int [{\rm d}k] ,\eeq
where $V$ is the spatial volume,
 $k^\mu=(i\omega_n, {\bf k})$
and $\omega_n = n\pi T$, with $n$ even (odd) for bosonic 
(fermionic) fields (the fermions will be discussed later).
We have introduced a condensed notation for the
the measure of the loop integrals (i.e., the sum over the
Matsubara frequencies $\omega_n$ and the integral over the
spatial momentum ${\bf k}$): 
\beq\label{mesure}
\int[{\rm d}k]\equiv T\sum_{n, even} \int\frac{{\rm d}^3k}{(2\pi)^3}
\, ,
\qquad\,\,\,\,
 \int\{{\rm d}k\}\equiv  T\sum_{n, odd}
\int\frac{{\rm d}^3k}{(2\pi)^3}\,.
\eeq
Strictly speaking, the sum-integrals in equations like  Eq.~(\ref{LW})
contain  ultraviolet divergences, which requires regularization 
(e.g., by dimensional continuation). Since, however,
most of the forthcoming calculations will be free
of ultraviolet problems (for the reasons
explained at the end of this subsection), we do not need
to specify here the UV regulator  (see however Sect.~\ref{secsimplemodel} for
explicit calculations).

For the purpose of developing approximations for  the entropy 
it is convenient to 
perform the  summations over the Matsubara frequencies. One obtains  then
integrals over real frequencies involving discontinuities of propagators or
self-energies which have a direct physical significance. Using standard
contour integration techniques, one gets:
\be\label{Omega(D)}
\Omega/V=\int\!\!{d^4k\0(2\pi)^4}\, n(\omega)\left(\Im \log(-\omega^2+k^2+\Pi)
-\Im\Pi D\right)+T\Phi[D]/V
\ee
where $n(\omega)=1/({\rm e}^{\beta \omega}-1)$.

 The analytic propagator $D(\omega,k)$ can be
expressed in terms of the spectral function: 
\be
D(\o,k)=\int_{-\infty}^\infty {dk_0\02\pi}{\rho(k_0,k)\0k_0-\o}.
\ee
and we define, for $\omega$ real, 
\beq
\Im D(\omega,k)\equiv \Im D(\omega+i\epsilon,k)=\frac{\rho(\omega,k)}{2}.
\eeq
 The imaginary parts of other quantities are defined
similarly. 

We are now in the position to calculate  the entropy density:
\be
{\cal S}=-{\6(\O/V)/\6T}\,.
\ee
The thermodynamic potential, as given by Eq.~(\ref{Omega(D)}) depends on the
temperature through the statistical factors $n(\omega)$ and the spectral
function $\rho$, which is determined entirely by the self-energy.
Because of  Eq.~(\ref{selfcons}) the
temperature derivative of the spectral 
density in the dressed propagator cancels
out in the entropy density and one obtains
\cite{Riedel,VB}:
\bea\label{Ssc}
{\cal S}&=&-\int\!\!{d^4k\0(2\pi)^4}{\6n(\o)\0\6T} \Im \log D^{-1}(\o,k) \nn
&&+\int\!\!{d^4k\0(2\pi)^4}{\6n(\o)\0\6T} \Im\Pi(\o,k) \Re D(\o,k)+{\cal S}'
\eea
with
\be\label{SP0}
{\cal S}'\equiv -{\6(T\Phi)\0\6T}\Big|_D+
\int\!\!{d^4k\0(2\pi)^4}{\6n(\o)\0\6T} \Re\Pi \Im D. 
\ee
We shall verify explicitly that for the two-loop skeletons, we have:
\beq\label{Sprime}
{\cal S}'=0.
\eeq
Loosely speaking, the first two terms in Eq.~(\ref{Ssc})
represent essentially the entropy of ``independent quasiparticles'', 
while ${\cal S}'$   
accounts for a residual interaction among these quasiparticles
\cite{VB}.

Since the condition (\ref{Sprime}) plays an important role in our work, we shall
derive it explicitly  in a scalar model
with interaction term 
$${\cal L}_{int}=(g/3!)\phi^3-(\lambda/4!)\phi^4,$$ which
is a simple toy model of the tri- and quadrilinear
self-interactions of gauge bosons. (Interactions with fermions
are already covered by the analysis contained in Ref.~\cite{VB}.)  
In the two-loop
approximation, where only the first two diagrams of the skeletons in
Eq.~(\ref{skeleton}) are kept,  the contribution involving two 3-vertices reads
\be
-T\Phi^{(a)}={g^2\012}T^2 \sum_{\o_1,\o_2}\int{d^3k_1\,d^3k_2\0(2\pi)^6}
D(\o_1,|\vec k_1|)D(\o_2,|\vec k_2|)D(-\o_1-\o_2,|-\vec k_1-\vec k_2|).
\ee
Expressing the propagators in terms of the spectral functions, 
and evaluating the
Matsubara sums by contour integration, one gets:
\bea\label{phiPhia}
-T\Phi^{(a)}&=&{g^2\012}\int {d^4k\,d^4k'\,d^4k''\0(2\pi)^{9}}
\d^3(\vec k+\vec k'+\vec k'')
\rho(k)\rho(k')\rho(k''){\bf P}{1\0k_0+k_0'+k_0''}\nn
&&\qquad\times \left\{ [n(k_0)+1][n(k_0')+n(k_0'')+1]+n(k_0')n(k_0'')\right\}
\eea
where {\bf P} denotes the principal value prescription and we have used
the identity:
\be
n(x+y)[1+n(x)+n(y)]=n(x)n(y)\,.
\ee

The two-loop skeleton involving the 4-vertex is given by the simpler
expression
\be\label{phiPhib}
-T\Phi^{(b)}=-{\l\08} \left[\sum_\o\int\!\!{d^3k\0(2\pi)^3}D(\o,k)\right]^2\nn
=-{\l\08}\int {d^4k\,d^4k'\0(2\pi)^{8}}\rho(k)\rho(k')
\left\{n(k_0)n(k'_0)\right\}.
\ee

According to Eq.~(\ref{SP0}),
the first
contribution to ${\cal S}'$ is given by differentiating Eqs.~(\ref{phiPhia})
and (\ref{phiPhib}) with respect to $T$ at fixed $\rho$. 
Because the integrand in front of the curly brackets in (\ref{phiPhia})
is symmetric, the arguments of the distribution functions can be freely
exchanged as long as the fact that their products come with distinct arguments
is preserved. $\6_T[-T\Phi]$ is therefore obtained by replacing
the terms in curly brackets 
in (\ref{phiPhia}) by $\{6n(k_0')\6_Tn(k_0)+3\6_Tn(k_0)]\}$
and that in (\ref{phiPhib}) by $2n(k_0')\6_T n(k_0)$.

The second contribution to ${\cal S}'$ involves the real part of the
self-energy as given by the two 
(dressed) one-loop diagrams following from opening up one line
in the first two diagrams in (\ref{skeleton}),
\bea
\Re \,\Pi^{(a)}(\o,q) &=& -{g^2\02} \int\!\! {d^3 k\0(2\pi)^3}
\int\! {dk_0\02\pi}{dk_0'\02\pi}
\rho(k_0,|\vec k|)\rho(k_0',|\vec k+\vec q|) \nn
&&\qquad\times [n(k_0)+n(k_0')+1]{\bf P}{1\0\o+k_0+k_0'}\\
\Re \,\Pi^{(b)}&=& {\l\02} \int\!\! {d^4k\0(2\pi)^4} n(k_0)\rho(k_0,k)
\eea

This gives
\bea
 \int\!\!{d^4k\0(2\pi)^4}{\6n(k_0)\0\6T} \Re\Pi^{(a)} \Im D &=&
-{1\04}\int\! {d^4k\,d^4k'\,d^4k''\0(2\pi)^{9}}
\d^3(\vec k+\vec k'+\vec k'')
\rho(k)\rho(k')\rho(k'')\nn
&&\qquad
\times{\bf P}{1\0k_0+k_0'+k_0''}[\6_T n(k_0)][n(k_0')+n(k_0'')+1]\\
 \int\!\!{d^4k\0(2\pi)^4}{\6n(k_0)\0\6T} \Re\Pi^{(b)} \Im D &=&
\int\!\!{d^4k\0(2\pi)^4}{\6n(k_0)\0\6T}{\rho(k)\02}{\lambda\02}
\int\!\!{d^4k'\0(2\pi)^4}n(k_0')\rho(k')
\eea
where we have used $\Im D=\rho/2$.
Indeed, this cancels precisely $-\6_T[T\Phi]$ as obtained above, verifying
the proposition that ${\cal S}'=0$ for the lowest-order (two-loop)
diagrams in $\Phi[D]$.

As the previous derivation shows, the vanishing of ${\cal S}'$  holds
whether the propagator are the self-consistent propagators or not. 
That is,
only the relation (\ref{PhiPi}) is used, and the proof does not require $D$
to satisfy the self-consistent Dyson equation (\ref{Dyson}).
A general analysis of the contributions to 
${\cal S}'$ and their physical 
interpretation can be found in Ref.~\cite{CP2}.  

We emphasize  now a few attractive features of Eq.~(\ref{Ssc}) 
with ${\cal S}'=0$,
which makes the entropy a privileged quantity to study the thermodynamics of
ultrarelativistic plasmas. We note first that  the formula for
$\cal S$ at 2-loop order
involves the self-energy only at 1-loop order.  Besides this important
simplification, this formula for
${\cal S}$, in contrast to the pressure, has the
advantage of manifest ultra-violet finiteness, since ${\6n/\6T}$
vanishes exponentially for both $\o\to\pm\infty$.
Also, any multiplicative renormalization $D\to ZD$, $\Pi\to Z^{-1}\Pi$
with real $Z$ drops out from Eq.~(\ref{Ssc}). Finally, the entropy has  a more
direct quasiparticle interpretation than the pressure. 
This will be illustrated 
explicitly in the simple model of the next subsection. 
More generally, Eq.~(\ref{Ssc}) can be
transformed with the help of the following identity:
\beq
\Im \log D^{-1}(\o,k)=\arctan\left( 
\frac{\Im \Pi}{\Re D^{-1}} \right)-\pi\epsilon(\o)\theta(-\Re D^{-1}),
\eeq
with $\epsilon(\o)$ the sign function and
$-{\pi\02}<\arctan(x)< {\pi\02}$. Using this identity we rewrite ${\cal S}$ as
${\cal S}= {\cal S}_{pole}+{\cal S}_{damp}$,
with
\beq\label{Spole}
{\cal S}_{pole}&=&\int\!\!{d^4k\0(2\pi)^4}{\6n(\o)\0\6T} \pi\epsilon(\o)
\theta(-{\Re}D^{-1}(\o,k))\nonumber\\
 &=& \int\!\!{d^3k\0(2\pi)^3}\Bigl\{(1+n_k)\log (1+n_k) \,-\,
n_k\log n_k\Bigr\}.
\eeq
To get the second line, we have made an integration by part, using
\beq\label{ID1}
 \frac{\del n(\o)}{\del T}=-\,{\6\sigma(\o)\0\6\o},\qquad\sigma(\o)
\equiv -n\log n + (1+n)\log(1+n), \eeq
and we  have set $n_k\equiv n(\epsilon_k)$, with $\epsilon_k$ solution
of 
$\Re D^{-1}(\o=\varepsilon_k,k)=0$.
The quasiparticles thus defined by the poles of the 
propagator are sometimes called
``dynamical quasiparticles''
\cite{CP2}.  The quantity ${\cal S}_{pole}$ is the entropy of a system of such
non-interacting quasiparticles, while the  quantity
\beq
{\cal S}_{damp} = \int\!\!{d^4k\0(2\pi)^4}{\6n(\o)\0\6T} \left[{\rm
Im}\Pi(\o,k)\Re D(\o,k)- \arctan\left( 
\frac{\Im \Pi}{\Re D^{-1}} \right)\right],
\eeq
which vanishes when $\Im \Pi$ vanishes, is a contribution coming from
the continuum part of the quasiparticle spectral weights.

\subsection{A simple model} 
\label{secsimplemodel}

In this section we shall present the  self-consistent solution
for the $(\lambda/4!) \phi^4$ theory, keeping in $\Phi$ only the two-loop
skeleton whose explicit expression is given in Eq.~(\ref{phiPhib}). 
Anticipating the fact that the fully dressed propagator will be that
of a massive particle, we write the spectral function as
$\rho(k_0,{\bf k})=2\pi \,\epsilon(k_0)\,\delta(k_0^2-{\bf k}^2-m^2)$,
and consider $m$ as a variational parameter. The thermodynamic potential
(\ref{LW}), or equivalently the pressure, becomes then a simple 
function of $m$.
By Dyson's equation, the self-energy is simply $\Pi=m^2$. We set:
\beq\label{IM}
I(m)\,\equiv \,\frac{1}{2}\int [{\rm d}k]\, D(k)\,= \2\int [{\rm d}k]
\,\frac{1}{\omega_n^2 + {\bf k}^2 + m^2}.
\eeq
Then  the pressure
can be written as:
\beq\label{Pscal}
-P=\frac{\Omega}{V}=\frac{1}{2}\int\frac{{\rm
d}^3k}{(2\pi)^3}\,\varepsilon_k+\frac{1}{\beta}\int\frac{{\rm
d}^3k}{(2\pi)^3}\,\log(1-{\rm e}^{-\beta\varepsilon_k})
-m^2 I(m)+\frac{\lambda_0}{2}\, I^2(m),
\eeq
where
  $\varepsilon_k^2\equiv k^2+m^2$.
By demanding that $P$ be stationary with respect to $m$
one obtains the self-consistency condition which takes
here the form of a ``gap equation'':
\beq
\label{gapp}
m^2\,=\,{\lambda_0}\,I(m).
\eeq
{The pressure in the two-loop $\Phi$-derivable approximation,
as given by Eqs.~(\ref{IM})--(\ref{gapp}), is formally the same
as the pressure per scalar degree of freedom in the (massless)
$N$-component model with the interaction term written as
${3\0N+2}(\lambda/4!)(\phi_i\phi_i)^2$ in the limit $N\to \infty$
\cite{DHLR2}. From the experience with this latter model, we know
that Eqs.~(\ref{IM})--(\ref{gapp}) admit an exact, renormalizable solution
which we recall now.
}

At this stage, we need to specify some properties of the loop integral $I(m)$ 
which  we can write as  the sum of a vacuum piece $I_0(m)$ and a finite
temperature piece $I_T(m)$ such that, at fixed $m$,
$I_T(m)\to 0$ as $T\to 0$.
We use dimensional regularization
to control the ultraviolet divergences present in $I_0$, 
which implies $I_0(0)=0$. Explicitly
one has:
\beq\label{I(m)regul}
\mu^{\epsilon}
I(m)=-\frac{m^2}{32\pi^2}
\left(\frac{2}{\epsilon}+\log\frac{\bar\mu^2}{m^2}+1\right)
+I_T(m)+{\rm O}(\epsilon),
\eeq
with
\beq\label{ITm}
I_T(m)=\int\!\!{d^3k\0(2\pi)^3}\,\frac{n(\varepsilon_k)}
{2\varepsilon_k}\,,
\eeq
and 
$\varepsilon_k\equiv (k^2+m^2)^{1/2}$. In Eq.~(\ref{I(m)regul}), $\mu$ is
the scale of dimensional regularization, introduced, as usual,
by rewriting the bare coupling $\lambda_0$
as $\mu^{\epsilon}\hat\lambda_0$, with dimensionless
$\hat\lambda_0$; furthermore, $\epsilon=4-n$, with $n$ the number of
space-time dimensions, and
$\bar\mu^2=4\pi{\rm e}^{-\gamma}\mu^2$. 

We use the modified minimal subtraction
scheme ($\overline{\hbox{MS}}$) 
and define a dimensionless renormalized coupling
$\lambda$ by:
\beq\label{RENL}
\frac{1}{\lambda}=\frac{1}{\lambda_0\mu^{-\epsilon}}+\frac{1}{16\pi^2\epsilon}.
\eeq 
When expressed in terms of the renormalized coupling, the gap
equation becomes free of ultraviolet divergences. It reads:
\beq\label{GAP2}
m^2\,=\,\frac{\lambda}{2}\int\frac{{\rm d}^3k}{(2\pi)^3}\,
\frac{n(\varepsilon_k)}{\varepsilon_k}\,+\,
\frac{\lambda  m^2}{32 \pi^2}\left(\log \frac{m^2}{\bar\mu^2}
\,-1\right),
\eeq
The renormalized coupling constant satisfies
\be\label{RGSC}
\frac{{\rm d}\lambda}{{\rm d}\log \bar\mu}\,=\,\frac{\lambda^2}{16\pi^2},
\ee
which ensures that the solution $m^2$ of Eq.~(\ref{GAP2})
is independent of $\bar\mu$. {Eq.~(\ref{RGSC}) coincides with the exact 
$\beta$-function in the large-$N$ limit, 
but gives only one third of the lowest-order perturbative 
$\beta$-function for $N=1$. 
This is no actual fault since the running
of the coupling affects the thermodynamic potential only at
order $\lambda^2$ which is beyond the perturbative accuracy of
the 2-loop $\Phi$-derivable approximation. In order to see the
correct one-loop $\beta$-function at finite $N$, the approximation
for $\Phi$ would have to be pushed to 3-loop order.}

Note also that, in the present approximation, the
renormalization (\ref{RENL}) of the coupling constant
is sufficient to make the pressure (\ref{Pscal}) finite.
Indeed, in dimensional regularization
the sum of the zero point energies ${\varepsilon_k}/2$
in Eq.~(\ref{Pscal}) reads:
\beq\label{zeropoint}
\mu^{\epsilon}\int\frac{{\rm d}^{n-1}k}{(2\pi)^{n-1}}\,\,
\frac{\varepsilon_k}{2}\,=\,-\frac{m^4}{64\pi^2}
\left(\frac{2}{\epsilon}+\log\frac{\bar\mu^2}{m^2}+{3\02}\right)
+{\rm O}(\epsilon),
\eeq
so that
\beq
\mu^\epsilon\int\frac{{\rm d}^{n-1}k}{(2\pi)^{n-1}}\,\,
\frac{\varepsilon_k}{2}\,-\,
\frac{\Pi^2}{2\hat\lambda_0}\,=\,-\,{m^4\02\lambda}\,-\,
\frac{m^4}{64\pi^2}
\left(\log\frac{\bar\mu^2}{m^2}+{3\02}\right)+{\rm O}(\epsilon)\eeq
is indeed UV finite as $n\to 4$. {After also using the
gap equation (\ref{GAP2}), one obtains the $\bar\mu$-independent
result}
\beq\label{PPHI}
P=-T\int\frac{{\rm
d}^3k}{(2\pi)^3}\,\log(1-{\rm e}^{-\beta\varepsilon_k})
+{m^2\02}I_T(m)+{m^4\0128\pi^2}.
\eeq

We now compute the entropy according to  Eq.~(\ref{Ssc}). Since 
 $\Im\Pi=0$ and $\Re\Pi=m^2$, we have simply:
\be\label{Sscphi4}
{\cal S}=-\int\!\!{d^4k\0(2\pi)^4}{\6n(\o)\0\6T}\Im\log (k^2-\o^2+m^2).
\ee
Using
\be\label{IM0}
 \Im\log(k^2-\o^2+m^2) \,=\,-\pi\epsilon(\o)\theta(\o^2-
\varepsilon_k^2),
\ee
and the identity (\ref{ID1}),  
one can rewrite Eq.~(\ref{Sscphi4})  in the  form  
(with $n_k\equiv
n(\varepsilon_k)$):
\be\label{SPHI}
{\cal S} =\int\!\!{d^3k\0(2\pi)^3}\Bigl\{(1+n_k)\log (1+n_k) \,-\,
n_k\log n_k\Bigr\}.\ee
This formula   shows that,
in the present approximation, the entropy of the interacting scalar
gas is formally identical to the entropy of an ideal gas of massive
bosons, with mass $m$. 

It is instructive to observe that such a simple
interpretation does not hold for the pressure. The pressure of
an ideal gas of massive bosons is given by:
\be\label{Pmassbosons}
P^{(0)}(m)=\int\!\!{d^3k\0(2\pi)^3}\int_{\epsilon_k}^\infty{\rm
d}\o\left(n(\o)+\frac{1}{2}\right)
=-\int\!\!{d^3k\0(2\pi)^3}\left\{T\log(1-{\rm
e}^{-\epsilon_k/T})+\frac{\epsilon_k}{2}\right\},
\ee
which differs indeed from Eq.~(\ref{Pscal}) by the term $m^4/\lambda$ which
corrects for the double-counting of the interactions included in the
thermal mass.  Note that since the mass depends on the temperature,
and since ${\cal S}={\rm d}P/{\rm d}T$,
it is not surprising to find such a mismatch.

{Moreover,} unlike the correct expression
(\ref{Pscal}), Eq.~(\ref{Pmassbosons}) is afflicted with
UV divergences which in dimensional regularization are proportional
to $m^4$ (cf. Eq.~(\ref{zeropoint})), and hence dependent upon the
temperature. 
This is precisely the kind of divergences which are met in the
one-loop HTL-resummed calculation of the pressure in QCD of Ref.~\cite{ABS}.

\subsection{Comparison with thermal perturbation theory}
\label{secscapprox}

In view of the subsequent application to QCD, where a  fully
self-consistent determination of the gluonic self-energy  seems
prohibitively difficult, {we shall be led to consider
approximations to the gap equation. 
These 
will be constructed such that they reproduce
(but eventually transcend)
the perturbative results up to and including 
order $\lambda^{3/2}$ or $g^3$, which
is the maximum perturbative accuracy allowed
by the approximation  ${\cal S}'=0$. 

In view of this it is important to understand the perturbative
content of the
self-consistent approximations
for $m^2$, $P$ and ${\cal S}$.
In this section we shall demonstrate that, when expanded in powers
of the coupling constant, these approximations reproduce 
the correct perturbative results up to order $\lambda^{3/2}$
\cite{Kapusta}. This will also elucidate how perturbation
theory gets reorganized by the use of the skeleton representation
together with the stationarity principle.

For the scalar theory with only $(\lambda/4!)\,\phi^4$ self-interactions,
we write\footnote{This normalization for $g$ is chosen in
view of the subsequent extension to QCD since it makes
the scalar thermal mass in Eq.~(\ref{SCMHTL}) equal to
the leading-order Debye mass in pure-glue QCD (Eq.~(\ref{MD0}) with $N=3$).}
$\lambda\equiv 24g^2$, and compute the corresponding
self-energy $\Pi= m^2$ by solving the gap equation (\ref{GAP2})
in an expansion in powers of $g$, up to order $g^3$.
Since we anticipate $m$ to be of order $gT$, we can ignore
the second term $\propto \lambda m^2 \sim g^4$ in the r.h.s.
of Eq.~(\ref{GAP2}), and perform a high-temperature expansion of 
the integral $I_T(m)$
in the first term (cf. Eq.~(\ref{ITm})) up to terms linear in $m$.
This gives the following, approximate, gap equation:
\beq\label{APPROXGAP}
 m^2\,\simeq\,g^2T^2-{3\0\pi}\,g^2Tm\,.\eeq
The first term in the r.h.s.\ arises as
\beq
\label{SCMHTL}
24 g^2I_T(0)\,=\,12g^2
\int\!\!{d^3k\0(2\pi)^3}\,\frac{n(k)}{k}\,=\,g^2T^2\,\equiv\,\hat m^2.
\eeq
This is also the leading-order result for $m^2$,
commonly dubbed
the ``hard thermal loop'' (HTL)\footnote{In the following, HTL
quantities will be marked by a hat.}
 \cite{BP,QCD} because the 
loop integral in Eq.~(\ref{SCMHTL}) is saturated by hard momenta
$k\sim T$. 

The second term, linear in $m$, in Eq.~(\ref{APPROXGAP}) comes from
\beq
12 g^2 \int\!\!{d^3k\0(2\pi)^3}\left(
\frac{n(\varepsilon_k)}{\varepsilon_k}-\frac{n(k)}{k}\right)
\,\simeq \,12 g^2 T\int\!\!{d^3k\0(2\pi)^3}\left(\frac{1}
{k^2+ m^2}-\frac{1}{k^2}\right)\,=\,
-\frac{3g^2}{\pi} mT\,,
\eeq
where we have used the fact that
the momentum integral is saturated by soft
momenta $k\sim gT$, so that to the order of interest
$n(\varepsilon_k)\simeq T/\varepsilon_k$ (and similarly
$n(k)\simeq T/k$).
This provides the next-to-leading order (NLO)
correction to the thermal mass
\beq\label{SCNLO}
\delta m^2\,\equiv\,-\frac{3g^2}{\pi}\hat mT\,=\,-\frac{3}{\pi}\,g^3T^2\,.
\eeq

Thus, to order $g^3$, one has $m^2=\hat m^2+\delta m^2$.
In standard perturbation theory\cite{Kapusta,MLB}, 
the first term arises as the
one-loop tadpole diagram evaluated with a bare massless propagator,
while the second term comes from the same diagram where 
the internal line is soft and dressed by the HTL, that is
$\hat D(\o,k) \equiv -1/(\o^2 - k^2 -\hat m^2)$.
(At {\it soft} momenta $k\sim \5m \sim gT$, 
$\hat m^2$  is of the same order as the free inverse propagator
$D_0^{-1}\sim k^2 \sim g^2T^2$, and thus cannot be expanded
out of the HTL-dressed propagator $\hat D(\o,k)$.)

Consider similarly the perturbative estimates for the pressure and 
entropy, as obtained
by evaluating Eqs.~(\ref{Pscal}) and (\ref{SPHI})
with the perturbative self-energy $\Pi =m^2 
\simeq \hat m^2+\delta m^2$, and further
expanding in powers of $g$, to order $g^3$.
The renormalized version of 
Eq.~(\ref{Pscal}) yields, to this order
(recall that $m\sim gT$ and $\lambda\sim g^2$),}
\beq\label{Pdem}
P\,\simeq\,
\frac{\pi^2 T^4}{90}-\frac{m^2 T^2}{24}+\frac{m^3T}{12\pi}+\cdots
+\frac{m^4}{2\lambda}.
\eeq
The first terms before the dots represent the pressure of massive bosons,
i.e. Eq.~(\ref{Pmassbosons}) expanded up to third order in powers of
$m/T$. From Eq.~(\ref{Pdem}), it can be easily verified that the 
above perturbative solution for $m^2$  ensures the stationarity of
$P$ up to order $g^3$, as it should. Indeed, if we denote
\beq
 P_2(m) \,\equiv\,-\frac{m^2 T^2}{24}\,+\,\frac{m^4}{2\lambda}\,,
\qquad\, P_3(m)\,\equiv\, \frac{m^3 T}{12\pi}\,,\eeq
then the following identities hold:
\beq\label{STAT2}
{\del P_2\0\del m}\bigg|_{\5m}\,=\,0,\qquad
{\del P_2\0\del m}\bigg|_{\5m+\delta m}\,+\,
{\del P_3\0\del m}\bigg|_{\5m}\,=\,0.\eeq
This shows that the NLO mass correction $\delta m^2\sim g^3T^2$ 
can be also obtained as
\beq
\delta m^2\,=\,-\,{(\del P_3/\del m)\0(\del^2 P_2/\del m^2)}\bigg|_{\5m}
\,=\,-\frac{3g}{\pi}\,\5m^2\,,\eeq
in agreement with Eq.~(\ref{SCNLO}).
Moreover, $P_2\equiv  P_2(\5m) = -g^2T^2/48$ and $P_3\equiv  P_3(\5m) =
\5m^3 T/12\pi$ are indeed the correct perturbative corrections
to the pressure, to orders $g^2$ and $g^3$, respectively
\cite{Kapusta}. In fact, the pressure to this order can be
written as:
\beq\label{pressurescal}
P&=&\frac{\pi^2 T^4}{90}-\frac{\hat m^2
T^2}{24}\,(1-\frac{3}{\pi}\,g)+\frac{\hat m^3T}{12\pi}+\cdots
\,+\,\frac{\hat m^4}{2\lambda}\,(1-\frac{3}{\pi}\,g)^2 +{\cal O}(g^4)\nonumber\\
&=&\frac{\pi^2 T^4}{90}-\frac{\hat m^2}{48} T^2+\frac{\hat m^3 T}{12\pi}\,.
\eeq
Note that the term of order $g^2$ is only {\it half} 
of that one would obtain from Eq.~(\ref{Pmassbosons}) 
by replacing $m$ by $\hat m$. This is due to the aforementioned
mismatch between  Eq.~(\ref{Pmassbosons}) and the correct expression
for the pressure, Eq.~(\ref{Pscal}). 
In fact, going back to Eq. (2.1), one observes that the net
order $g^2$ contribution to the pressure comes from $\Phi$ evaluated with
bare propagators: the order $g^2$ contributions in the other two terms
mutually cancel indeed. This is to
be expected: there is a single diagram of order $g^2$; this is a skeleton
diagram, counted therefore once and only once in $\Phi$.
Observe also that the
terms of order $g^3$ originating from  the terms  $\hat m^2$ and 
$\hat m^4$ mutually cancel; that is, the NLO mass correction
$\delta m$ drops out from the pressure up to order
$g^3$. This is no accident: the cancellation 
results from the stationarity of $P$ at order $g^2$, the first equation
(\ref{STAT2}).

Consider now the entropy density. The correct perturbative result up to order
$g^3$ may be obtained directly by taking the total derivative of the
pressure, Eq.~(\ref{pressurescal}) with respect to
$T$. One then obtains:
\beq\label{SPERTSC}
{\cal S}=\frac{4}{T}\left( \frac{\pi^2 T^4}{90}
-\frac{\hat m^2 T^2}{48}+\frac{\hat m^3 T}{12\pi}\right)+{\cal O}(g^4).
\eeq

We wish, however, to proceed differently, using Eq.~(\ref{SPHI}), or
equivalently, 
 since $\del P/\del m=0$ when $m$ is a solution of the
gap equation, by writing:
\beq
{\cal S}=\left.\frac{\del P}{\del T}\right|_m.
\eeq
This yields:
\beq\label{Sdem}
{\cal S}=\frac{4}{T}\left( \frac{\pi^2 T^4}{90}-\frac{m^2 T^2}{48}+\frac{m^3
T}{48\pi}\right)+{\cal O}(m^4/T),
\eeq
which coincides as expected with the expression obtained
by expanding the entropy of massive bosons, Eq.~(\ref{SPHI}),
up to order $(m/T)^3$. 
If we now 
replace $m$ by its leading order value
$\hat m$, the resulting approximation 
for ${\cal S}$
reproduces the perturbative effect of order $\sim g^2$,
but it underestimates the correction of order $g^3$ by a factor of 4. 
This is corrected  by changing $m$ to
$\hat m+\delta m$ with
$\delta m=-3g\hat m/2\pi$ in the second order term of Eq.~(\ref{Sdem}). Note
that  although it makes no difference to enforce the gap equation to order
$g^3$ in the pressure (because of the cancellation discussed above), 
there is no such cancellation in the entropy.

In view of the forthcoming application to QCD, we shall now rephrase  the
 previous discussion in slightly more general terms, though still
restricted to the main simplification that the present simple model
offers: a self-energy that is constant and real. 

Because of the stationarity of the thermodynamic potential, 
Eq.~(\ref{selfcons}),
the order $g^3$ term in
the pressure is coming entirely from the $\log $ term in the thermodynamic
potential with $\Pi=\hat \Pi$, which reads:
\beq\label{P3scal}
P_3&=&-\int\!\!{d^4k\0(2\pi)^4}\frac{T}{\omega} \Im\left[
\log(1+D_0(\o,k)\hat\Pi)-D_0(\o,k)\hat\Pi\right] \nonumber\\
 &\approx&-\, 
\frac{T}{2}\int\!\!{d^3k\0(2\pi)^3}\left[
\log\left(1+\frac{\hat\Pi}{k^2}\right)-\frac{\hat\Pi}{k^2}\right]\,=\,
\frac{T}{12\pi}\, \hat\Pi^{3/2},
\eeq
where we have subtracted the order-$g^2$ contribution and used the fact that the
remaining integrand is dominated by soft momenta to replace 
$n(\omega)$ by $T/\omega$.  The corresponding contribution to the entropy
follows as:
\bea\label{SP3}
{\cal S}_3\,=\,\frac{{\rm d} P_3}{{\rm d} T}\,=\,
\frac{{\del} P_3}{{\del} T}\bigg|_{\5\Pi}\,+\,
\frac{{\del} P_3}{{\del}\5\Pi }\bigg|_T\,\frac{{\rm d}\5\Pi}
{{\rm d}T}\,\equiv\,{\cal S}_3^{(a)}+{\cal S}_3^{(b)},\eeq
where ${\cal S}_3^{(a)}$, the derivative of $P_3$ at constant
$\hat\Pi$, equals 1/4 of the total order-$g^3$ entropy. 
The remaining 3/4 come from the derivative of $\hat\Pi$.

Alternatively, the entropy can be obtained from our master
equation (\ref{Ssc}) which, in the
present model where $\Im \,\Pi=0$, simplifies into:
\bea\label{Sscscal}
{\cal S}&=&-\int\!\!{d^4k\0(2\pi)^4}{\6n(\o)\0\6T} \Im \log D^{-1}(\o,k) 
\eea
The term of order $g^2$ is obtained by writing $\log D^{-1}=\log D_0^{-1}+\log
(1+D_0\Pi)$, setting $\Pi=\hat\Pi$ and expanding the logarithm to first order in
$\hat\Pi$. One then obtains:
 \bea\label{Sscscal2}
{\cal S}_2=-\int\!\!{d^4k\0(2\pi)^4}{\6n(\o)\0\6T}\,\hat\Pi\, \Im D_0(\o,k). 
\eea
Since $\Im D_0(\o,k)=\pi\epsilon(\o)\delta(\o^2-k^2)$, the integrand in
(\ref{Sscscal2}) is concentrated on the unperturbed mass-shell. 
The ensuing momentum integral immediately yields
${\cal S}_2=-T\hat\Pi/12$, in agreement with Eq.~(\ref{SPERTSC}).

According to Eq.~(\ref{SP3}), the contribution of order
$g^3$ involves two pieces,
${\cal S}_3={\cal S}_3^{(a)}+{\cal S}_3^{(b)}$ (cf. Eq.~(\ref{SP3})).
These can be also understood as the contributions to Eq.~(\ref{Sscscal})
from different momentum regimes.
Specifically, the soft momenta in the latter yield:
\beq\label{S3soft}
{\cal S}_3^{\mathrm soft}&=&
-\int\!\!{d^4k\0(2\pi)^4}\,\frac{1}{\omega}\,\Im
\Bigl[\log(1+ D_0\hat\Pi) -  D_0\hat\Pi\Bigr]\,,\eeq
which is the same as ${\cal S}_3^{(a)}$ in Eq.~(\ref{SP3}).
The second contribution of order $g^3$ comes from hard momenta 
in Eq.~(\ref{Sscscal}), and is obtained by replacing $\hat\Pi \rightarrow 
\delta\Pi$ in Eq.~(\ref{Sscscal2}). This yields
\beq\label{S3hard}
{\cal S}_3^{\mathrm hard}
&=&{1\02}\,\delta\Pi\int\!\!{d^3k\0(2\pi)^3}\,{1\0k}\,{\6n(k)\0\6T}\,
=\,{1\0\lambda}\,\delta\Pi\,\frac{{\rm d}\5\Pi}{{\rm d}T}\nn
&=&{T\02}\,\frac{{\rm d}\5\Pi}{{\rm d}T}
\int\!\!{d^3k\0(2\pi)^3}\left(\frac{1}{k^2+\hat\Pi}-\frac{1}{k^2}\right)\nn
&=&
\int\!\!{d^4k\0(2\pi)^4}\,{T\0\o}\,\Im\biggl[
\frac{{\rm d}\hat \Pi}{{\rm d} T}\,(\hat D -D_0)\biggr]={\cal S}_3^{(b)},\eeq
where we have used Eq.~(\ref{SCMHTL})
for $\hat \Pi$ in the first line and Eq.~(\ref{SCNLO}) for 
$\delta\Pi$ in the second line.

\subsection{Approximately self-consistent solutions}
\label{secscnum}

{
As we have seen, the 2-loop $\Phi$-derivable approximation
provides an expression for the entropy ${\cal S}$ as a functional
of the self-energy $\Pi$ --- namely, Eq.~(\ref{Ssc}) with
${\cal S}'=0$ --- which has a simple quasiparticle
interpretation and is manifestly ultraviolet finite for any
(finite) $\Pi$. These attractive
features of Eq.~(\ref{Ssc}) are independent of the specific form of
the self-energy, and will be shown to hold in QCD as well. 
Of course, within this approximation, the self-energy
is uniquely specified: by the stationarity principle, this is 
given by the self-consistent solution to the one-loop gap equation.
In the scalar $\phi^4$-model, it was easy to give the exact 
solution to this equation (cf. Sect. 2.B), which coincides
with the well-known solution of a scalar O($N$)-model 
in the limit $N\to\infty$ \cite{DHLR2}.
In QCD, however, it will turn out that a fully self-consistent 
solution is both prohibitively difficult (because of the 
non-locality of the gap equation), and not really desirable
(for reasons to be discussed in Sect. 3.B below). This leads 
us to consider
{\it approximately self-consistent} resummations, which are
obtained in two steps: (a) An approximation is constructed for
the solution $\Pi$ to the gap equation, and (b) the entropy
(\ref{Ssc}) is evaluated {\it exactly} (i.e., numerically)
with this approximate self-energy. While step (b) above is
unambiguous and inherently nonperturbative, 
step (a), on the other hand, will be constrained
primarily by the requirement of preserving the
maximum possible perturbative accuracy, of order $g^3$ 
(cf.\ Sect.\ 2.C). In addition to that, we shall add the qualitative
requirement that the approximation for $\Pi$, and the
ensuing one for ${\cal S}$, are well
defined and physically meaningful for all the values of $g$ of
interest, and not only for small $g$---that is, for all the values
of $g$ where the fully self-consistent calculation makes sense a priori.
As we shall shortly see, this last requirement generally excludes 
a strictly perturbative solution to the gap equation.

Of course, even with this last requirement, there is still a
large ambiguity in the choice of the approximate self-energy.
In this respect the scalar $\phi^4$-model
provides an opportunity for testing the quality of
these approximations against the exact solution of the gap equation
of the fully self-consistent two-loop calculation.
Similar approximations will be subsequently used in QCD.
}

\begin{figure}
\epsfxsize=8.5cm
\centerline{\epsfbox[70 180 540 540]{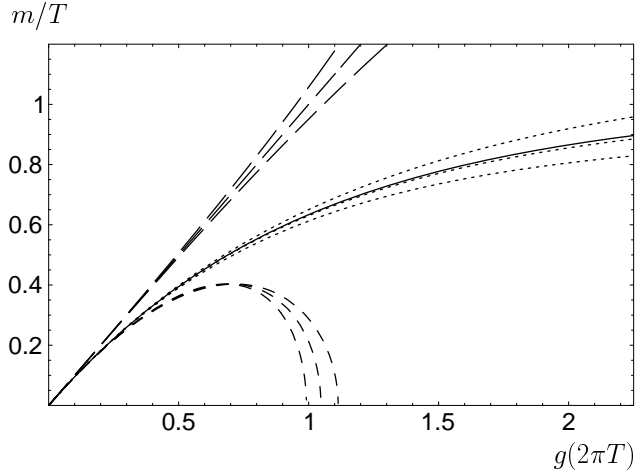}}
\caption{Comparison of the exact thermal mass in the large-$N$
scalar O($N$)-model as a function of $g(\bar\mu=2 \pi T)$ 
(full line) with the leading-order (HTL) perturbative result 
(long-dashed lines), the NLO one corresponding to
$\hat m^2+\d m^2$ (shorter-dashed lines),
and the perturbatively equivalent {(NLA)} version
(\ref{masnlo}) (dotted lines). Except for the exact result,
all these are renormalization scale dependent, the central lines
corresponding to $\bar\mu'=2\pi T$, the adjacent ones
to $\pi T$ and $4\pi T$.
\label{figdm}}
\end{figure}

The exact solution\footnote{More precisely, as 
discussed in detail in Ref.~\cite{DHLR2}, Eq.~(\ref{GAP2})
has two solutions, a fact that is frequently overlooked.
The larger of the two is
exponentially larger than $T$ for small coupling and
has to be ruled out because our scalar model is consistent only
as an effective (cut-off) theory.} 
of the gap equation
is determined
by the transcendental
Eq.~(\ref{GAP2}) with $\lambda\equiv 24g^2$. {With
$\bar\mu=2\pi T$, the} result $m/T$ as
a function of $g$ is given
by the full line in Fig.~\ref{figdm}. {As an exact result, it
is independent of the renormalization scale:
a change of  $\bar\mu \to \bar\mu'$ has to be followed
by a change of the renormalized coupling $g(\bar\mu)\to g(\bar\mu')$
according to (cf.\ Eq.\ (\ref{RGSC}))\footnote{So 
the scalar theory is fully defined by giving
both a dimensionful scale $\bar\mu$ and the associated coupling
strength $g(\bar\mu)$. Equivalently, as usually done in QCD,
we could just give a scale
$\Lambda_\phi$ and agree e.g.\ that $g( \Lambda_\phi)=+\infty$.
In this section we shall take the former point of view, 
so for any given temperature
$T$, different values of $g(2\pi T)$ parametrize differently coupled
theories.}
}
\be\label{largeNgrun}
g^2(\bar\mu')=g^2(\bar\mu)\left[1+g^2(\bar\mu)(3/2\pi^2)\log(\bar\mu/\bar\mu')
\right]^{-1}.
\ee

\firstfigfalse

{All perturbative results on the other hand
suffer from the problem of
renormalization scheme dependence, the more so
the stronger the coupling. 
Having settled for the $\overline{\hbox{MS}}$-scheme,
all of the remaining ambiguity is in the choice of the
renormalization scale $\bar\mu'$. 
Throughout this paper,}
we shall 
choose $\bar\mu=2\pi T$ as our fiducial scale
and consider the range $\bar\mu'=\pi T\ldots 4 \pi T$ to
test for the scheme dependence of the various approximations.\footnote{In
Ref.~\cite{DHLR2}, from which we deviate slightly in taking $\bar\mu=2\pi T$
rather than $\bar\mu=T$ as the fiducial scale, 
the scheme dependence of thermal perturbation
theory has been studied in the above scalar model in great detail
with the result that at least at high orders of perturbation theory
$\bar\mu' \sim 2\pi T$ seems to be an optimal renormalization point,
corroborating the expectations expressed in Ref.~\cite{BN}.}

The leading-order (HTL) result, Eq.~(\ref{SCMHTL}), is simply
$m/T=\hat m/T=g$. For $g=g(2\pi T)$, this is the straight long-dashed line
in Fig.~\ref{figdm}. For the different choices $\bar\mu'=\pi T$ and
$4\pi T$, $g$ is instead the function of
$g(2\pi T)$ given by Eq.~(\ref{largeNgrun}) and $m/T$ is given by
the long-dashed lines below and above the central one.

The NLO correction (\ref{SCNLO}) is negative, 
eventually making the
perturbative result for $m^2=\hat m^2+\d m^2$ negative, in fact
already at moderately
large coupling $g\approx 1$
(shorter-dashed lines in Fig.~\ref{figdm}, individually corresponding
to $\bar\mu'=\pi T,2\pi T,4\pi T$ again).
Clearly, using this strictly perturbative result would make the
thermodynamic potentials fall back to the free result 
at $g=\pi/3$ where $\hat m^2+\d m^2$ vanishes,
and give rise to tachyonic singularities beyond.

However, there is no unique ``strictly perturbative'' result.
Defining a NLO mass through $m=\hat m+\d m$ 
would involve $\delta m \equiv \delta m^2/2\hat m$. This would
lead to an obvious breakdown of perturbation theory only for twice
as large values of $g$,
$g>2\pi/3\approx 2$, with 
negative rather than imaginary values for $m$.

{But this does not mean that there is no physical content in the
NLO effects beyond $g\sim 1$. Rather, the physical content is 
unnecessarily lost
by the restriction to a polynomial result for $m^2$ (or $m$)
which does not preserve the monotonous behavior of $m/T$ as a function of $g$ 
that is observed both at leading order and in the exact result.

\begin{figure}
\epsfxsize=8.5cm
\centerline{\epsfbox[70 200 540 640]{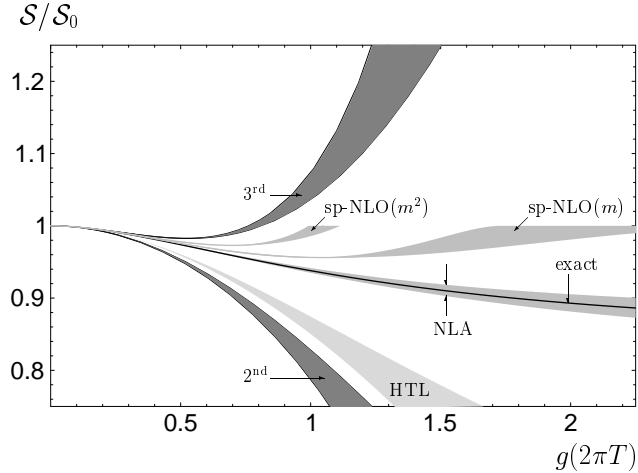}}
\vskip0.5cm
\caption{Comparison of perturbative and HTL-improved approximations to
the entropy in the large-$N$ scalar O($N$)-model. {The shaded areas
denote the variation under changes of the renormalization scale
from $\bar\mu'=\pi T$ to $4\pi T$. The band marked ``HTL'' refers
to using the leading-order (HTL) mass in the 2-loop $\Phi$-derivable
entropy, ``NLA'' to using the approximately self-consistent
NLO mass (\ref{masnlo}). Also given are the corresponding
results for a naive strictly perturbative NLO mass when defined
through $m^2$ or $m$, respectively.}
\label{figsccomp}}
\end{figure}

In order to ensure such a monotonous behavior,
in Refs.~\cite{PRL,PLB} we
have considered the simple Pad\'e approximant
$\hat m^2+\d m^2 \to \hat m^2/[1-\d m^2/\hat m^2]=g^2T^2/[ 1+3g/\pi]$,
which 
already achieves
a dramatic improvement for $g\gtrsim 1$.
An alternative, which is in fact more in the spirit of
approximate self-consistency, is to return to the approximate gap
equation (\ref{APPROXGAP})
\be\label{m2corr}
m^2\,=\,g^2T^2-{3\0\pi}\,g^2Tm\,,
\ee
and solve this quadratic equation for $m$ exactly, yielding
\be\label{masnlo}
m_{\rm NLA}/T=\sqrt{g^2+(3g^2/2\pi)^2}-3g^2/2\pi.
\ee
In what follows, this will be referred to as our
``next-to-leading approximation'' (NLA) for the scalar thermal mass.
Also this approximation} preserves the propertym of being
a monotonously growing function of $g$. {For very
large $g$ it saturates at $m_{\mathrm NLA}\to (\pi/3)T$.} 
The corresponding
results for the various renormalization scale choices are given by the
dotted lines in Fig.~\ref{figdm}, showing
a {striking} improvement over the standard perturbative results
also for {very large} coupling.

With $m$ approximated either by its leading-order (HTL) value 
$\hat m=g(\bar\mu)T$ or by the NLA result (\ref{m2corr}), the
correspondingly approximated entropy is obtained by
evaluating numerically the expression (\ref{Sscphi4}).
In Fig.~\ref{figsccomp} this is compared
with the strictly perturbative expressions for ${\cal S}/{\cal S}_0$
up to and including order $g^2$, and $g^3$, respectively.\footnote{This
plot differs from the corresponding one presented in Ref.~\cite{PRL}
in that in the latter the fiducial renormalization scale $\bar\mu=T$
has been used, so the abscissae are non-linearly related.}
The shaded bands indicate the 
variation of the results with $\bar\mu'=\pi T\ldots 4\pi T$.
Evidently, the perturbative 3rd-order result fails to be
a better approximation than the 2nd-order one for $g\gtrsim1$.
{The semi-perturbatively evaluated HTL result is already an appreciable
improvement over the 2nd-order perturbative result, whereas
the NLA follows closely the exact ($N\to\infty$) result.
Also shown are the results
corresponding to the two ``strictly perturbative'' NLO mass definitions
mentioned above when used in the same manner.
}

\section{QCD: Approximately self-consistent resummations}

We turn now to our main case of interest, the QCD plasma.
In this section, we shall concentrate on a purely gluonic plasma,
deferring the addition of quarks to the next section.
Although the thermodynamic potential in QCD is a gauge independent
quantity, in writing down its skeleton representation
we have to specify a gauge.
In formulating the two-loop $\Phi$-derivable approximation
we find it convenient to start with the temporal axial gauge.
While this approximation is by itself gauge dependent, when
supplemented by perturbative approximations on the generalized
gap equation it results in a gauge invariant resummation scheme
for the entropy.

\subsection{The skeleton representation of the entropy}

\begin{figure}
\epsfxsize=9cm
\centerline{\epsfbox[40 310 505 550]{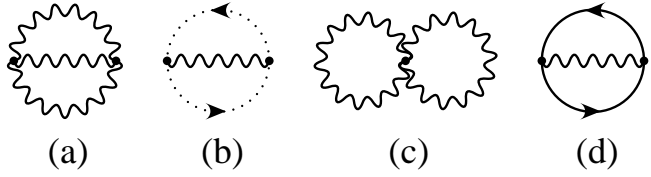}}
\vskip0.5cm
\caption{Diagrams for $\Phi$ at 2-loop order in QCD.
Wigly, plain, and dotted lines refer respectively to
gluons, quarks, and ghosts.
\label{figphiqcd}}
\end{figure}

In QCD, the thermodynamic potential is a functional of the full
gluon ($D$), quark ($S$), and Faddeev-Popov ghost ($D_{gh}$) propagators,
\bea\label{LWQCDwgh}
\b \O[D,S,D_{gh}]&=&\2 \Tr \log D^{-1} - \Tr \log S^{-1} - \Tr \log D^{-1}_{gh}
\nn&&
- \2 \Tr \Pi D + \Tr \S S + \Tr \Pi_{gh} D_{gh} + \Phi[D,S,D_{gh}],
\eea
where Tr now includes traces over color indices, and also over
Lorentz and spinor indices when applicable. The self-energies
for gluons, quarks and ghosts are denoted respectively by
$\Pi$, $\S$ and $\Pi_{gh}$.
In Fig.~\ref{figphiqcd}, the lowest-order (two-loop) skeleton diagrams
for $\Phi$ are displayed.

In gauges which do not break rotational invariance, the gluon propagator
at finite temperature contains up to four different structure functions
\cite{KKHKT}.
Only two of them correspond to degrees of freedom which are transverse
in 4 dimensions; the remaining ones are unphysical,
constrained by a Ward identity \cite{KKM}, and compensated
for by the Faddeev-Popov ghost degrees of freedom.

In general, the gluon self-energy $\Pi_{\mu\nu}(k)$
is a tensor which is not transverse
with respect to the 4-momentum $k^\mu=(\o,{\bf k})$, but also contains up to
4 structure functions. There are however gauges where ghosts
decouple and where as a consequence $\Pi_{\mu\nu}$ is strictly 
transverse\footnote{This property can nevertheless be lost in approximations
which do not preserve gauge symmetry; cf. the discussion after
Eq.~(\ref{SP0QCD}).}:
axial gauges $n_\mu A^{a\mu}=0$, with $n_\mu$ a constant 4-vector.

A particularly convenient choice appears to be the temporal axial gauge,
where $n_\mu$ coincides with the rest-frame
velocity of the heat bath and thus preserves rotational invariance.
Ignoring the well-known difficulties with this gauge in the
imaginary-time formalism \cite{JL}, the temporal axial gauge
would lead to great simplifications of the structure of Eq.~(\ref{LWQCDwgh}):
The ghost self-energy $\Pi_{gh}$ vanishes and the ghost propagator does not
appear in $\Phi$. Secondly, there are only two independent
structure functions in the gluon self-energy, which can then be
written as (suppressing the color labels)
\bea
&&\label{Piij}
\Pi_{ij}(\o,k)=\( \d_{ij}-{k_i k_j\0k^2}\)\Pi_T(\o,k) 
-{k_i k_j \o^2 \0 k^4}\Pi_L(\o,k),\\
\label{Pitransv}
&&\Pi_{00}(\o,k)=-\Pi_L(\o,k),\qquad
\Pi_{0i}(\o,k)=-{\o k_i \0 k^2}\Pi_L(\o,k).
\eea
With these definitions, the propagator in temporal axial gauge
reads
\be\label{DijTAG}
D_{ij}^{\rm TAG}(\o,k)=\( \d_{ij}-{k_i k_j\0k^2}\) D_T(\o,k)
+{k_i k_j\0k^2} {k^2\0\o^2} D_L(\o,k)
\ee
where
\beq\label{DLT}
D_T(\o,k)\,\equiv\,\frac{-1}{\o^2-k^2-\Pi_T(\o,k)}\,,\qquad
D_L(\o,k)\,\equiv\,\frac{-1}{k^2+\Pi_L(\o,k)}\,.\eeq
Note that because $D^{\rm TAG}_{0\mu}=D^{\rm TAG}_{\mu0}\equiv0$,
only the spatial components $\Pi_{ij}$ of the polarization
tensor enter Eq.~(\ref{LWQCDwgh}) in temporal axial gauge.

For later use we introduce the following spectral representations:
\beq\label{Dspec}
D_T(\omega, k)&=&\int_{-\infty}^{\infty}\frac{{\rm d}k_0}{2\pi}
\,\frac{\rho_T(k_0, k)}{k_0-\omega}\,,\nonumber\\
D_L(\omega, k)&=&-\,\frac{1}{k^2}
\,+\int_{-\infty}^{\infty}\frac{{\rm d}k_0}{2\pi}
\,\frac{\rho_L(k_0, k)}{k_0-\omega}\,.\eeq
Here $\rho_T$ and $\rho_L$ are the spectral densities:
\beq\label{rhos}
\rho_{L,T}(k_0,k) \,\equiv\, \lim_{\eta\to0}
2\,{\rm Im}\,D_{L,T}(k_0+i\eta ,k)\,.\eeq
[Note the subtraction performed in the spectral representation
of $D_L(\omega,k)$: this is necessary since
$D_L(\omega,k)\to -1/k^2$ as $|\omega|\to \infty$.
At tree-level, $\rho_L^{(0)}=0$ and
$\rho_T^{(0)}=2\pi\epsilon(k_0)\delta(k^2)$, and therefore
$D_T^{(0)}=-1/(\o^2-k^2)$ and $D_L^{(0)}=-1/k^2$.]

Concentrating on the gluonic contributions for now and postponing
the inclusion of fermions to the next section, we obtain
in analogy to Eq.~(\ref{Omega(D)})
\bea\label{Omegag}
\Omega_g/V&=&
N_g \int\!\!{d^4k\0(2\pi)^4}\, n(\omega) \biggl\{
2 \left(\Im \log(-\omega^2+k^2+\Pi_T)-
\Im \Pi_T D_T \right) \nonumber\\
&&\qquad\qquad + \left(\Im \log(k^2+\Pi_L)+
\Im \Pi_L D_L \right) \biggr\}
+T\Phi_g[D_T,D_L]/V
\eea
where $N_g$ is the number of gluons ($N^2-1$ for SU($N$), 
i.e.\ 8 for QCD).\footnote{Here
we have assumed a principal-value treatment of the factor
$k^2/\o^2$ in Eq.~(\ref{DijTAG}) for the contour integration.
Because this factor is real and positive, it can be dropped
from within the imaginary part of the logarithm
involving $\Pi_L$.}
The entropy of purely gluonic
QCD can then be written in complete analogy to the derivation of
Eq.~(\ref{Ssc}) as
\be\label{SQCD}
{\cal S}={\cal S}_T\,+\,{\cal S}_L+{\cal S}'\,\ee
where 
\begin{mathletters}
\bea\label{SQCDT}
{\cal S}_T&=& -2N_g\int\!\!{d^4k\0(2\pi)^4}{\6n(\o)\0\6T}
\Bigl\{\Im\log(-\o^2+k^2+\Pi_T) -\Im\Pi_T\Re D_T\Bigr\},\\
\label{SQCDL}
{\cal S}_L&=& -N_g\int\!\!{d^4k\0(2\pi)^4}{\6n(\o)\0\6T}
\Bigl\{\Im\log(k^2+\Pi_L)+\Im\Pi_L \Re D_L\Bigr\},
\eea
\label{SQCDD}
\end{mathletters}
and
\be\label{SP0QCD}
{\cal S}'\equiv -{\6(T\Phi)\0\6T}\Big|_D+N_g
\int\!\!{d^4k\0(2\pi)^4}{\6n(\o)\0\6T} \(
2\Re\Pi_T \Im D_T-\Re\Pi_L \Im D_L\). 
\ee

As in the scalar case, we are interested in the $\Phi$-derivable approximation
obtained by keeping only the two-loop skeletons of Fig.~\ref{figphiqcd}.
In gauge theories, however, the $\Phi$-derivable approximations
have in general the drawback of violating gauge symmetry, because
vertex functions are not treated on equal footing with self-energies
(in particular, in the two-loop
approximation to $\Phi$ there are no vertex corrections at all). 
Thus the corresponding approximation to the
polarization tensor $\Pi_{\mu\nu}$ needs not be
transverse. Nevertheless, in the temporal axial gauge, the previous expressions
are not affected by a loss of 4-dimensional transversality, because
they involve only the spatial components $\Pi_{ij}$, or equivalently
$\Pi_T$ and $\Pi_L$ (cf. Eq.~(\ref{Piij})).

Therefore, in this gauge, the property that ${\cal S}'=0$
in the two-loop approximation to $\Phi$ still holds,
for the same, essentially combinatorial reasons as in the scalar
field theory with cubic and quartic interactions of the previous
section. In this approximation,
the self-energies $\Pi_T$, $\Pi_L$ and propagators
$D_T$, $D_L$ are to be determined {\it self-consistently},
by solving the generalized ``gap equations''
\be\label{Dyson1}
D^{-1}_T\,=\,-\o^2+k^2+\Pi_T[D_T,D_L]\,,\qquad
D^{-1}_L\,=\,-\,k^2\,-\,\Pi_L[D_T,D_L]\,,\ee
i.e., the Dyson equations where $\Pi_s[D_T,D_L]$
($s=T,\,L$) are the {\it one-loop} self-energies built out of $D_T$
and $D_L$.

Whereas the entropy expressions (\ref{SQCDD})
themselves are manifestly UV finite, Eqs.~(\ref{Dyson1})
contain UV divergences which require renormalization. 
Because of the
simple Ward identities of axial gauges, (wave function) renormalization of the
gluon self-energy at lowest order in $g$ 
contains the correct one-loop coefficient of the beta function
\cite{Kummer,FrJCT}.
Beyond lowest order, however,
it is not clear that the gap equations  (\ref{Dyson1}) can
be renormalized in a simple manner
(in contrast to the scalar toy model of Sect.~\ref{secsimplemodel}).

{At any rate, in general gauges the 2-loop $\Phi$-derivable approximation
misses the correct perturbative running of the coupling constant. Indeed,
the latter is an order-$g^4$ effect in the thermodynamic potentials and
is thus beyond the perturbative accuracy of a 2-loop 
$\Phi$-derivable approximation.}

\subsection{Approximately self-consistent solutions}

Unlike the scalar theory with
$\lambda\phi^4$ interactions, in QCD the ``gap equations'' (\ref{Dyson1})
are non-local, which makes their exact solution prohibitively difficult. 
But in fact, as we have just explained, uncertainties concerning
gauge symmetry and renormalization beyond order $g^3$ make 
such a fully self-consistent solution not really desirable.

For this reason we shall construct {\em approximately self-consistent}
solutions which maintain equivalence with conventional perturbation
theory up to and including order $g^3$ (the maximum perturbative 
accuracy allowed by two-loop approximations for $\Phi$), and
which are manifestly gauge-independent and UV finite.
After such approximations---where
the gluon polarization tensor is transverse and the ghost self-energy
(in gauges with ghosts) is neglected---, 
Eqs.~(\ref{SQCDD})
have the same formal structure in any other gauge,
and ${\cal S}'=0$ to the same accuracy. We can therefore drop the
restriction to the somewhat problematic temporal axial gauge.
For instance, in the more commonly used Coulomb gauge 
the gauge propagator is given by
\be\label{DCG}
D_{00}^{\rm CG}(\o,k)=D_L(\o,k),\quad
D_{ij}^{\rm CG}(\o,k)=\( \d_{ij}-{k_i k_j\0k^2}\) D_T(\o,k)
\ee
and the ghost propagator does not contribute as long as there is
no nontrivial ghost self-energy;
in covariant gauges under the same circumstances,
the then propagating ghosts just compensate for an additional
massless pole that is present in the gluon propagator.

With the gauge-independent
approximations for $\Pi_{L,T}$ that we shall 
obtain from (HTL) perturbation theory,
the effectively one-loop expressions for the entropy, 
Eqs.~(\ref{SQCDD}),
constitute a gauge-invariant approximation to the full entropy.
By then computing
exactly these expressions, we shall obtain a gauge-invariant result
which is nonperturbative in the coupling $g$, while
being equivalent
to conventional resummed perturbation theory up to and including
order $g^3$. 

As generally with thermal field theories \cite{BIO,MLB}, the perturbative
solution of Eqs.~(\ref{Dyson1}) requires to distinguish between soft 
($k\lesssim gT$) and hard ($k\sim T$) fields, which are dressed differently
by thermal fluctuations.
In (purely gluonic) QCD, and in the Coulomb gauge,
the hard fields are always transverse, while the soft fields
--- which may be seen as
collective excitations of the former \cite{QCD,BIO} 
--- can be either longitudinal, or transverse. 

Because of the limited phase-space, the leading order (LO)
contribution of the {\it soft} modes to the thermodynamical
functions is already
of order $g^3$ \cite{Kapusta}, so the corresponding self-energies are
needed only to leading order in $g$. These are the so-called
{\it hard thermal loops} $\5\Pi_L$ and $\5\Pi_T$
\cite{KKW,BP}, which in the present
formalism appear as the solutions to Eqs.~(\ref{Dyson1}) to LO in $g$
and for soft $(k\sim gT)$ external momenta. They read:    
\bea
\label{PiL}
\5\Pi_L(\o,k)&=&\5m_D^2\left[1-{\o\02k}\log{\o+k\0\o-k}\right],\\
\label{PiT}
\5\Pi_T(\o,k)&=&\frac{1}{2}\left[\5m_D^2+\,\frac{\o^2 - k^2}{k^2}\,
\5\Pi_L\right],
\eea
with the Debye mass 
\beq\label{MD0}
\hat m^2_D&=&-\,\frac{g^2N}{\pi^2}\int_{0}^\infty 
{\rm d}k \,k^2\,\frac{\6n}{\6k}\,=\,{g^2 T^2 N\03}\,.\eeq
The HTL's (\ref{PiL}) are manifestly UV finite: they derive from
one-loop Feynman graphs, but involve only the contribution
of the {\it thermal} fluctuations in the latter (as opposed to
the {\it vacuum} fluctuations, which are responsible
for UV divergences).
The corresponding propagators are then defined via the Dyson equations
(\ref{Dyson1}): 
\beq\label{DHTL}
\5D_T^{-1}(\o,k)\,=\,-\o^2+k^2+\5\Pi_T(\o,k)\,,\qquad
\5D_L^{-1}(\o,k)\,=\,-k^2 -\5\Pi_L(\o,k)\,,\eeq 
Note that, for $k\sim gT$, the self-energy corrections
in Eqs.~(\ref{PiL})--(\ref{DHTL}) are as important as the corresponding 
tree-level inverse propagators $D_0^{-1} \sim k^2 \sim g^2T^2$.
Thus, at soft momenta, the self-energies cannot be expanded
out of the HTL-resummed propagators.
The HTL spectral densities consist of quasiparticle poles
at time-like momenta and Landau damping cuts for $|\o|<k$.
When  $k\gg gT$, the transverse pole describes the usual single-particle
excitations (hard transverse gluons), while the additional pole
associated to the collective longitudinal excitation has exponentially
vanishing residue \cite{P}.

\begin{figure}
\epsfxsize=12cm
\centerline{\epsfbox[150 400 500 520]{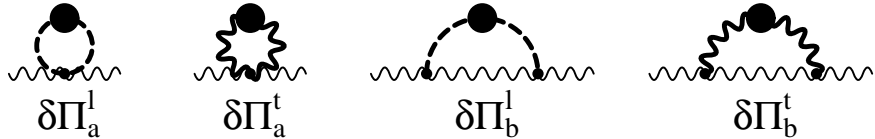}}
\vskip0.5cm
\caption{NLO contributions to $\d\Pi_T$ at hard momentum. Thick
dashed and wiggly lines with a blob represent HTL-resummed longitudinal
and transverse propagators, respectively.
\label{figdPit}
}
\end{figure}

For {\it hard}, transverse, fields, we need the solution $\Pi_T(k\sim T)$
of Eqs.~(\ref{Dyson1}) to leading, and next-to-leading order (NLO).
This is obtained as:
\be\label{PERTDH}
\Pi_T(k\sim T) \simeq \Pi_T^{(2)} + \delta \Pi_T,\ee
where $\Pi_T^{(2)}\equiv
\Pi_T[D_0]\sim g^2$ is the {\it bare} one-loop self-energy 
(i.e., the standard one-loop diagrams with tree-level
propagators $D_0=(D_T^{(0)},D_L^{(0)})$ on the internal lines), and
$\delta \Pi_T \sim g^3T^2$ is an {\it effective} one-loop self-energy
where one of the internal lines is hard (and transverse), while the other
one is soft (longitudinal or transverse) and dressed by the HTL. 
Thus, $\delta \Pi_T$ is the sum of the four diagrams depicted
in Fig.~\ref{figdPit};
these are explicitly computed in App.\ \ref{appdPit}.

A priori, the one-loop self-energy involves also
vacuum fluctuations, and therefore UV divergences, which call
for renormalization. The UV divergences could be absorbed by a
wave-function renormalization constant, which drops out from the entropy
expressions (\ref{SQCDD}). 
As it will turn out presently, only the light-cone
limit of $\Pi_T[D_0]$ will contribute to the order of interest.
In line with our strategy of restricting to gauge-invariant
approximations to the self-energy, we shall altogether drop the
gauge-dependent vacuum pieces, which in fact vanish on the light-cone.

Because from standard HTL perturbation theory we take
UV finite approximations for $\Pi_{L,T}$, we shall in fact have no
inherent beta function\footnote{A refinement
of the present approach which is accurate at and above order $g^4$ and
which has correct (lowest-order) coupling constant 
renormalization would require
at least a 3-loop approximation to the thermodynamic potentials.}
prescribing the scale dependence of the
coupling $g$. 
When numerically evaluating the results,
we shall simply adopt the standard running
coupling constant of the $\overline{\hbox{MS}}$ scheme
and consider the resulting renormalization-scale
dependence of our results as an estimate of our theoretical error
(cf.\ Sect.\ \ref{secscnum}).

\subsection{Perturbation theory: Lowest orders}
\label{secg2}

In this and the following subsections, we shall
consider the perturbative expansion
of our master equation for the entropy, 
Eqs.~(\ref{SQCDD}), and recover in the process
the standard perturbative results up to order $g^3$.
This is useful not only as a cross check of the various 
approximations, but also as an illustration of the rather non-trivial 
way that perturbation theory gets reorganized by this equation.
Moreover, the perturbative expansion will shed more light on the
physical interpretation of the various terms in Eqs.~(\ref{SQCDD}),
and give us hints for better approximations to be used in
the non-perturbative, numerical calculations to come.

The leading-order result is obtained
by putting $\Pi_T=\Pi_L=0$ in Eqs.~(\ref{SQCDD}). This
is the Stefan-Boltzmann entropy of a free gas of massless transverse gluons:
\beq\label{SB}
{\cal S}_{SB}&=& -2N_g\int\!\!{d^4k\0(2\pi)^4}{\6n(\o)\0\6T}
\,\Im\log(-\o^2+k^2) \nn
&=& -2N_g\int\!\!{d^3k\0(2\pi)^3} {\6\0\6T}
\Bigl[T\log(1- e^{-k / T})\Bigr] =\frac{4\pi^2}{45}\, N_gT^3.
\eeq
Here the retarded prescription ($\omega \to \omega +i\epsilon$)
is implicit in the first integral, which is
evaluated with the help of the identities (\ref{IM0})
and (\ref{ID1}).

The order-$g^2$ contribution to the entropy comes
also exclusively from hard
transverse gluons, via one-loop corrections. Specifically,
by  expanding Eq.~(\ref{SQCDT}) to order $g^2$, one obtains:
\bea\label{SG2}
{\cal S}_2 &=&-2N_g\int\!\!{d^4k\0(2\pi)^4}\,{\6n\0\6T}\left\{-
\Im\frac{\Pi_T^{(2)}}{\o^2-k^2}+\Im{\Pi_T^{(2)}}\Re\frac
{1}{\o^2-k^2}\right\}\nonumber\\
 &=&2N_g\int\!\!{d^4k\0(2\pi)^4}\,{\6n\0\6T}\,\Re{\Pi_T^{(2)}}
\Im\frac{1}{\o^2-k^2}\nonumber\\
&=&-2\pi N_g\int\!\!{d^4k\0(2\pi)^4}\,{\6n\0\6T}\,
\e(\o)\d(\o^2-k^2)
\Re\Pi_T^{(2)}(\o,k),\eeq
where the integral is indeed dominated by hard
momenta $k\sim T$.
Note that ${\cal S}^{(2)}$ involves only the light-cone
projection $\Re\Pi_T^{(2)}(\o=k)$ of the one-loop self-energy
for (hard) transverse gluons $\Pi_T^{(2)}(\o,k)$. This projection
is a priori UV finite: indeed, gauge symmetry guarantees that the
vacuum contribution to $\Re\Pi_T^{(2)}(\o=k)$ must vanish.
Moreover, quite remarkably, this projection turns out
to be also momentum-independent \cite{KKR},
\be\label{minfty}
\Pi_T^{(2)}(\o^2=k^2)=g^2NT^2/6 \equiv m_\infty^2,
\ee
and thus defines a (thermal) mass correction, also known 
as the {\it asymptotic mass}. Thus, finally,
\bea\label{S2}
{\cal S}_2 &=&-N_g{m_\infty^2T\06}=-{NN_g\036}g^2T^3,
\eea
which is indeed the correct result \cite{Kapusta}.
Note also that at leading order
the asymptotic mass is simply related to the (HTL)
Debye mass: $m_\infty^2 = \hat m^2_D/2$.

It is worth emphasizing that Eq.~(\ref{SG2}) is the same as
the entropy of an ideal gas of massive particles (with constant
masses equal to $m_\infty$) when expanded to leading order in 
$m_\infty^2$. As was the case in the scalar model discussed in
Sect. II, such a simple identification is specific to the
entropy, and does not hold for the order-$g^2$ effect in
the pressure.

In the scalar case we have seen that the HTL-resummed one-loop
pressure over-includes the LO interaction term by a factor of
two. For gluons, Ref.~\cite{ABS} reported instead a factor of
three. Inspecting the corresponding calculation reveals that this
arises because of an incomplete implementation of dimensional regularisation.
While in the latter
$2-2\e$ transverse polarisations of the gluons are considered,
the HTL expressions for $\Pi_{\mu\nu}$ have not been modified accordingly.
However in $d\not=3$ spatial dimensions, Eqs.~(\ref{PiT},\ref{PiL})
become
\be
\hat\Pi_T={1\0d-1}\left[\5m_D^2+\frac{\o^2 - k^2}{k^2}\,
\5\Pi_L\right],\quad
\hat\Pi_L=\5m_D^2\left[1- \,
{}_2F_1(\2,1;{d\02};{k^2\0\omega^2}) \right],
\ee
where 
\be
\hat m_D^2 
=(d-1)g^2 N T^2({T\0\mu})^{d-3}
 {\zeta(d-1) \Gamma({d+1\02}) \0 \pi^{(d+1)/2} } \qquad (d>2)
\ee
as determined by the $d$-dimensional analog of Eq.~(\ref{MD0}).
This gives a real and constant $\hat\Pi^\mu_\mu=\hat m_D^2=(d-1)m_\infty^2$
such that the order-$g^2$ contribution to the 1-loop HTL-resummed
pressure $P_{HTL}=-{1\02}\Tr\log(D_0^{-1}+\hat\Pi)$ is
\be
P_{HTL}^{(2)}=N_g\int\!\!{d^{d+1}k\0(2\pi)^{d+1}}\,n(\o)
\Im\frac{\hat\Pi^\mu_\mu}{\o^2-k^2}=
N_g \hat m_D^2\int\!\!{d^{d+1}k\0(2\pi)^{d+1}}\,n(\o)
\Im\frac{1}{\o^2-k^2} \to 2\times P_2
\ee
as $d\to 3$, with dimensional regularization eliminating the
quadratic divergence for $\o\to-\infty$.
This is then consistent with momentum cut-off regularization,
where $d=3$ can be kept throughout,
after dropping a divergence $\propto \hat m_D^2 \Lambda^2$.
Presumably, the numerical results reported in Ref.~\cite{ABS}
will change significantly when corrected accordingly.

This sensitivity to (a consistent usage of) regularization schemes
is related in fact to the UV behavior of HTL-screened perturbation
theory; it is not present in our
UV-finite HTL-resummation of (two-loop)
entropy and density.

\subsection{Perturbation theory: Order $g^3$}
\label{secglueg3}

The extraction of the order-$g^3$ contribution to the entropy
in Eq.~(\ref{SQCDD}) turns out to be more intricate than the
standard calculation of the plasmon effect in the pressure
\cite{Kapusta}. 

\subsubsection{The order $g^3$ in the pressure}

Let us
briefly discuss first the plasmon effect in the pressure, as 
obtained from the skeleton representation (\ref{LW}).
As explained for the scalar case in Sect. \ref{secscapprox}, 
the order-$g^3$ contribution to the
pressure comes entirely from soft momenta, and reads
(cf. Eq.~(\ref{P3scal})):
\beq\label{P3QCD}
P_3&=&
-\int\!\!{d^4k\0(2\pi)^4}\,\frac{T}{\omega}\,\Im
\Bigl[\log(1+ D_0\hat\Pi) -  D_0\hat\Pi\Bigr]\,.\eeq
In QCD, $D=(D_T,D_L)$, $\5\Pi=(\5\Pi_T,-\5\Pi_L)$, and a sum over
color and polarization states is implicit in (\ref{P3QCD}).
[Note the minus sign in front of $\Pi_L$ in these
compact notations; this reflects our conventions
in Eqs.~(\ref{Piij})--(\ref{DLT}).]
The integral over $\omega$ yields:
\beq
\int\frac{d\o}{\pi\o}\,\Im
\Bigl[\log(1+ D_0\hat\Pi) - \hat\Pi D_0\Bigr]&=&
\log\Bigl[
1+ D_0(\o=0)\hat\Pi(\o=0)\Bigr] - \hat\Pi(\o=0) D_0(\o=0)\nonumber\\
&=&\log\left(
1+\frac{\5 m_D^2}{k^2}\right)-\frac{\5 m_D^2}{k^2}\,,\eeq
where the non-vanishing contribution in the second line comes
from the longitudinal sector alone \cite{P3}, since $\5\Pi_L(\o=0)=\5 m_D^2$,
while $\5\Pi_T(\o=0)=0$. Thus,
\bea\label{P31}
P_3&=&-N_gT\int\!\!{d^3 k\0(2\pi)^3}\,\left[\log\left(
1+\frac{\5 m_D^2}{k^2}\right)-\frac{\5 m_D^2}{k^2}\right]
\,=\,N_g \frac{\hat m_D^3 T}{12\pi},\eea
where the color factor $N_g=N^2-1$ has been reintroduced.
Eq.~(\ref{P31}) is indeed the standard result for $P_3$,
generally obtained by summing the ring diagrams in the
imaginary-time perturbation theory \cite{Kapusta}.

The order-$g^3$ effect in the entropy can be now directly calculated as
the total derivative of $P_3$ with respect to $T$. We thus obtain
${\cal S}_3={\cal S}_3^{(a)}+{\cal S}_3^{(b)}$, where
\begin{mathletters}
\label{SDERPs}
\bea \label{SDERPa}
{\cal S}_3^{(a)}\equiv \frac{{\del} P_3}{{\del} T}\bigg|_{\5m_D}
\,=\,-N_g\int\!\!{d^4k\0(2\pi)^4}\,{\6n(\o)\0\6T}\,\Im
\Bigl[\log(1+ D_0\hat\Pi) - \hat\Pi D_0\Bigr]\,=\,N_g
\frac{\hat m_D^3}{12\pi}\,,
\eea
is the derivative at fixed $\Pi=\hat\Pi$ (recall that the HTL's
depend upon the temperature only via the Debye mass;
cf. Eqs.~(\ref{PiL}) and (\ref{MD0})), and
\bea \label{SDERPb}
{\cal S}_3^{(b)}\equiv \frac{{\del} P_3}{{\del}\5m_D }\,\frac{{\rm d}\5m_D}
{{\rm d}T}\,=\,-N_g\int\!\!{d^4k\0(2\pi)^4}\,n(\o)\,\Im
\left[\frac{{\rm d}\hat \Pi}{{\rm d} T}(\hat D -D_0)\right]=\,N_g
\frac{\hat m_D^3}{4\pi}\,.
\eea
\end{mathletters}
This decomposition of ${\cal S}_3$ is interesting in view of
the comparison with the perturbative expansion of Eqs.~(\ref{SQCDD}),
to which we now turn.

\subsubsection{The order $g^3$ in the entropy}

Unlike what happens for the pressure, 
the order-$g^3$ effects of the hard modes do {\it not} cancel 
in Eqs.~(\ref{SQCDD}), similarly to what we have
observed in the scalar case in Sect.~\ref{secscapprox}. 
Rather, we get a non-zero such
contribution by replacing $\Re\Pi_T^{(2)}\longrightarrow
\Re\delta \Pi_T$ in Eq.~(\ref{SG2}), with $\delta \Pi_T 
\sim g^3T^2$ the NLO self-energy correction of hard
transverse gluons (cf. Eq.~(\ref{PERTDH})). This yields:
\bea\label{ST32}
{\cal S}_3^{\mathrm hard}&=&
- N_g\int\!\!{d^3k\0(2\pi)^3}\,{1\0k}\,{\6n(k)\0\6T}\,
\Re\delta \Pi_T(\o=k).\eea 
Once again,  we need only the light-cone projection
of the self-energy of the hard particles. What is, however,
new as compared to the situation at order $g^2$
is that $\Re\delta \Pi_T(\o=k)$ is not
a constant ``mass correction'', but rather a complicated
function of $k$ (see Eqs.~(\ref{b11}) 
and (\ref{b21})). 
The calculation of  ${\cal S}_3^{\mathrm hard}$ is deferred
to the Appendix, but the final result can be anticipated,
as we shall see shortly.

The other contributions of order $g^3$ come from
the soft gluons, which can be longitudinal or
transverse, and we write ${\cal S}_3^{\mathrm soft}=
{\cal S}_L^{(3)}+{\cal S}_T^{(3)}$.
We have (with $n(\o)\simeq T/\o$):
\bea\label{SL3}
{\cal S}_L^{(3)}&=& -N_g\int\!\!{d^4k\0(2\pi)^4}\,{1\0\o}\,
\Bigl\{\Im\log(k^2+\5\Pi_L)+\Im\5\Pi_L \Re \5 D_L \Bigr\},\\
\label{ST31}
{\cal S}_T^{(3)}&=& -2N_g\int\!\!{d^4k\0(2\pi)^4}\,{1\0\o}
\biggl\{\Im\left[\log\left(1-\frac{\5\Pi_T}{\o^2-k^2}\right)
+\frac{\5\Pi_T}{\o^2-k^2}\right] \nonumber\\ &&\qquad\qquad\qquad\qquad-
\Im\5\Pi_T \Re \Bigl(\5 D_T - D_T^{(0)}\Bigr)\biggr\},\eea
where in the transverse sector, the
contribution of order $g^2$ has been subtracted (cf. Eq.~(\ref{SG2})).
More precisely, Eq.~(\ref{SG2}) involves the full one-loop
self-energy $\Pi_T^{(2)}$, while
the subtracted terms in Eq.~(\ref{ST31}) 
involve only  $\5\Pi_T$, the HTL.
This is nevertheless correct since $\Pi_T^{(2)}$ and $\5\Pi_T$
coincide on the light-cone:
\be\label{ID2}
\5\Pi_T(\o^2=k^2)=\Pi_T^{(2)}(\o^2=k^2)= m_\infty^2.
\ee

Ultimately, all the contributions of order $g^3$ displayed
in Eqs.~(\ref{ST32})--(\ref{ST31}) are {\it soft} field effects:
the quantities ${\cal S}_L^{(3)}$ and ${\cal S}_T^{(3)}$ are
the LO entropies of the soft gluons, while
${\cal S}_3^{\mathrm hard}$ is the NLO correction to the entropy
of the hard gluons induced by their coupling to
the soft fields (cf. Fig.~\ref{figdPit}).
We expect these three contributions to add
to the standard result for the plasmon effect
in the entropy, namely (cf. Eqs.~(\ref{SDERPs})):
\beq\label{STOT3}
{\cal S}_3^{\mathrm soft}+{\cal S}_3^{\mathrm hard} \,=\,
{\cal S}_3\,\equiv\,N_g \hat m_D^3/(3\pi).\eeq
This is  verified in the Appendix, where the quantities
in Eqs.~(\ref{ST32})--(\ref{ST31}) are explicitly computed,
but it can be also understood on the basis
of the following argument.

Eqs.~(\ref{ST32})--(\ref{ST31}) can be compactly rewritten as
\beq\label{S3P3}
{\cal S}_3&=&
-\int\!\!{d^4k\0(2\pi)^4}\,\frac{1}{\omega}\left\{\Im
\Bigl[\log(1+ D_0\hat\Pi) - \hat\Pi D_0\Bigr]\,-\,
\Im\hat \Pi\Re(\hat D-D_0)\right\}-\nonumber\\
&{}&\,\,\,\,-\,\int\!\!{d^4p\0(2\pi)^4}{\6n(p_0)\0\6T} \Re\delta
\Pi\Im D_0\,,\eeq
where the sum over color and polarization states is again implicit.
The first term within the (soft) integral over $k$ is 
obviously the same as ${\cal S}_3^{(a)}$, the
temperature derivative of $P_3$ at fixed $\5m_D$ (cf. 
Eq.~(\ref{SDERPa})). It thus remains to show that the other
terms in Eq.~(\ref{S3P3}) add to ${\cal S}_3^{(b)}$, the piece
of the entropy involving the derivative of the Debye mass
(cf. Eq.~(\ref{SDERPb})). That is, one has to prove the following
relation:
\beq\label{S-H}
\int\!\!{d^4p\0(2\pi)^4}{\6n(p_0)\0\6T} \Re\delta
\Pi\Im D_0=\int\!\!{d^4k\0(2\pi)^4}\biggl\{
{\6n(\o)\0\6T} \Im\hat\Pi\Re(\hat D-D_0)\nn
\qquad \,\,\,\,+n(\o)\Im\biggl[
\frac{{\rm d}\hat \Pi}{{\rm d} T}(\hat D -D_0)\biggr]\biggr\}.\eeq
Eq.~(\ref{S-H}) is nothing but the general 2-loop
identity ${\cal S}'=0$ expanded to the
order $g^3$. Indeed, to order $g^3$, Eq.~(\ref{SP0})
implies:
\beq\label{S3prime}
{\6(T\Phi_3)\0\6T}\Big|_{D}=
\int\!\!{d^4k\0(2\pi)^4}{\6n(\o)\0\6T} \Re\hat\Pi\Im(\hat D-D_0)
+\int\!\!{d^4p\0(2\pi)^4}{\6n(p_0)\0\6T} \Re\delta
\Pi\Im D_0,\eeq
where the first integral is saturated by soft momenta $k\sim gT$,
while the second one is dominated by $p$ hard, $p\sim T$.
On the other hand, $\Phi_3[D]$ has the explicit 
expression\footnote{This follows by expanding $\Phi[D]$
in powers of $g$ as follows:
$\Phi[D] =\Phi[D_0] + (\d\Phi[D]/\d D)|_{D_0} \,
(D-D_0)+\,\cdots\equiv\Phi_2 + \Phi_3 +\,\cdots$.}
\beq\label{Phi30}
T\Phi_3[D] \,\equiv\,{T\02}\Tr[\Pi[D_0]\,(D-D_0)]\,\simeq\,
\int\!\!{d^4k\0(2\pi)^4}\,n(\o)\, \Im[\hat\Pi\, (\hat D-D_0)]
\eeq
which implies:
\beq\label{dPhi3}
{\6(T\Phi_3)\0\6T}\Big|_{D}=
\int\!\!{d^4k\0(2\pi)^4}\left\{
{\6n(\o)\0\6T}\Im[\hat\Pi(\hat D-D_0)]+n(\o)
\Im\biggl[
\frac{{\rm d}\hat \Pi}{{\rm d} T}(\hat D -D_0)\biggr]\right\}.
\eeq
A comparison of Eqs.~(\ref{S3prime}) and (\ref{dPhi3})
immediately leads to Eq.~(\ref{S-H}).

Moreover, the soft longitudinal and transverse
sectors are decoupled at this order: $\Phi_3[D]$ in Eq.~(\ref{Phi30})
is simply the sum of two two-loop diagrams, one with a soft
{\it electric} gluon, the other one with a soft {\it magnetic} gluon.
The condition ${\cal S}'=0$ can be applied to any of these
two diagrams separately. It follows that
Eq.~(\ref{S-H}) must hold separately in the electric, and the 
magnetic sector. This is explicitly verified in the Appendix,
via a lengthy calculation.
Remarkably, Eq.~(\ref{S-H}) provides a relation between the effects
of thermal fluctuations on the hard and soft excitations, which are both
encoded in the two-loop diagrams for $\Phi_3$:
By opening up the soft line in $\Phi_3$, one obtains the
hard one-loop diagram responsible for the HTL $\5\Pi$;
by opening up one of the hard lines, one gets the
effective one-loop diagrams for $\delta\Pi$
displayed in Fig.~\ref{figdPit}. In the case of the scalar
theory, this relation is explicitly verified
in Eqs.~(\ref{Sscscal2})--(\ref{S3hard}).

Let us conclude this subsection on perturbation theory with
a comment on the higher-order contributions to ${\cal S}_L\,$:
By inspection of Eq.~(\ref{SQCDL}), it is easy to verify that not
only the LO contribution $\sim g^3$ discussed above, but also
the corrections of order $g^4$ and $g^5$, come exclusively
from {\it soft} momenta. Indeed, one can 
estimate the contribution of hard momenta by expanding
the integrand in Eq.~(\ref{SQCDL}) in powers of $\Pi_L/k^2$, to obtain:
\beq\label{SLexp}
\Im\log(k^2+\Pi_L)&=&{\Im\Pi_L\0k^2}-{1\02}{\Im(\Pi_L)^2\0k^4}+\,\cdots
\,=\,{\Im\Pi_L\0k^2}-{\Im\Pi_L \Re\Pi_L\0k^4}+\,\cdots\nonumber\\
-\Im\Pi_L \Re{1\0k^2+\Pi_L}&=&-{\Im\Pi_L\0k^2}+{\Im\Pi_L \Re\Pi_L\0k^4}
+\,\cdots\,,\eeq
up to terms of order $(\Pi_L/k^2)^3$. Remarkably, not only the LO
terms, but also the NLO ones, of order $g^4$, mutually cancel in the sum
of the above equations. Thus, as anticipated, the hard modes
contribute to ${\cal S}_L$ only at order $g^6$ or higher.
This shows that our approximation scheme is rather insensitive
to the unphysical, hard longitudinal modes. This is to be
contrasted to the direct HTL resummation of the pressure
where, to one-loop order, the longitudinal sector is sensitive to
hard momenta already at order $g^4$, as indicated by the presence
of UV divergences at this order \cite{ABS}.

\subsubsection{The HTL entropy}
\label{secSHTL}

Since $\delta \Pi_T(\o=k)$ is a complicated, non-local function,
whose numerical treatment is difficult, it is interesting
to explore first approximations
where $\delta \Pi_T$ is set to zero. Specifically,
let us define the following approximation to the entropy,
which is obtained from Eqs.~(\ref{SQCDD})
by replacing all propagators and self-energies by their HTL counterparts:
\bea\label{SHTL}
{\cal S}_{HTL}&=& -N_g\int\!\!{d^4k\0(2\pi)^4}{\6n(\o)\0\6T}\,\,
\Bigl\{2\Im\log(-\o^2+k^2+\5\Pi_T) \nonumber\\
&{}&\qquad\quad -2\Im\5\Pi_T\Re \5D_T
+\Im\log(k^2+\5\Pi_L)+\Im\5\Pi_L \Re\5 D_L\Bigr\}.
\eea
We shall succinctly refer to this as {\it the HTL entropy}.
Clearly, this is still a non-perturbative approximation,
since its expansion contains all orders in $g$. 

A priori, Eq.~(\ref{SHTL}) is not
doing justice to the hard particles, since it uses the HTL
corrections for {\it both} hard and soft momenta (while we know that
the HTL's are the LO self-energies for {\it soft} momenta alone). But
it turns out that the order-$g^2$ effect, which is entirely due
to the hard fields, is nevertheless correctly reproduced by
Eq.~(\ref{SHTL}): ${\cal S}_{HTL}^{(2)}={\cal S}_2$. The
point, as emphasized in Sect. \ref{secg2}, is that ${\cal S}_2$ is
sensitive only to the light-cone projection of the self-energy,
where the HTL $\5\Pi_T$ is a good LO approximation for the hard
modes (cf. Eq.~(\ref{ID2})).\footnote{This is to be contrasted
with a direct HTL resummation of the one-loop expression for the
{\em pressure} in QCD along the lines of Ref.~\cite{ABS}---there the
HTL corrections contribute throughout the hard momentum phase space,
while no longer being the right approximation. Instead they give rise
to artificial UV problems.}

\begin{figure}
\epsfxsize=9cm
\centerline{\epsfbox[65 145 550 725]{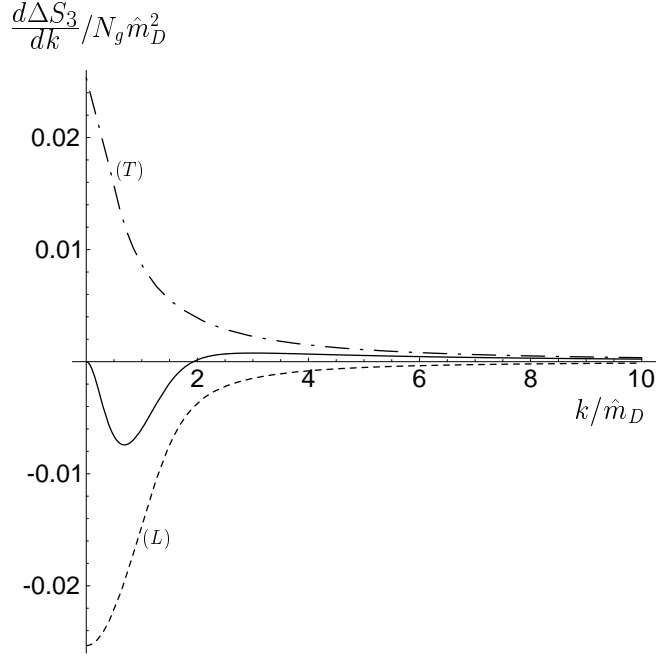}}
\vskip0.5cm
\caption{The integrand of Eq.~(\ref{DELTASS}) after performing
the energy integral. The transverse $(T)$ and longitudinal $(L)$ contributions
do not cancel for each value of $k$; their sum (full line) vanishes
only upon integration over all $k$.
\label{figDS3}}
\end{figure}

On the other hand, ${\cal S}_{HTL}$ contains only a part of the
$g^3$ effect, namely that part which is associated with the entropy of
soft gluons: indeed, it is obvious that the order-$g^3$ contribution 
to Eq.~(\ref{SHTL}) comes from soft momenta alone, where it
coincides with ${\cal S}_3^{\mathrm soft}={\cal S}_L^{(3)}+{\cal S}_T^{(3)}$,
cf. Eqs.~(\ref{SL3})--(\ref{ST31}). Let us therefore study this
quantity in more detail (it is the same as the first integral
in Eq.~(\ref{S3P3})):
\bea\label{SHTLD}
{\cal S}_{HTL}^{(3)}&=&
-\int\!\!{d^4k\0(2\pi)^4}\,\frac{1}{\omega}\left\{\Im
\Bigl[\log(1+ D_0\hat\Pi) - \hat\Pi D_0\Bigr]\,-\,
\Im\hat \Pi\Re(\hat D-D_0)\right\}\nonumber\\
&=&{\cal S}_3^{(a)}+\Delta {\cal S}_3,\eeq
where ${\cal S}_3^{(a)}=({\del} P_3/\del T)|_{\5m_D}={\cal S}_3/4$
(cf. Eq.~(\ref{SDERPa})), and the remainder is
\bea\label{DELTASS}
\Delta {\cal S}_3&\equiv &
N_g\int\!\!{d^4k\0(2\pi)^4}\,{1\0\o}\,
\left\{2\Im\5\Pi_T \Re \Bigl(\5 D_T - D_T^{(0)}\Bigr)
-\Im\5\Pi_L \Re\Bigl(\5 D_L-D_L^{(0)}\Bigr)\right\}\nn
&\equiv &\Delta {\cal S}_T^{(3)}+\Delta {\cal S}_L^{(3)}\,.\eea
Remarkably, we have found that the transverse and
longitudinal contributions to
$\Delta {\cal S}_3$ cancel within the accuracy 
that we have reached in 
a numerical integration of
Eq.~(\ref{DELTASS}) (more than 8 significant digits). 
With $\Delta {\cal S}_3=0$, ${\cal S}_{HTL}^{(3)}$ is
precisely equal to one fourth of the total $g^3$ effect, as it was
also the case in the scalar theory with $g^2\phi^4$ self-interactions 
(cf. Sect. \ref{secscapprox}):
\begin{mathletters}\label{S3quarter}
\vspace{-18pt}
\beq
{\cal S}_{HTL}^{(3)}\equiv {\cal S}_T^{(3)}+{\cal S}_L^{(3)}
=\,\frac{{\del} P_3}{{\del} T}\bigg|_{\5m_D}\,= {\cal S}_3/4.\eeq
In QCD, however, this property is much more subtle: In the
scalar theory, the quantity which we call here $\Delta {\cal S}_3$
was trivially zero, since $\Im\5\Pi=0$ in that case. Here,
$\Delta {\cal S}_3=0$ only because a compensation takes place in
between the transverse and longitudinal contributions to
Eq.~(\ref{DELTASS}), both of which arise from Landau-damping
contributions at space-like momenta.
Moreover, this cancellation occurs only after integrating
over all energies and momenta (for generic $k$, the result of
the energy integral in Eq.~(\ref{DELTASS}) is non-zero, see Fig.~\ref{figDS3}).
Numerically, the contributions to 
${\cal S}_{HTL}^{(3)}\equiv {\cal S}_T^{(3)}+{\cal S}_L^{(3)}$
turn out to be
\be
{\cal S}_T^{(3)}+{\cal S}_L^{(3)}=(0.34008738\ldots-0.09008738\ldots)
{\cal S}_3. 
\ee
\end{mathletters}

Let us summarize here the various cancellations which
take place at order $g^3$ in the complete two-loop entropy:
The straightforward perturbative expansion of our
master equations (\ref{SQCDD}) leads us to 
Eqs.~(\ref{ST32})--(\ref{ST31}), and thus to the following
expression for ${\cal S}_3$ [recall the compact notation introduced
after Eq.~(\ref{P3QCD})]:
\beq\label{S3SUMMARY}
{\cal S}_3&=&{\cal S}_3^{\mathrm soft}\,+\,{\cal S}_3^{\mathrm hard}\,,\nn
{\cal S}_3^{\mathrm soft}&=&
\frac{{\del} P_3}{{\del} T}\bigg|_{\5m_D}\,+\,
\Delta {\cal S}_L^{(3)}+\Delta {\cal S}_T^{(3)}\,,\nn
{\cal S}_3^{\mathrm hard}&=&
\frac{{\del} P_3}{{\del}\5m_D }\,\frac{{\rm d}\5m_D}
{{\rm d}T}\,-\,N_g\int\!\!{d^4k\0(2\pi)^4}\,
{\6n(\o)\0\6T} \Im\hat\Pi \Re(\hat D-D_{0})\nn
&=&\frac{{\del} P_3}{{\del}\5m_D }\,\frac{{\rm d}\5m_D}
{{\rm d}T}\,-\Delta {\cal S}_L^{(3)}-\Delta {\cal S}_T^{(3)}\,.\eeq
In these equations, $\Delta {\cal S}_L^{(3)}$ and $\Delta {\cal S}_T^{(3)}$
have been defined in Eq.~(\ref{DELTASS}), and the second line
in the above expression for ${\cal S}_3^{\mathrm hard}$ follows
either by using ${\cal S}'=0$ (cf. Sect. III.D.2), or by
explicitly computing Eq.~(\ref{ST32}) within HTL-resummed
perturbation theory (cf. the Appendix; see especially 
Eqs.~(\ref{DS3}) and (\ref{S3T0}) there). Furthermore, 
by construction, ${\cal S}_{HTL}^{(3)}$ is the same as
${\cal S}_3^{\mathrm soft}$.

According to these equations, the quantities 
$\Delta {\cal S}_L^{(3)}$ and $\Delta {\cal S}_T^{(3)}$
cancel in ${\cal S}_3^{\mathrm soft}\,+\,{\cal S}_3^{\mathrm hard}$
{\it independently} in the longitudinal and transverse sectors,
thus yielding the correct result for ${\cal S}_3$, 
cf. Eqs.~(\ref{SDERPa}) and (\ref{SDERPb}).
This is what we have been able to prove {\it analytically}
(cf. Sect. III.D.2 and the Appendix).
On the other hand, we have found {\it numerically} that
$\Delta {\cal S}_3=\Delta {\cal S}_L^{(3)}+\Delta {\cal S}_T^{(3)}=0$, so that
the actual results for ${\cal S}_3^{\mathrm soft}$ and
${\cal S}_3^{\mathrm hard}$ are even simpler:
\beq\label{SSH}
{\cal S}_3^{\mathrm soft}\,=\,
\frac{{\del} P_3}{{\del} T}\bigg|_{\5m_D}\,,\qquad
{\cal S}_3^{\mathrm hard}\,=\,
\frac{{\del} P_3}{{\del}\5m_D }\,\frac{{\rm d}\5m_D}
{{\rm d}T}\,.\eeq
At this stage, we have no fundamental understanding of the
``sum rule'' $\Delta {\cal S}_3=0$. But this 
serendipitous result will have
important consequences in practice, as we shall see below,
because it determines the magnitude of ${\cal S}_3^{\mathrm hard}$ to
be $3/4$ of ${\cal S}_3$, as was the case in the simple scalar
model of Sect. \ref{secscapprox}, 
while being an incomparably more complicated expression
than (\ref{S3hard}).

A full numerical evaluation of the HTL entropy, non-perturbative
in $g$, will be presented
in Sect. \ref{SecnumHTL} below,
and estimates of the effects of including $\d \Pi_T\sim g^3 T^2$
in Sect. \ref{SecnumNLO}.

\section{QCD: Adding the fermions}

It is now straightforward to add fermions to our theory.
We consider $N_f$ flavors of massless fermions with equal
chemical potential $\mu$; we choose $\mu \ge 0$, which
corresponds to an excess of fermions over antifermions
for all flavors. Adding the fermions will have two effects:
first, this will modify the parameters of the gluonic sector,
namely the Debye mass $\hat m_D^2$, and therefore 
also the asymptotic mass $m_\infty^2 = \hat m^2_D/2$;
second, there will be new contributions to the entropy.
In addition, at finite $\mu$,
there is a new thermodynamic function of interest, namely
the density ${\cal N}$, which shares many of the interesting
properties found for $\cal S$.

The full (leading-order) Debye mass in the QGP reads \cite{QCD} :
\beq\label{MD}
\hat m^2_D&=&-\,\frac{g^2}{2\pi^2}\int_{0}^\infty 
{\rm d}k \,k^2\,\left\{
2N\,\frac{\6n}{\6k}\,+\,N_f\left(\frac{\6f_+}{\6k}
\,+\,\frac{\6f_-}{\6k}\right)\right\}\nn
&=&(2N+N_f)\,\frac{g^2T^2}{6}\,+\,N_f\,{g^2\mu^2\02\pi^2}\,.\eeq
We have introduced here the statistical distribution functions
for fermions ($f_+$) and  antifermions ($f_-$),
\beq\label{fpm}
f_\pm(k)\,\equiv\,\frac{1}{{\rm e}^{\beta(k\mp\mu)}+1}\,,\eeq
and we have used the following integral:
\beq\label{INT1}
\int {\rm d}k \,k\Bigl(f_+(k) + f_-(k)\Bigr)\,=\,
{\pi^2T^2\06}+{\mu^2\02}\,.\eeq

\subsection{Entropy and density from the skeleton expansion}

To construct the fermion contribution to the entropy,
let us return to the full skeleton representation of the
thermodynamic potential (in a ghost-free gauge)
and add fermions to it. This becomes
\beq\label{LWQCD}
\b \O[D,S]\,=\,\2 \Tr \log D^{-1}-\2 \Tr \Pi D
- \Tr \log S^{-1} + \Tr \Sigma S + \Phi[D,S],
\eeq
where $S$ and $\Sigma$ denote respectively the fermion propagator
and self-energy, and the sum over the gluon polarization states
(two transverse and one longitudinal) is implicit.
$\Phi[D,S]$ is the sum of the 2-particle-irreducible ``skeleton''
diagrams constructed out of the propagators $D$ and $S$.
Below, we shall be mainly interested in the 2-loop approximation
to $\Phi[D,S]$, where the only new diagram is the one
represented in Fig.~\ref{figphiqcd}d.
The self-energies $\Sigma$ and $\Pi$ in Eq.~(\ref{LWQCD}) are
themselves functionals of the propagators,
defined as
\beq\label{SPi}
\Sigma \,\equiv\, {\d\Phi[D,S]\0\d S},\qquad \Pi\,\equiv\,
2\,{\d\Phi[D,S]\0\d D}\,.
\eeq
The self-consistent propagators $D$ and $S$ are obtained by solving
the Dyson equations
\beq\label{DS}
D^{-1}\,=\,D^{-1}_0+\Pi,\qquad S^{-1}\,=\,S^{-1}_0+\Sigma.\eeq
Then, the functional $\Phi[D,S]$ is stationary under variations
of $D$ and $S$ around the solutions to Eqs.~(\ref{DS}):
\be\label{selfcons1}
{\d\O[D,S] / \d S}=0,\qquad{\d\O[D,S] / \d D}=0.
\ee
The entropy ${\cal S}(T,\mu)$ and the density ${\cal N}(T,\mu)$
are obtained as the derivatives
of the thermodynamic potential with respect to the temperature,
and the chemical potential, respectively:
\beq\label{SNdef}
{\cal S}\,=\,-\,\frac{\6 ({\O/V})}{\6 T}\Big|_\mu\,,\qquad
{\cal N}\,=\,-\,\frac{\6 ({\O/V})}{\6 \mu}\Big|_T\,.\eeq
Because of the stationarity property (\ref{selfcons1}),
we can ignore the $T$ and $\mu$ dependences of the spectral
densities of the propagators when differentiating
$\Phi[D,S]$. That is, we have to differentiate only
the statistical factors $n(\o) = 1/({\rm e}^{\beta \o}-1)$
and $f(\o)=1/({\rm e}^{\beta (\o-\mu)}+1)$ which arise after
performing the Matsubara sums in Eq.~(\ref{LWQCD}). This yields,
for the entropy,
\beq
{\cal S}\,=\,-\,\frac{\6 ({\O/V})}{\6 T}\Big|_{\mu,D,S}
\,\equiv\, {\cal S}_b+{\cal S}_f+{\cal S}',\eeq
where
${\cal S}_b={\cal S}_T+{\cal S}_L$
is the purely gluonic part of the entropy,
as shown in Eqs.~(\ref{SQCD})--(\ref{SQCDL}), ${\cal S}_f$
is the corresponding fermionic piece, which reads
(the trace below refers to Dirac indices)
\beq\label{SQCDF}
{\cal S}_f &\equiv& -2\int\!\!{d^4k\0(2\pi)^4}{\6f(\o)\0\6T}\,
\tr\left\{\Im\log(\gamma_0 S^{-1}) \,-\,\Im(\gamma_0 \Sigma)
\Re(S\gamma_0)\right\},\eeq
and 
\bea\label{SP1}
{\cal S}'\equiv -{\6(T\Phi)\0\6T}\Big|_{D,S}+
\int\!\!{d^4k\0(2\pi)^4}\left\{{\6n(\o)\0\6T}
\Re\Pi \Im D + 2{\6f(\o)\0\6T}
\tr\Bigl[\Re(\gamma_0\Sigma)\Im(S\gamma_0)\Bigr]\right\}
\eea
has the important property to vanish at 2-loop order \cite{VB}.
That is, ${\cal S} \simeq {\cal S}_b+{\cal S}_f$
to the order of interest.

The corresponding expression for the density is obtained
by replacing $(\6 f/\6 T)\to(\6 f/\6\mu)$ in all the formulae
above. This gives ${\cal N}={\cal N}_f+{\cal N}'$, with 
${\cal N}'=0$ in the 2-loop approximation. Thus,
to the order of interest,
\beq\label{NQCD}
{\cal N} &\simeq & -2\int\!\!{d^4k\0(2\pi)^4}{\6f(\o)\0\6\mu}\,
\tr\left\{\Im\log(\gamma_0 S^{-1}) \,-\,\Im(\gamma_0 \Sigma)
\Re(S\gamma_0)\right\}.\eeq

For simplicity, all the previous formulae have been 
written for only one fermionic degree of freedom; the 
corresponding formulae for $N$ colors and $N_f$ flavors
can be obtained by multiplying the fermionic contributions
above by $N N_f$.

Note finally the following {\it Maxwell relations},
\beq\label{MAXWELL}
\frac{\6{\cal S}}{\6\mu}\Big|_T\,=\,
\frac{\6{\cal N}}{\6 T}\Big|_\mu\,,\eeq
which express the equality of the mixed, second order derivatives
of the thermodynamic potential.
In our subsequent, self-consistent construction of ${\cal S}$
and  ${\cal N}$, these relations will be satisfied at the
same order as the requirement of self-consistency.

\subsection{The structure of the fermion propagator}

In the previous formulae we have always associated
a factor of $\gamma_0$ with the fermion propagator and self-energy.
This was possible since $\gamma_0^2 =1$ and det$\,\gamma_0 = 1$;
it is also convenient since, e.g., $S^\dagger=\gamma_0
S\gamma_0$, and it is preferable to work with hermitian Dirac
matrices.

In order to compute the Dirac traces in 
Eqs.~(\ref{SQCDF})--(\ref{NQCD}), it is useful to recall the
structure of the fermion propagator at finite temperature
and density: The most general form of the self-energy
$\Sigma$ which is compatible with the rotational and chiral
symmetries is: 
\beq\label{Sigma1}
\Sigma(\omega, {\bf  k})\,=\,a(\omega, k)\,\gamma^0\,+\,b(\omega, k)
{\hat{\bf  k}}\cdot{\bfgamma}.\eeq
(For a massive fermion, this would also include a mass correction,
i.e., $\Sigma = a(\omega, k)\gamma^0+b(\omega, k)
{\hat{\bf  k}}\cdot{\bfgamma}+c(\omega, k)$.)
This can be rewritten as:
\beq\label{SIGL}
\gamma_0\Sigma(\omega, {\bf  k})\,=\,\Sigma_+(\omega,k)
\Lambda_+(\hat {\bf k})\,-\,\Sigma_-(\omega,k)
\Lambda_-(\hat {\bf k}),\eeq
where $\Sigma_\pm(\omega,k)\equiv b(\omega, k) \pm \,a(\omega, k)$,
and the spin matrices 
\beq
\Lambda_{\pm}(\hat {\bf k})&\equiv& \frac{1 \pm \gamma^0
 \bfgamma\cdot\hat{\bf k}}{2},\qquad \Lambda_++\Lambda_-=1,\nn
\Lambda_{\pm}^2&=&\Lambda_{\pm},
\qquad \Lambda_+\Lambda_-\,=\,\Lambda_-\Lambda_+=0,
\qquad \tr \Lambda_\pm = 2, \eeq
project onto spinors whose chirality is equal ($\Lambda_+$),
or opposite ($\Lambda_-$), to their  helicity.
Dyson's equation $S^{-1}= -{\not\! k} + \Sigma$ then implies:
\beq\label{SINV}
\gamma_0 S^{-1} (\omega, {\bf  k})\,=\,\Delta_+^{-1}(\omega,k)
\Lambda_+ \,+\,\Delta_-^{-1}(\omega,k)\Lambda_-,\eeq
with $\Delta_\pm^{-1} \equiv - [\omega\mp(k+\Sigma_\pm)]$.
This is trivially inverted to yield the fermion propagator:
\beq\label{SFL}
S\gamma_0(\omega, {\bf  k})\,=\,
\Delta_+(\omega, k)\Lambda_+ +\Delta_-(\omega, k)\Lambda_-.\eeq

The presence of the projection operators $\Lambda_{\pm}$
in Eqs.~(\ref{SIGL}), (\ref{SINV}) and (\ref{SFL}) allows one
to easily compute the Dirac traces in
Eqs.~(\ref{SQCDF}) and (\ref{NQCD}), and thus obtain:
\beq\label{SF}
{\cal S}_f &=& -4\int\!\!{d^4k\0(2\pi)^4}{\6f(\o)\0\6T}\,
\Bigl\{\Im\log\Delta_+^{-1}\,+\,\Im\log(-\Delta_-^{-1})\,+\nn
&{}&\qquad\qquad \,-\,\Im\Sigma_+\Re\Delta_+\,+\,
\Im\Sigma_-\Re\Delta_-\Big\}.\eeq
The corresponding expression for ${\cal N}$ is obtained
by replacing $(\6 f/\6 T)\to(\6 f/\6\mu)$ in the equation above.

\subsection{Perturbation theory for $S_f\,$: order $g^2$}

Eq.~(\ref{SF}) will be now supplied with certain approximations for
the quark self-energies $\Sigma_\pm$. As before, we aim at
reproducing the results of perturbation theory up to order $g^3$.
This will be achieved by approximations analogous to those
employed for the gluons,
namely the HTL approximation $\hat\Sigma_\pm$,
supplemented by the NLO correction $\delta\Sigma_\pm$ to the
{\it hard} fermion self-energy on the light cone.

Note, however, an important difference with respect to the
gluon case: unlike the soft gluons, which contribute to the entropy
already at order $g^3$, the soft fermions contribute only at
order $g^4$ or higher, because their contribution
is not enhanced by the statistics. Nevertheless, in our numerical
calculation below, we shall carefully include the contribution
of the soft fermions, appropriately dressed by the HTL.
This is in line with our general strategy of constructing
non-perturbative approximations for the entropy (or other
thermodynamic quantities) which include as much as possible
the dominant collective effects in the plasma.

In the HTL approximation, the fermion self-energies
read as follows \cite{KlW,MLB} :
\beq\label{SIGHTL}
\hat\Sigma_\pm(\omega,k)\,=\,{\hat M^2\0k}\,\left(1\,-\,
\frac{\omega\mp k}{2k}\,\log\,\frac{\omega + k}{\omega - k}
\right),\eeq
where $\hat M^2$ is the plasma frequency for fermions,
i.e., the frequency of long-wavelength ($k\to 0$) fermionic
excitations ($C_f=(N^2-1)/2N$):
\beq\label{MF}
\hat M^2\,=\,\frac{g^2 C_f}{4\pi^2}\int_{0}^\infty 
{\rm d}k \,k\Bigl(2n(k)+f_+(k) + f_-(k)\Bigr)\,=\,
{g^2 C_f\08}\left(T^2+{\mu^2\0\pi^2}\right).\eeq

We are now in position to evaluate the fermionic entropy 
and density up to order $g^2$: To zeroth order, i.e.,
for an ideal gas of massless fermions
at temperature $T$ and chemical potential $\mu$, we obtain
the well known results \cite{Kapusta} (the color-flavor
factor $N N_f$ is here reintroduced):
\beq
{\cal S}_f^{(0)}\,=\,N N_f\left(\frac{7\pi^2 T^3}{45}\,+\,
\frac{\mu^2 T}{3}\right),\qquad
{\cal N}^{(0)}\,=\,N N_f\,{\mu\0 3}\left(T^2+{\mu^2\0\pi^2}\right).
\eeq
The correction of order $g^2$ involves 
the fermion self-energies to one loop order, $\Sigma_\pm^{(2)}$ :
\bea\label{SF20}
{\cal S}_f^{(2)}/NN_f &=&-4\int\!\!{d^4k\0(2\pi)^4}\,{\6f\0\6T}\,\Bigl\{
\Re\Sigma_+^{(2)}\Im\frac{-1}{\o-k}\,-\,
\Re\Sigma_-^{(2)}\Im\frac{-1}{\o+k}\Bigr\}\nn
&=&-2\int\!\!{d^3k\0(2\pi)^3}\,\left\{{\6f(k)\0\6T}
\Re\Sigma_+^{(2)}(\omega=k)\,-\,{\6f(-k)\0\6T}\
\Re\Sigma_-^{(2)}(\omega=-k)\right\}.\eeq
As in the gluon case (cf. Eq.~(\ref{SG2})), the correction
of order $g^2$ is sensitive only to the light-cone projection of the 
one-loop self-energy, which is correctly reproduced by the HTL
approximation (\ref{SIGHTL}) \cite{KKR}. That is,
\beq\label{MASIMP}
\Re\Sigma_\pm^{(2)}(\omega=\pm k)\,=\,\hat\Sigma_\pm
(\omega=\pm k)\,=\,{\hat M^2\0k}\,.\eeq
Eqs.~(\ref{SFL}) and (\ref{MASIMP})
show that, to order $g^2$, the hard fermions (or antifermions) 
propagate as massive particles, with dispersion relation
$\varepsilon_k^2 = k^2 + 2\hat M^2$. This identifies the
fermionic asymptotic mass as $M^2_\infty = 2\hat M^2$.
By also using the properties $f(k) = f_+(k)$ and $f(-k)=1 - f_-(k)$
(cf. Eq.~(\ref{fpm})), together with Eq.~(\ref{INT1}), 
we finally deduce
\beq\label{SF2} {\cal S}_f^{(2)}/NN_f \,=\,-\,{\hat M^2\0 \pi^2}\,
{\6\0\6 T}\,\left[
{\pi^2T^2\06}+{\mu^2\02}\right]\,=\,-\,{\hat M^2 T\0 3}
\,=\,-\,{M^2_\infty T\0 6}\,.\eeq
The leading-order correction to the density ${\cal N}_f^{(2)}$
is obtained similarly:
\beq\label{NF2} {\cal N}^{(2)}_f/NN_f\,=\,-\,{\hat M^2\0 \pi^2}\,
{\6\0\6 \mu}\,\left[
{\pi^2T^2\06}+{\mu^2\02}\right]\,=\,-\,{\mu \hat M^2 \0 \pi^2}
\,=\,-\,{\mu M^2_\infty \0 2\pi^2}\,.\eeq

The above results for ${\cal S}_f^{(2)}$ and
${\cal N}^{(2)}_f$, together with the previous ones for
scalars, Eq.~(\ref{Sscscal2}), or gluons, Eq.~(\ref{S2}),
can be generalized to the following, remarkably simple, formulae,
which hold for an arbitrary field theory involving massless 
bosons (with zero chemical potentials) and fermions:
\beq\label{SNBF2}
{\cal S}_2 =-  T\left\{\sum_B { m_{\infty\,B}^2 \0 12}\,+\,
\sum_F { M_{\infty\,F}^2\0 24}\right\},\qquad
{\cal N}_2=-\,{1\0 8\pi^2}\sum_F \mu_F M_{\infty\,F}^2.\eea
Here the sums run over all the bosonic ($B$) and fermionic ($F$)
degrees of freedom (e.g. 4 for each Dirac fermion), 
which are allowed to have different asymptotic
masses and, in the case of fermions, different chemical potentials.
According to Eq.~(\ref{SNBF2}), the leading-order interaction term
in the entropy as well as in the density has a very simple physical origin:
it is entirely due to the thermal masses acquired by the 
hard plasma particles, i.e., directly given by the spectral properties
of the dominant degrees of freedom.

To conclude this discussion of the order $g^2$, let us
summarize here the respective contributions to entropy
(${\cal S}_2\equiv {\cal S}^{(2)}_b+{\cal S}^{(2)}_f$) and
density (${\cal N}_2$) 
in hot SU($N$) gauge theory with $N_f$ quark flavors:
these follow from Eqs.~(\ref{S2}), (\ref{SF2}), (\ref{NF2})
(with the thermal masses (\ref{MD}) and (\ref{MF})), and read:
\beq\label{QCD2}
{\cal S}_2&=&-\,{g^2N_gT\048}\left\{
\frac{4N+5N_f}{3}\,T^2\,+\,{3N_f\0\pi^2}\,\mu^2\right\},\quad
{\cal N}_2=-\,{g^2\mu N_g N_f\016\pi^2}\left(T^2\,+\,
{\mu^2\0\pi^2}\right),\nn
{P}_2&=&-\,{g^2N_g\032}\left\{
\frac{4N+5N_f}{18}\,T^4\,+\,{N_f\0\pi^2}\,\mu^2T^2
\,+\,{N_f\02\pi^4}\,\mu^4\right\}.\eeq
In writing these equations, we have also added the
corresponding expression of the pressure  (${P}_2$),
as taken from Ref. \cite{Kapusta}. Clearly, our above
results for ${\cal S}_2$ and ${\cal N}_2$ are consistent
with this expression for ${P}_2\,$: ${\cal S}_2=
\6 P_2/\6 T$, ${\cal N}_2=\6 P_2/\6 \mu$.

\subsection{Perturbation theory for $S_f\,$: order $g^3$}

Unlike the $g^2$ corrections in Eq.~(\ref{QCD2}), --- which apply
to the whole area of the $\mu\!\!-\!\!T$ plane where the coupling
constant is small (i.e., such that max($\mu,\,T$) is much
larger than $\Lambda_{QCD}$) ---, the corrections of order $g^3$
that we shall discuss now apply only to the high temperature 
regime\footnote{If $\mu=0$, then $\hat m_D \sim gT$,
and this condition is equivalent to weak coupling;
for $\mu > 0$, however, there is a new scale in
the problem, and the high-$T$ condition becomes an independent
condition.} $T \gg \hat m_D$. This restriction is obvious in the 
imaginary time
formulation of thermal perturbation theory, where the effects
of order $g^3$ arise entirely from the sector with zero Matsubara
frequency \cite{Kapusta}. In the present calculation, these effects
are obtained by approximating
$n(k)\simeq T/k$ for $k\sim \5m_D$, which is valid
provided $\5m_D\ll T$. Assuming this condition to be satisfied, we
shall now show how the ``plasmon effect'' arises in our
formalism when the fermions are also included.
This is similar to the previous discussion 
of the pure glue case (cf. Sect.~\ref{secglueg3}), 
so we shall indicate here only 
the relevant differences.

\begin{figure}
\epsfxsize=7cm
\centerline{\epsfbox[70 405 285 525]{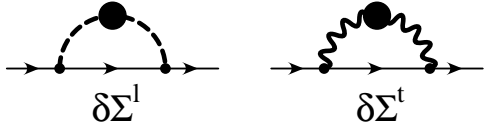}}
\vskip0.5cm
\caption{NLO contributions to $\d\Sigma$ at hard momentum. Thick
dashed and wiggly lines with a blob represent HTL-resummed longitudinal
and transverse propagators, respectively.
\label{figdSigma}
}
\end{figure}

There are two types of contributions
of order $g^3$ to the entropy: ({\it i}) the direct contribution
of the soft gluons, 
${\cal S}^{\mathrm soft}_3={\cal S}_L^{(3)}+{\cal S}_T^{(3)}$,
which is still given by Eqs.~(\ref{SL3}) and (\ref{ST31}),
and ({\it ii}) the NLO correction ${\cal S}^{\mathrm hard}_3$
to the entropy of the hard particles, which now includes
contributions from both transverse gluons and fermions, via 
the NLO corrections to the corresponding
self-energies on the light cone (cf. Eq.~(\ref{ST32}) and
(\ref{SF20})):
\beq\label{deltaSQCD}
{\cal S}^{\mathrm hard}_3&=&
- \int\!\!{d^3k\0(2\pi)^3}\,\biggl\{N_g\,{1\0k}\,{\6n(k)\0\6T}\,
\Re\delta \Pi_T(\o=k)\,+\nn
&{}&\qquad\,\,\,\,+\,2N N_f\biggl(
{\6f_+(k)\0\6T}\Re\delta\Sigma_+(\omega=k)\,+\,{\6f_-(k)\0\6T}\
\Re\delta\Sigma_-(\omega=-k)\biggr)\biggl\}.\eeq
The diagrams pertinent to $\delta \Pi_T$ have been shown
in Fig.~\ref{figdPit}. The corresponding diagrams for $\delta\Sigma_\pm$
are similar, and are displayed in Fig.~\ref{figdSigma}. Their evaluation
proceeds along the same lines, and is briefly discussed
in App.\ \ref{AppPlasmon}. Let us summarize here the final results:

As in the pure glue
case, it can be verified that there is no net contribution
from the soft {\it transverse} gluons: the direct contribution
${\cal S}_T^{(3)}$ in Eq.~(\ref{ST32}) is precisely cancelled
by the corresponding contributions to the self-energies
of the hard particles, $\delta \Pi_T^t$ and $\delta\Sigma_\pm^t$
(cf. Figs.~\ref{figdPit} and \ref{figdSigma}). As expected, the whole
contribution of order $g^3$ comes from soft {\it longitudinal}
gluons (either directly, via ${\cal S}_L^{(3)}$, or indirectly,
via their contribution to ${\cal S}^{\mathrm hard}_3$), and reads:
\beq\label{S3F}
{\cal S}_3\,=\,\frac{N_g\hat m_D^3}{12\pi}\,+\,
T\,{\6\5m_D^2\0\6T}\,\frac{N_g\5m_D}{8\pi}\,=\,
\frac{N_g}{12\pi}\,\Bigl(\5m_D^3\,+\,
3\5m_D m_T^2\Bigr),\eeq
where we have introduced the notation
\beq\label{MTM}
{\hat m}_D^2\,=\,m_T^2\,+\,m_\mu^2,\qquad
m_T^2\,\equiv\,(2N+N_f)\,\frac{g^2T^2}{6},\qquad
m_\mu^2\,\equiv\,N_f\,{g^2\mu^2\02\pi^2}\,,\eeq
so that $T\6_T\5m_D^2 = 2m_T^2$.
Note that, formally, Eq.~(\ref{S3F}) would predict a
non-vanishing entropy in the zero temperature limit,
coming from the term $\5m_D^3$; this is, however, wrong, since,
as already mentioned, this expression has been obtained on the 
basis of a high temperature expansion and cannot be
extrapolated to small temperatures. 

Still as in the pure glue case, the two terms in 
the r.h.s. of Eq.~(\ref{S3F})
are the same as ${\cal S}^{\mathrm soft}_3$ and
${\cal S}^{\mathrm hard}_3$, respectively, because of 
the ``sum rule'' $\Delta {\cal S}_3=0\,$.
(Cf. the discussion in Sect. \ref{secSHTL}; the arguments 
leading to Eq.~(\ref{SSH}) are not changed by the addition of fermions,
since they hold for any value $\5m_D^2$ of the Debye mass.)
The only difference with respect Sect. \ref{secSHTL} is that, 
for $\mu\ne 0$, the two terms
in Eq.~(\ref{S3F}) are no longer equal to 1/4 and, 
respectively,  3/4 of the
total result (compare to Eq.~(\ref{S3quarter}));
indeed, the identity
$T\6_T\5m_D^2 = 2\5m_D^2$ is valid only at $\mu=0$.

Consider now the order-$g^3$ effect in the quark density:
since soft fermions do not contribute to order $g^3$, the
only such contribution comes from the NLO corrections
$\delta\Sigma_\pm$ to the hard fermion self-energies.
This is calculated explicitly in App.\ \ref{AppPlasmon} along the same
lines as for the entropy (cf. Eqs.~(\ref{N3A}) and (\ref{N3B}))
with the result
\beq\label{N3F}
{\cal N}_3\,=\,\frac{N_gT \5m_D m_\mu^2}{4\pi \mu}
\,\equiv\,{g^2N_g N_f\08\pi^3}\,\mu\5m_D T\,.\eeq

The previous expressions for ${\cal N}_3$ and
${\cal S}_3$ verify the Maxwell relation,
\beq
\frac{\6{\cal S}_3}{\6\mu}\,=\,
\frac{\6{\cal N}_3}{\6 T}\,=\,{g^2N_g N_f\08\pi^3}\,
\frac{\mu(\5m_D^2 + m_T^2)}{\5m_D}\,,\eeq
which is as expected, since our calculational scheme has preserved
self-consistency up to order $g^3$. These are also consistent with 
the well-known result for the sum of the ring diagrams
\cite{Kapusta},
${P}_3\,=\,{N_gT \5m_D^3}/{12\pi}\,.$
As emphasized already, this result is valid only
for high enough temperatures,  $T \gg \5m_D$.
In the opposite limit $T=0$, it is well known \cite{FM,TOIMELA}
that the sum of the ring diagrams gives a result
${P}_{ring} \sim g^4  \mu^4\log g$.

\section{QCD: Numerical evaluations}

In the following, we shall turn to a full numerical evaluation of
the entropy and the density in the approximation
${\cal S}'=0={\cal N}'$ when further approximated, firstly
by the HTL approximation (cf. Sect. \ref{secSHTL}),
secondly by also including NLO corrections
to the self-energy of hard excitations.

\subsection{HTL/HDL approximation}
\label{SecnumHTL}

We have seen that the HTL approximation (or in the case of $T=0$ and
high $\mu$ the hard-dense-loop [HDL] approximation) is sufficient
for a correct leading-order interaction term in entropy and/or density---in
contrast to a direct HTL approximation of the one-loop pressure.
On the other hand, the so-called plasmon effect of order $g^3$
is included only partly, namely only in the form of ``direct'' contributions
from soft modes; a (larger) ``indirect'' contribution
is due to NLO corrections to the self-energy of
hard particles on the light-cone as given by 
standard HTL perturbation theory.

Since we have found in our scalar toy model of Sect.~\ref{secscapprox}
that already the HTL approximation in the entropy expression with ${\cal S}'=0$
is an improvement over the leading-order
perturbative result, we shall first concentrate on
numerically including all the higher-order effects of
HTL/HDL propagators in entropy and density.

Concerning the contributions of the gluonic quasiparticles,
the task is to evaluate Eq.~(\ref{SHTL}) without expanding
out the integrand in powers of $\hat m_D/T\propto g$.

${\cal S}_{HTL}(T,\hat m_D)$ involves
two physically distinct contributions. One corresponds to the transverse
and longitudinal gluonic quasiparticle poles,
\be\label{SQP}
{\cal S}^{\mathrm QP}_{HTL}=- N_g\int\limits_0^\infty{ k^2 dk \0 2\pi^2}
 {\6\0\6T} \Bigl[2T\log(1-e^{-\o_T(k) / T}) 
+ T\log{1-e^{-\o_L(k) / T}\01-e^{-k/T}} \Bigr],
\ee
where only the explicit $T$ dependences are to be differentiated,
and not those implicit in the HTL dispersion laws $\o_T(k)$ and $\o_L(k)$.
The latter are given by the solutions of $\o_T^2-k^2=\5\Pi_T(\o_T,k)$
and $k^2=-\5\Pi_L(\o_L,k)$ with $\5\Pi_L$ and $\5\Pi_T$ given by
Eqs.~(\ref{PiL},\ref{PiT}).

Secondly, there are contributions associated with the continuum
part of the spectral weights. These read
\bea\label{SLD}
{\cal S}^{\mathrm LD}_{HTL}&=&
-N_g \int\limits_0^\infty{ k^2 dk \0 2\pi^3} 
\int\limits_0^k \! d\o {\6n(\o)\0\6T} 
\Bigl\{ 2 \arg [ k^2-\o^2+\5\Pi_T ] \nn
&&-2\Im \5\Pi_T \Re[k^2-\o^2+\5\Pi_T]^{-1} 
+ \arg [ k^2+\5\Pi_L ] - \Im \5\Pi_L \Re[k^2+\5\Pi_L]^{-1} \Bigr\}.
\eea

\begin{figure}
\epsfxsize=8.5cm
\centerline{\epsfbox[70 180 540 640]{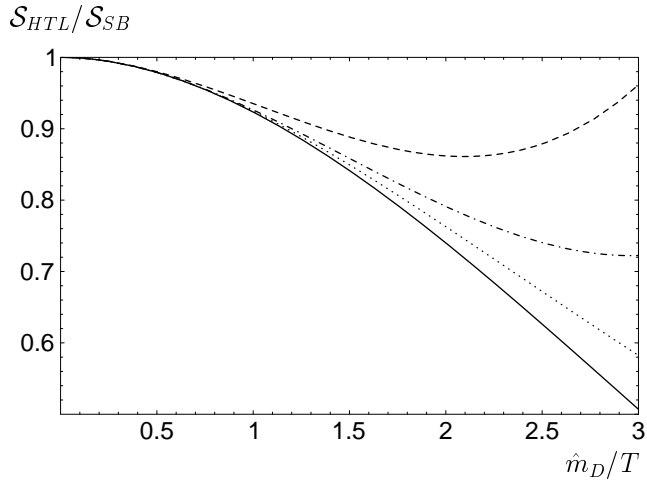}}
\vskip0.2cm
\caption{The HTL entropy per gluonic degree of freedom normalized to
its Stefan-Boltzmann value as a function of the Debye mass 
$\hat m_D(T,\mu)/T$. 
The full line gives the complete numerical
result corresponding to Eq.~(\ref{SHTL}); the dashed line corresponds
to the perturbative result to order $(\hat m_D/T)^3\sim g^3$. The
dotted line gives the entropy for scalar degrees of freedom with
momentum-independent mass $m=m_\infty=\hat m_D/\sqrt2$; its
perturbative approximant is given by the dash-dotted line.
\label{figSS0g}
}
\end{figure}

Both the Stefan-Boltzmann part ${\cal S}_{SB}$ and 
the standard perturbative $g^2$-contribution 
${\cal S}_2$ of Eq.~(\ref{S2}) are contained in the first
term of Eq.~(\ref{SQP}); all the other terms in Eqs.~(\ref{SQP}),(\ref{SLD})
are of order $g^3$ in a small-$g$ expansion.
However, if such an expansion were truncated beyond order $g^3$,
the resulting entropy would be a function of $g$ that initially
decreases with $g$, but eventually grows without bound to values larger
than ${\cal S}_{SB}$ (dashed line in Fig.~\ref{figSS0g}).

On the other hand, the full numerical result for the HTL entropy
(full line in Fig.~\ref{figSS0g}) turns out to be a monotonously decreasing
function of $\hat m_D/T$. In the case of Eq.~(\ref{SQP}), the numerical
evaluation involves solving first numerically the transcendental
equations for $\o_T(k)$ and $\o_L(k)$, and a numerical integration,
in which it is advisable to separate off the Stefan-Boltzmann value
through the replacement
$\log[1-e^{-\o_T(k) / T}] \to \log[(1-e^{-\o_T(k) / T}/(1-e^{-k/T})]$;
Eq.~(\ref{SLD}) requires two successive numerical integrations.

It is interesting to compare the rather complicated 
expression ${\cal S}_{HTL}$
with the simple scalar expression Eq.~(\ref{Sscphi4}) of the entropy
of an ideal gas of massive bosons, $2N_g{\cal S}_0(m)$,
which is basically what is
considered in the simple massive quasiparticle models of
Ref.~\cite{Peshier,LH}. If in the latter the boson masses
are identified with the asymptotic mass of the gluons, $m=m_\infty=
\hat m_D/\sqrt2$, then this reproduces the correct leading-order
interaction term in the entropy. The plasmon effect (i.e.\ the
order-$g^3$ contribution) is included
only partially, but not as $1/4$ of the complete plasmon effect,
but as $1/(4\sqrt2)$. This is because a constant thermal mass 
equal to its asymptotic value underestimates
the Debye mass by a factor of $1/\sqrt2$ and therefore the plasmon
effect by $(1/\sqrt2)^3$, which is only partially compensated by now
having $2N_g$ degrees of freedom exhibiting the analog of Debye screening
instead of only the $N_g$ longitudinal ones.

Numerically, $2N_g{\cal S}_0(m_\infty)$ 
reproduces the HTL entropy very accurately (within $\lesssim 0.1\%$)
up to $\hat m_D
\approx T$. For larger values of $\hat m_D$,
the HTL entropy leads to significantly larger
deviations from ${\cal S}_{SB}$. 
This latter fact is somewhat surprising since the plasmon effect,
which in the HTL entropy is 30\% {\em greater} than in the simple massive 
quasiparticle entropy, always counteracts the leading-order
interaction contribution, as can be seen from the
perturbative approximant of $2N_g{\cal S}_0(m_\infty)$ 
(dash-dotted line in Fig.~\ref{figSS0g}) and that of the
HTL entropy (dashed line) through order $g^3$.
This is partly due to the fact that ${\cal S}_{HTL}$ contains
also a term $\sim g^4 \log(c/g)$,  
which is not present in
the simple massive quasiparticle entropy $2N_g{\cal S}_0(m_\infty)$.

\begin{figure}
\epsfxsize=8.75cm
\centerline{\epsfbox[70 180 540 640]{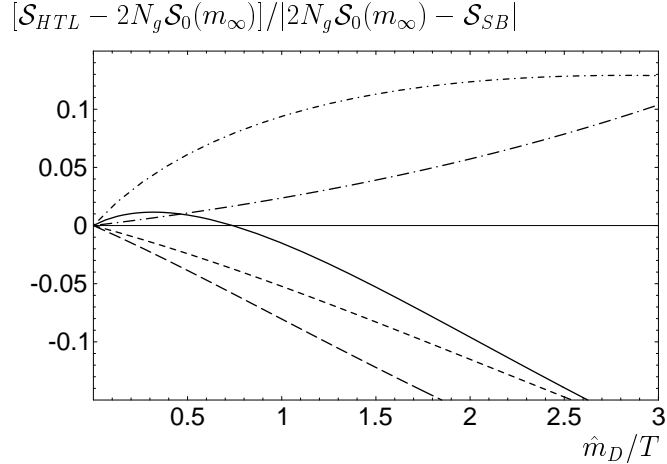}}
\caption{Relative deviation of the HTL entropy from that of a gas of massive
bosons with (constant) mass $m_\infty$ (full line).
The relative deviation of just the transverse quasiparticle contribution
is given by the uppermost dash-dotted line; the transverse Landau-damping
contribution is given by the lower dash-dotted line. The short-dashed
line gives the longitudinal quasiparticle contribution; the long-dashed
line the longitudinal Landau-damping one.
\label{figdSdetail}
}
\end{figure}

Inspecting in more detail the numerical deviation of the HTL entropy from that
of a massive gas of bosons,
one finds that the quasiparticle contribution
from the transverse modes, which is always the dominant contribution
to the entropy, by itself 
is always above $2N_g{\cal S}_0(m_\infty)$.
The transverse Landau-damping contribution is
also positive, but relatively smaller. On the other hand, both the longitudinal
quasiparticle and Landau-damping contributions are negative, resulting
in a small net deviation from the simple massive boson entropy
for small values of  $\hat m_D/T$. When normalized to the {\em deviation}
of $2N_g{\cal S}_0(m_\infty)$ from the Stefan-Boltzmann result,
the deviation of ${\cal S}_{HTL}$ from $2N_g{\cal S}_0(m_\infty)$ 
is less than
about +1\% for $\hat m_D/T < 0.739$, while negative and rapidly
growing for larger
values of $ \hat m_D/T$, as shown in Fig.~\ref{figdSdetail}.

The formulae for the fermionic contributions to the entropy 
are quite analogous to the gluonic contributions.
They read ${\cal S}_{f,HTL}
= {\cal S}_{f,HTL}^{\rm QP}+{\cal S}_{f,HTL}^{\rm LD}$
with
\bea\label{SfQP}
{\cal S}_{f,HTL}^{\rm QP}&=&NN_f \int\limits_0^\infty{ k^2 dk \0 \pi^2}
{\6\0\6T} \Bigl\{ T\log(1+\e^{-[\o_+(k)-\mu]/T}) \nn
&&\qquad\qquad\qquad+ T\log{1+\e^{-[\o_-(k)-\mu]/T} \0 1+\e^{-(k-\mu)/T}} 
+(\mu\to-\mu) \Bigr\}
\eea
where again only the explicit $T$ dependences are to be
differentiated and not those
implicit in the dispersion laws $\o_+(k)$ and $\o_-(k)$ of the fermionic
quasiparticles, which are given by the solutions of
$\o_\pm=\pm[p+\5\Sigma_\pm(\o_\pm,k)]$ with $\5\Sigma_\pm$ given
by Eq.~(\ref{SIGHTL}).

\begin{figure}
\epsfxsize=8.5cm
\centerline{\epsfbox[70 180 540 640]{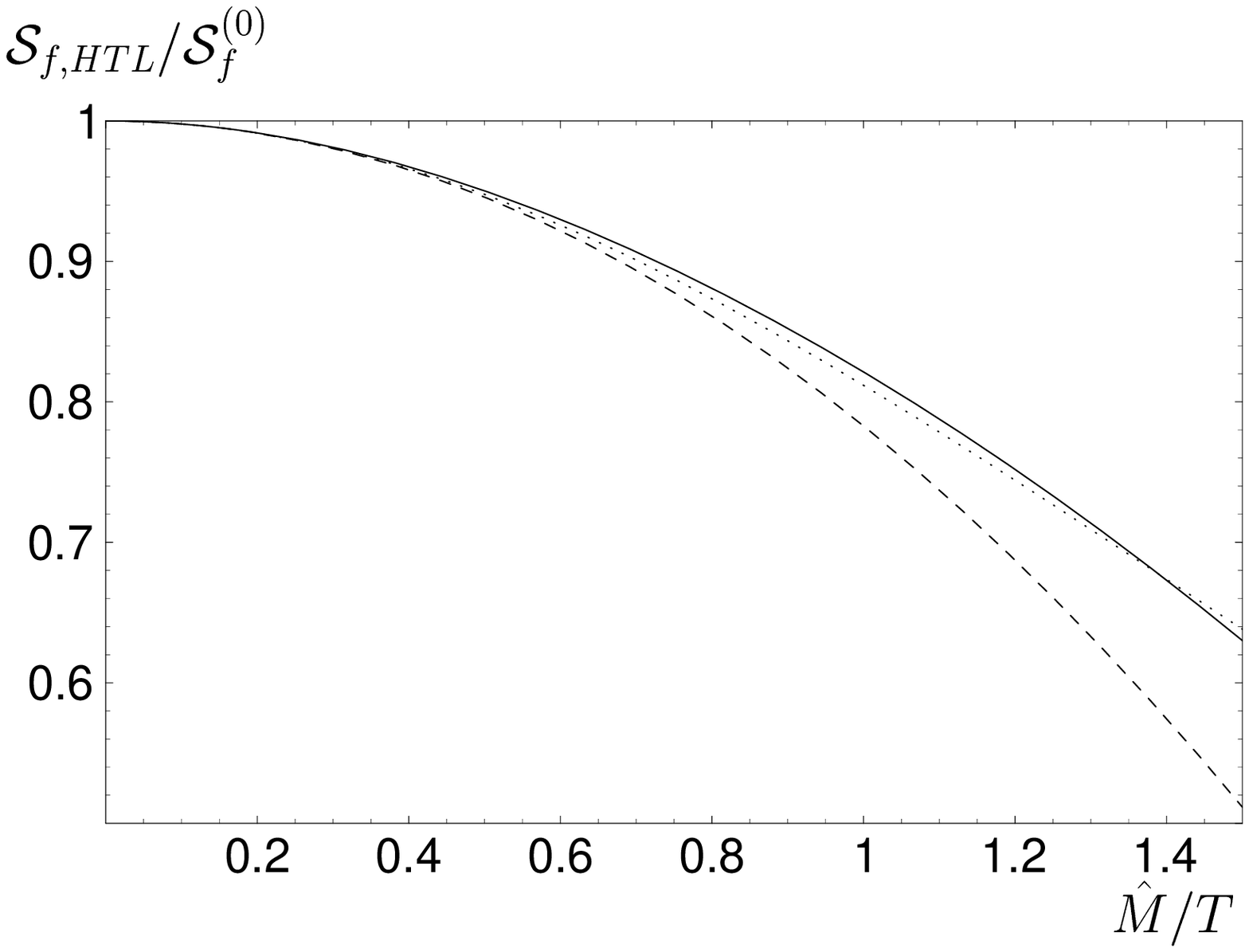}}
\vskip0.1cm
\caption{The HTL entropy per quark degree of freedom at $\mu=0$ normalized to
its free value as a function of the fermionic plasma
frequency $\hat M/T$. 
The full line gives the complete numerical
result corresponding to Eq.~(\ref{SQCDF})
in the HTL approximation; the dashed line corresponds
to the perturbative result to order $(\hat M/T)^2\sim g^2$; the
dotted line gives the entropy for a fermionic degree of freedom with
momentum-independent mass $M=M_\infty=\sqrt2\hat M$, which has 
the same perturbative approximant to order $g^2$.
\label{figSS0q}
}
\end{figure}

The fermionic Landau-damping contribution to the entropy is
\bea\label{SfLD}
&&{\cal S}_{f,HTL}^{\rm LD}=-NN_f \int\limits_0^\infty{ k^2 dk \0 \pi^3}
\int\limits_0^k \! d\o \left[{\6f_+(\o)\0\6T}+{\6f_-(\o)\0\6T}\right]  
\Bigl\{ \arg[k-\o+\hat\Sigma_+(\o,k)] \nn
&&\qquad\qquad\qquad\qquad\qquad\quad
-\Im\hat\Sigma_+(\o,k) \Re[k-\o+\hat\Sigma_+(\o,k)]^{-1} 
\nn&&\qquad\quad+\,
\arg[k+\o+\hat\Sigma_-(\o,k)]-\Im\hat\Sigma_-(\o,k) 
\Re[k+\o+\hat\Sigma_-(\o,k)]^{-1}
\Bigr\}
\eea

In the case of the gluonic contributions to the HTL entropy,
there was no difference between vanishing and non-zero chemical
potential other than the resulting different value of $\hat m_D$,
which depends on $\mu$ according to Eq.~(\ref{MD}).
For the quark contributions to the entropy, the chemical
potential enters both explicitly through the Fermi-Dirac
distribution function $f$ and through the magnitude of
the fermionic plasma frequency $\hat M$.

In Fig.~\ref{figSS0q} the results of a numerical evaluation of
the fermionic contribution to the HTL entropy
normalized to its free value is given as a function of $\hat M/T$ for $\mu=0$.
When compared with the free entropy of simple massive fermions
of mass $M=M_\infty=\sqrt2\hat M$, one finds that the HTL entropy
exceeds the latter by at most +1.2\% for $\hat M/T\approx1$, 
coincides with it at $\hat M/T\approx1.39$, and becomes lower
for larger $\hat M/T$.\footnote{Again, this good agreement requires
all quasiparticle and Landau-damping contributions together; for instance, the
normal (+) quasiparticle pole contributions alone
would give deviations which go up to about +7\% for
the range of $\hat M/T$ considered.} On the other hand,
the strictly perturbative result up to order $g^2$ is significantly
lower, but compared to the gluonic contribution the discrepancy 
is much smaller because
there is no (direct) plasmon effect in the fermionic
contributions [all order $g^3$ contributions eventually
arise from NLO corrections to $M_\infty$]. 

Turning now to the quark density, its quasiparticle and Landau-damping
contributions are
obtained by replacing $\6/\6T$
in the above formulae (\ref{SfQP},\ref{SfLD}) by $\6/\6\mu$. 

In the limit $T\to0$, the resulting
expressions can be simplified to read (for $\mu>0$)
\be\label{NHDLQP}
{\cal N}_{HDL}^{\rm QP}\Big|_{T=0} = NN_f 
\int\limits_0^{\mu}
{k^2 dk\0\pi^2} [  \theta(\mu-\o_+(k)) - \theta(\o_-(k)-\mu) ].
\ee
and
\bea
{\cal N}_{HDL}^{\rm LD}\Big|_{T=0}&=&-NN_f 
\int\limits_\mu^\infty {k^2\,dk\0\pi^3}
  \Bigl\{ \arg[k-\mu+\hat\Sigma_+(\mu,k)] \nn
&&\qquad\qquad\qquad\qquad
-\Im\hat\Sigma_+(\mu,k) \Re[k-\mu+\hat\Sigma_+(\mu,k)]^{-1} 
\nn&&\qquad+\,
\arg[k+\mu+\hat\Sigma_-(\mu,k)]-\Im\hat\Sigma_-(\mu,k) 
\Re[k+\mu+\hat\Sigma_-(\mu,k)]^{-1}
\Bigr\}.
\eea

For $\mu>\hat M$, the quasiparticle contribution (\ref{NHDLQP})
can be more explicitly written as
\be\label{NHDLQPi}
{\cal N}_{HDL}^{\rm QP}/NN_f \Big|_{T=0,\mu>\hat M} =
{\mu^3\03\pi^2}-{1\03\pi^2} \left[ \mu^3-k_+^3(\mu) \right]
-{1\03\pi^2} \left[ \mu^3-k_-^3(\mu) \right]
\ee
where $k_\pm(\mu)$ is the solution of
$\o_\pm(k_\pm)=\mu$.

The first term on the right-hand side of Eq.~(\ref{NHDLQPi})
represents the free contribution of one massless Dirac fermion, the
two bracketed terms are the corrections from the nontrivial
dispersion laws of the two fermionic quasiparticle branches.\footnote{Because
of the ``plasmino dip'', Eq.~(\ref{NHDLQPi}) becomes more
complicated for $\mu<\hat M$, but this case corresponds
to much too strong coupling to be taken seriously anyway.}

For comparison, the fermion density of a free massive Dirac fermion
with mass $M$ is given by
\be\label{NT0}
{\cal N}_0(M)\Big|_{T=0}=\left\{ \begin{array}{c@{\quad\rm for\quad}l}
{1\03\pi^2}(\mu^2-M^2)^{3/2} & \mu > M \\
0 & \mu < M
\end{array}
\right.
\ee
Identifying\footnote{Occasionally \cite{BR}, in simple quasiparticle models
of the {\em pressure} of 
fermions at high density the identification $M=\hat M$ is made.
This happens to give the correct leading-order
interaction term of order $M^2/\mu^2 \sim
g^2$ there, but only because of compensating
errors. At high densities the mass of quasiparticles at the Fermi
surface is actually
$M_\infty^2=2 \hat M^2$, but in the pressure the leading-order
interaction term is over-included by precisely a factor 2 when considering only
the expression for free particles and inserting a constant thermal mass.}
 $M=M_\infty=\sqrt2\hat M$ gives the correct leading-order
interaction term of order $g^2$,
while leading to somewhat larger values for ${\cal N}$ than
the perturbative order-$g^2$ result for all $\hat M/\mu$.

\begin{figure}
\epsfxsize=8.5cm
\centerline{\epsfbox[70 180 540 640]{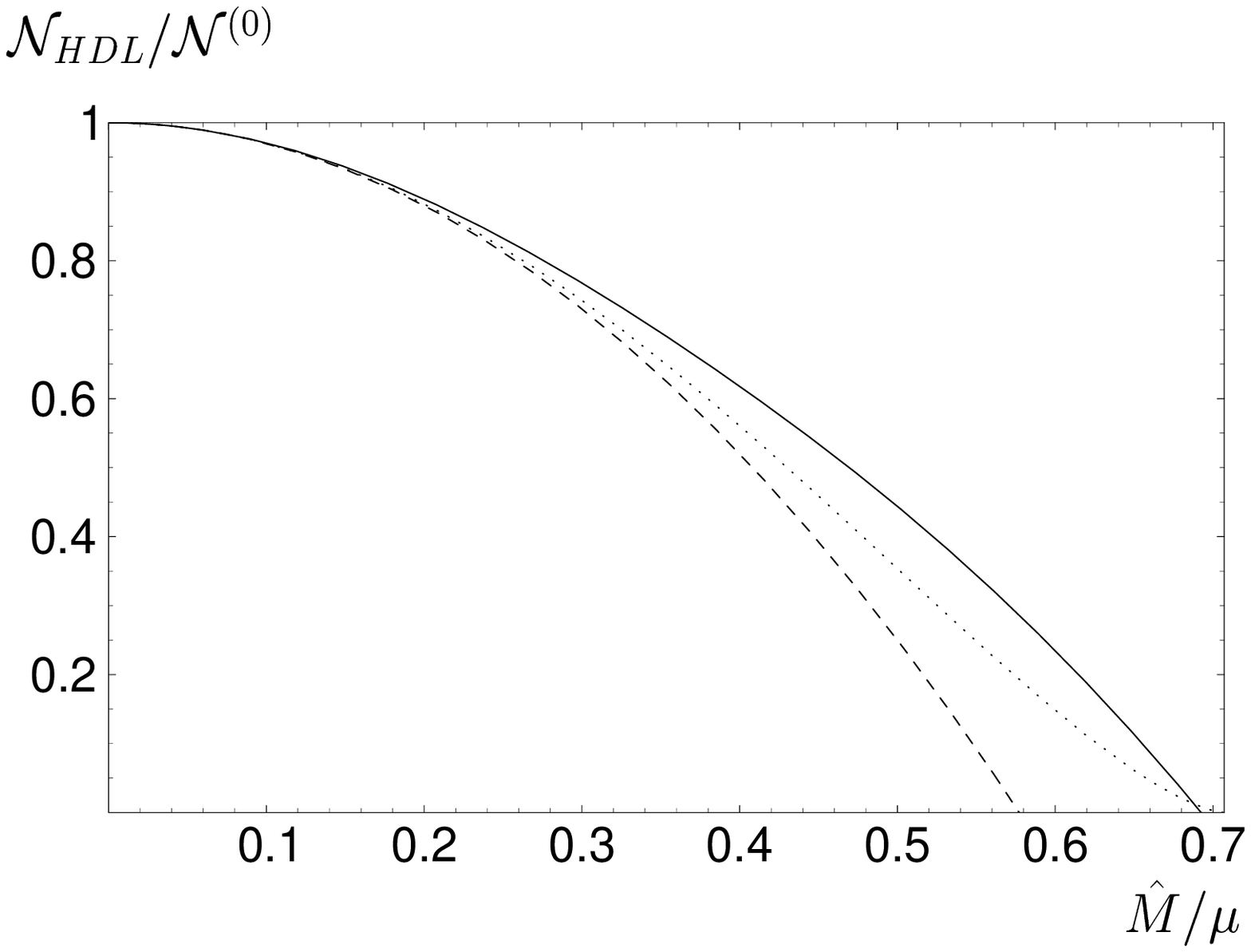}}
\vskip0.2cm
\caption{The HDL quark density per quark degree of freedom at $T=0$
normalized to
its free value as a function of the fermionic plasma
frequency $\hat M/\mu$. 
The full line gives the complete numerical
result corresponding to Eq.~(\ref{NQCD}) in the
HDL approximation; the dashed line corresponds
to the perturbative result to order $(\hat M/\mu)^2\sim g^2$; the
dotted line gives the density for a fermionic degree of freedom with
momentum-independent mass $M=M_\infty=\sqrt2\hat M$, which has 
the same perturbative approximant to order $g^2$, and vanishes
for $M\ge\mu$, i.e. $\hat M/\mu\ge 1/\sqrt2$.
\label{figNN0}
}
\end{figure}

In Fig.~\ref{figNN0} the numerical result for ${\cal N}_{HDL}$ at $T=0$
is plotted for $\hat M/\mu$ up to $1/\sqrt2$, where the fermion density
of Eq.~(\ref{NT0}), displayed by the dotted line, vanishes. The HDL result, 
which is given by the full line, is seen to drop to zero almost at
the same ratio, to wit, $\hat M/\mu \approx 0.69264$. Beyond this
point the result becomes negative, showing that
the approximation is breaking down at such high values of $\hat M/\mu$.
[Note that, since $\hat M^2=g^2\mu^2/6\pi^2$ for $N=3$ and $T=0$
(cf. Eq.~(\ref{MF})),
$\hat M/\mu \approx 0.69$ corresponds to a relatively large
coupling 
$g\approx 5.3$.]

For comparison, the strictly perturbative
result to order $g^2$ is given by the dashed line in Fig.~\ref{figNN0},
which is seen to approach zero faster than the HDL density as well as
that of a simple massive quasiparticle.

\subsection{Estimate of NLO contributions}
\label{SecnumNLO}

As we have discussed at length in the previous sections, the HTL approximation
in the entropy contains only part of the plasmon effect, 
a different source of order $g^3$ contributions comes from
NLO corrections to the gluonic and fermionic self-energies at
hard momenta on the light-cone as given by Eq.~(\ref{deltaSQCD}). From
the result (\ref{S3quarter}) we know that this NLO contribution
corresponds precisely to the second term of the right-hand side of 
Eq.~(\ref{S3F}).

In the case of the density, it is clear from the absence of
a bosonic distribution function in Eq.~(\ref{NQCD}) that
${\cal N}$ in the HTL/HDL approximation does not contain any
$g^3$ contribution, so all of ${\cal N}_3$ as given by Eq.~(\ref{N3F})
arises from the NLO correction to the quark self-energy at
hard momenta on the light-cone.

As can be seen e.g.\ from Eqs.~(\ref{b11}) and (\ref{b21}),
the NLO self-energy corrections are complicated and nonlocal
quantities. Even when evaluated on the light-cone, they do not
simply give a constant correction to the asymptotic mass, but
a nontrivial function of the (hard) momentum. In fact, there are
even contributions of the form $g^2 \hat m_D p$, which grow
larger than $g^3 T^2$ for $p \gg T$, eventually causing a break-down
of standard
HTL perturbation theory, but fortunately such contributions
are irrelevant thanks to the fact that $n(p)$ shuts off exponentially then.

Because a full inclusion of the NLO self-energy corrections
is rather complicated and computationally expensive, and
because in the applications below the magnitude of the NLO corrections,
when treated along the lines of the scalar toy model in Sect.~\ref{secscnum},
turns out to be comparatively small, we shall in the following consider
the approximation of an effective constant NLO asymptotic mass.
The complete evaluation of $\delta \Pi$ and $\delta \Sigma$, which
involves a number of technical intricacies, will
be reserved for a separate publication.
Their eventual numerical effects on the thermodynamic
potentials is work in progress, though we
do not expect them to deviate too much from the estimates
derived in this subsection.

{}From the requirement that a replacement of $m_\infty^2$ and
$M_\infty^2$ in Eqs.~(\ref{S2}), (\ref{SF2}), and (\ref{NF2})
by effective constant (i.e.\ averaged) corrections
$\bar\delta m_\infty^2$ and $\bar\delta M_\infty^2$ equals
${\cal S}_3^{\mathrm hard}$ and ${\cal N}_3^{\mathrm hard}={\cal N}_3$
(cf. Eqs.~(\ref{S3F}) and (\ref{N3F}), respectively), we have
\bea
-{1\06}N_g\bar\delta m_\infty^2 T-{1\06}NN_f\bar\delta M_\infty^2 T
&=& {1\04\pi}N_g \hat m_D m_T^2 \\
-{1\02\pi^2}NN_f \bar\delta M_\infty^2 \mu &=& N_g {T\04\pi\mu}
\hat m_D m_\mu^2
\eea
with $\hat m_D$, $m_T$, and $m_\mu$ as defined in Eq.~(\ref{MTM}).
This has the remarkably simple unique solution
\be
\label{deltamas}
\bar\delta m_\infty^2=-{1\02\pi}g^2NT\hat m_D,\quad
\bar\delta M_\infty^2=-{1\02\pi}g^2C_fT\hat m_D,
\ee
where in the latter $C_f=N_g/(2N)$.
Indeed, in Eq.~(\ref{deltamas}) both the dependence on the Casimirs
$N$ and $C_f$ as well as their proportionality to $\hat m_D$
is in accordance to one's expectations from 
the form of the corresponding HTL-resummed
one-loop diagrams of Figs.~\ref{figdPit} and \ref{figdSigma}, respectively.

However, in complete analogy to the scalar toy model of Sect.~\ref{secscnum},
we find that the magnitude of the corrections to the asymptotic masses
are such that $m_\infty^2+\bar\delta m_\infty^2$ drops to negative
values for $g\gtrsim 1$, {which 
would give rise to tachyonic singularities in the
semi-perturbative entropy result} (for $N=3$ and $\mu=0$ the {naive
strictly perturbative mass is again given by}
the shorter-dashed line in Fig.~\ref{figdm}).
For slightly higher values of $\hat m_D/T\sim g$, the same
phenomenon occurs with $M_\infty^2+\bar\delta M_\infty^2$.

In the scalar model we have seen that including the NLO correction
to the thermal mass in the approximately self-consistent
form (\ref{m2corr}) gives instead a monotonously
growing function in $g$ and very good agreement with
the exact result in the $N\to\infty$ limit even for large $g$.
For QCD, we define in analogy to Eq.~(\ref{m2corr}) the
NLA asymptotic mass through the quadratic equation
\be\label{barmascorr}
\bar m_\infty^2 = 
{g^2(N+N_f/2)T^2\06} - {g^2NT\0\sqrt2 \pi}\bar m_\infty 
\ee
and similarly for $\bar M_\infty^2$.

In contrast to the scalar case, however, where the thermal mass
and its NLO correction was momentum-independent and therefore
applicable for all momenta, the results (\ref{deltamas}) apply only at
hard momenta. Indeed, NLO corrections to thermal masses in
QCD as far as they have been calculated turn out to be rather
different at soft momenta: In Ref.~\cite{Sch}, the NLO correction
for the plasma frequency of pure-glue QCD in the long-wavelength limit
has been calculated with the result $\delta m^2_{pl.}/\hat m^2_{pl.}\approx
-0.18\sqrt{N}g$, which is only about a third of the relative
(averaged) correction of $m^2_\infty$. Moreover, the Debye mass
turns out to even receive {\em positive} corrections \cite{RDeb}
$\delta m^2_D/\hat m^2_D = +\sqrt{3N}/(2\pi) \times g \log(c/g)$, with
recent lattice simulations \cite{LP} yielding a rather large constant $c$.

For this reason we choose to leave the HTL results for the
soft gluonic propagators and self-energies completely untouched, and we
implement the NLO correction to the asymptotic mass by
introducing a cutoff scale that separates hard from soft momenta
at a scale $\Lambda=\sqrt{2\pi T\hat m_D c_\Lambda}$ which
is proportional to the geometric mean of the hard Matsubara scale $2\pi T$
and a soft scale $c_\Lambda \hat m_D$. For momenta $k\le \Lambda$
we keep the HTL approximation and for $k \ge \Lambda$ we
take the thermal gluons to have the constant asymptotic mass $\bar m_\infty^2$
of Eq.~(\ref{barmascorr}).
{This completes the definition of our present
next-to-leading approximation to the entropy ${\cal S}_{NLA}$.}

\begin{figure}
\epsfxsize=8.5cm
\centerline{\epsfbox[70 180 540 640]{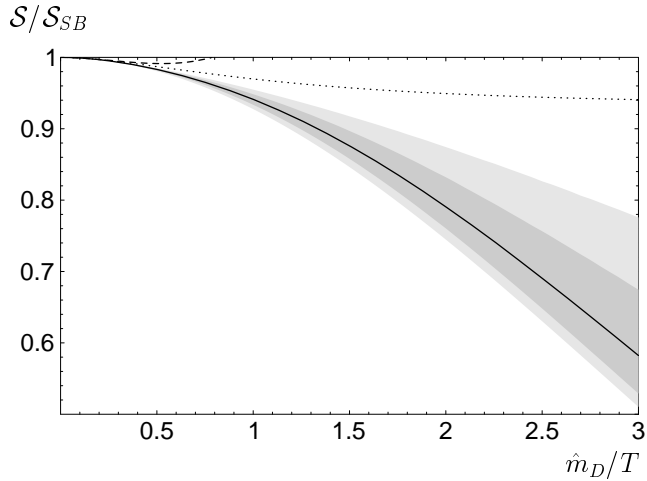}}
\vskip0.2cm
\caption{The NLA entropy obtained by including $\bar\delta m_\infty$ 
according to Eq.~(\ref{barmascorr}) for
hard momenta $k\ge\Lambda=\sqrt{2\pi T\hat m_D c_\Lambda}$
in the pure-glue entropy. The
central line in the shaded areas
corresponds to $c_\Lambda=1$, the two differently shaded areas
to the bands $c_\Lambda=\2\ldots 2$ and $c_\Lambda=\4\ldots 4$,
respectively. The dotted line corresponds
to a simple scalar model with constant mass $m$ modified according to
Eq.~(\ref{mscnlo}) such that it also contains the perturbative
pure-glue result up to and including order $g^3$; the latter is displayed
by the dashed line that leaves the plot already at $\hat m_D/T\approx 0.785$.
\label{fignSS0g}}
\end{figure}

{In Fig.~\ref{fignSS0g}, the numerical result for pure-glue QCD
with $c_\Lambda=1$ is given by the full line. The effect of varying
$c_\Lambda$ in the range $\2\ldots 2$, which keeps $\Lambda$
well in between the interval $(\hat m_D,2\pi T)$ for all values of 
$\hat m_D/T$, is displayed by the dark-grey band; the more extreme
variation $c_\Lambda=\4\ldots 4$ extends the latter by the
light-grey areas.}
We shall see, however, that in the eventual applications
to QCD at temperatures a few times the transition temperature
the resulting increase in our ``theoretical error'' will be still
moderate when compared to the renormalization scheme dependences.

{For the sake of comparison with a simpler quasiparticle model,
the dotted line in
Fig.~\ref{fignSS0g} shows the entropy
of two interaction-free scalar bosons with constant mass
\be\label{mscnlo}
\bar m^2=\2 \hat m_D^2 \( 1 - {4\sqrt2 -1\0\pi}{\bar m\0T} \)
\ee
which mimicks the NLA result (\ref{barmascorr}), but adjusted such
as to reproduce 
the perturbative QCD result up to and including order
$g^3$ in this simpler model}.

In the fermionic quantities ${\cal S}_f$ and ${\cal N}$, which
because of the absence of Bose enhancement are less sensitive
to the soft scale, we implement the fermionic analog of
Eq.~(\ref{barmascorr}) by rescaling $\hat M^2$ at {\it all}
momenta for simplicity (as in the scalar case in Sect. II.D).

\subsection{Renormalization-group improvement}

In the HTL/HDL approximation, all the gluonic and fermionic
contributions above depend on the
numerical value of the HTL/HDL masses $\hat m_D^2$ and $\hat M^2$,
respectively, which are proportional to $\alpha_s=g^2/4\pi$.
The latter is a renormalization scheme and renormalization scale $(\bar\mu)$
dependent quantity, and so are therefore our results for entropy
and density. Following Ref.~\cite{BN} we adopt modified minimal subtraction and
assume that an optimal
choice of the renormalization scale should be found around the scale of the
Matsubara frequencies, $2\pi T$, or in the case of zero
temperature and finite density around the scale of the
diameter of the Fermi sphere, $2\mu$. After all, the hard thermal/dense
loops are generated by hard excitations, as are in fact
the NLO contributions to the asymptotic masses
within HTL/HDL perturbation theory.

Exactly as done in Ref.~\cite{ABS} in a direct HTL resummation
of the thermodynamic pressure, we put in the running of the coupling
by hand and choose it to be determined by the 2-loop renormalization group 
equation according to 
\be\label{alphas}
\alpha_s(\bar\mu)={4\pi \0 \beta_0 \bar L(\bar\mu)}
\(1-{2 \beta_1 \log(\bar L(\bar\mu))\0 \beta_0^2 \bar L(\bar\mu)} \)
\ee
with $\bar L(\bar\mu)=\log(\bar\mu^2/\Lambda^2_{\overline{\mathrm MS}})$
and
\be
\beta_0=(11 N - 2 N_f)/3,\quad
\beta_1=(34 N^2 - 13 N N_f + 3 N_f/N)/6.
\ee

\subsubsection{Entropy}

At least at zero density, lattice results relate the QCD scale parameter
$\Lambda_{\overline{\mathrm MS}}$ to the critical temperature $T_c$,
which in accordance with Ref.~\cite{TcLa} we choose as $T_c=1.14 
\Lambda_{\overline{\mathrm MS}}$, both for pure-glue QCD and
also for $N_f\not=0$,
since lattice data indicate only a weak dependence of the {\em ratio} $T_c/
\Lambda_{\overline{\mathrm MS}}$ on the number of quark flavor.

Putting $\bar\mu=c_{\bar\mu}2\pi T$ in Eq.~(\ref{alphas}) and
assuming $c_{\bar\mu}\sim 1$ prescribes
reasonably small values for $\alpha_s$ and thus for $\hat m_D/(2\pi T)$ and
$\hat M/(\pi T)$ 
for all $T>T_c$ so as to make it interesting to
compare the above HTL and NLA expressions with nonperturbative results
from lattice gauge theory. Indeed, we have found that, for
$\hat m_D \ll 2\pi T$ and $\hat M \ll \pi T$, the deviation from
the free Stefan-Boltzmann result is small enough to make a
semi-perturbative picture minimally tenable, although it is clear that
the physics of the phase transition itself is completely beyond reach.
On the other hand, the strictly perturbative results up to and including
the order $g^3$ are such that entropy and pressure would be much higher
than their Stefan-Boltzmann values, indicating a complete loss of
convergence of strict thermal perturbation theory.

In order to have some indication of the theoretical uncertainty involved,
we consider, again as done in Ref.~\cite{ABS}, a variation of the
renormalization scale by a factor of $c_{\bar\mu}=\2\ldots2$.

\begin{figure}
\epsfxsize=8.5cm
\centerline{\epsfbox[70 180 540 640]{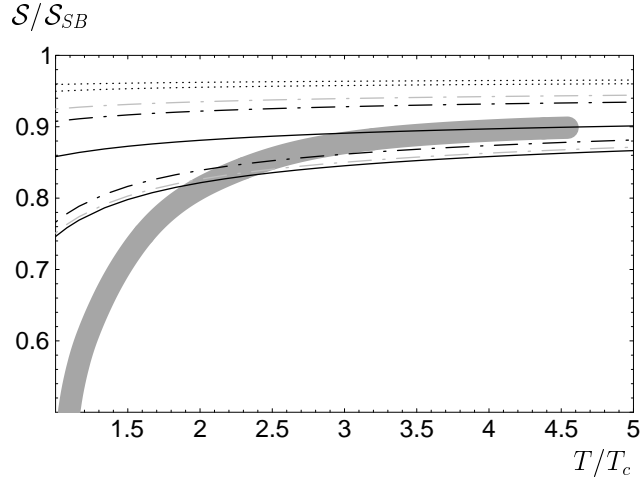}}
\vskip0.2cm
\caption{Comparison of the HTL entropy (full lines), the NLA results
for $c_\Lambda=\2\ldots2$ (dash-dotted lines)
as well as $c_\Lambda=\4\ldots4$ (gray dash-dotted lines), and the free
entropy of bosons with mass (\ref{mscnlo}) such as to reproduce
the correct perturbative plasmon effect (dotted lines), all with
$\overline{\hbox{MS}}$ renormalization scale $\bar\mu=\pi T\ldots
4\pi T$, with the lattice result of Ref.~\protect\cite{Boyd}
for pure SU(3) gauge theory (dark-gray band).
\label{figSg}}
\end{figure}

For purely gluonic QCD, the lattice results involve the least uncertainties.
In Ref.~\cite{Boyd}, the thermodynamic potentials of pure SU(3) gauge
theory have been calculated from plaquette action
densities on lattices up to $8\times 32^3$ for temperatures
up to about $4.5T_c$ and extrapolating
to the continuum limit by comparing different lattice sizes.
The lattice result for the entropy density is rendered in Fig.~\ref{figSg}
by a grey band whose thickness is meant to give a rough idea of
the errors reported in Ref.~\cite{Boyd}.

Our result for the HTL entropy as displayed in Fig.~\ref{figSS0g}
translates into a range of values bounded by the choices $\bar\mu=\pi T$
(lower full line) and $\bar\mu=4 \pi T$ (upper full line).
This already gives a remarkably good approximation of the lattice
result for $T\gtrsim 2 T_c$, somewhat underestimating the values
at higher temperatures. In all of this the parameter $\hat m_D/T$
takes on values in the range $\sim 1\ldots 2$, where we have found
the entropy of simple massive bosons to give only slightly larger
results (cf. Fig.~\ref{figSS0g}).

Now the HTL entropy
contains only part (here 1/4) of the plasmon effect. 
{The latter is completely included in the NLA entropy as
defined after Eq.~(\ref{barmascorr}). In Fig.~\ref{figSg},
${\cal S}_{NLA}$ is represented by the area
bounded by the black}
dash-dotted lines, where the lower one corresponds
to the choice $\bar\mu=\pi T$ and $c_\Lambda=2$, and the higher
one to $\bar\mu=4\pi T$ and $c_\Lambda=1/2$. For this range of 
$c_\Lambda$ the scale $\Lambda$ 
{remains well separated from both $\hat m_D$ and $2\pi T$
for all $T>T_c$. 
Extending the range of $c_\Lambda$ to $c_\Lambda=1/4\ldots4$
gives the area bounded by the gray dash-dotted lines. Although
$\Lambda$ now varies all the way from $\hat m_D$ to $2\pi T$, the
error band is only moderately enlarged}.\footnote{Although 
in Fig.~\ref{fignSS0g}
there was a noticeable increase in the error band for the NLA results
when increasing the range of $c_\Lambda$,
this does not affect so much the {\em total} error because the lower bound,
which corresponds to higher values of $\hat m_D/T$, is moved further down
only by increasing $c_\Lambda$, where the addition in the error
band is small; the upper bound on the other hand corresponds to
smaller values of $\hat m_D/T$, where the upward increase in the
error is correspondingly smaller.
} 

Evidently, the NLA estimates based upon Eq.~(\ref{barmascorr})
do not move
away too much from the HTL results, which first of all is what is
required to make our
semi-perturbative procedure tenable.
What is more, the results show a surprisingly good
agreement with the lattice results for temperatures
greater than 2\ldots3 times the critical temperature.

Recently, in Ref.~\cite{Okamoto} the results of Ref.~\cite{Boyd} have
been reproduced within errors by using a renormalization-group improved
lattice
action for temperature up to $3.5T_c$. The results of Ref.~\cite{Okamoto}
for the pressure
are systematically higher by about 5\ldots2\% for temperatures 2\ldots3.5$T_c$.
For the entropy, which has not been extracted explicitly in Ref.~\cite{Okamoto},
this would imply a result that is centered around the upper boundary
of the grey band in Fig.~\ref{figSg} for $T\approx 3 T_c$, 
and slightly flatter around $T\sim 2 T_c$, all with slightly reduced error
bars.
If anything, the agreement with our HTL and NLA results appears to be
even a bit improved.

Comparing finally with the entropy of free massive bosons with
mass according to Eq.~(\ref{mscnlo}) such as to reproduce the correct
perturbative plasmon effect, this is included in Fig.~\ref{figSg}
as the band bounded by the dotted lines corresponding to
$\bar\mu=\pi T\ldots 4\pi T$. Since the renormalization scale
dependence decreases with decreasing deviation from the Stefan-Boltzmann
value, this band is rather narrow. It is also clearly in lesser
agreement with the lattice data, {which thus favor the
momentum-dependent inclusion of NLO corrections to the thermal masses
that follows from NLO perturbation theory and that we have
modelled in our NLA estimates.}

\begin{figure}
\epsfxsize=8.75cm
\centerline{\epsfbox[70 180 540 640]{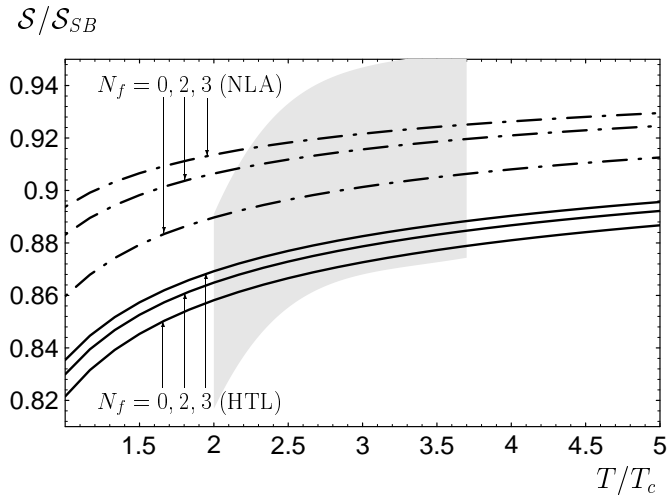}}
\vskip0.2cm
\caption{Comparison of the HTL entropy (full lines) and the NLA result
with $c_\Lambda=1$ (dash-dotted lines) for flavor numbers
$N_f=0,2,3$, all with the central choice of
$\overline{\hbox{MS}}$ renormalization scale $\bar\mu=2\pi T$. 
The estimate of a continuum extrapolation of the 
lattice result for $N_f=2$ as
reported in Ref.~\protect\cite{Karsch,KLP} and its estimated error,
converted to ${\cal S}/{\cal S}_{SB}$, is given by the gray band.
Notice the blown-up 
scale of the ordinate compared to Fig.~\ref{figSg}.
\label{figSNf023}}
\end{figure}

In Fig.~\ref{figSNf023}, the central results for the HTL and NLA entropy
($\bar\mu=2\pi T$ and $c_\Lambda=1$) are displayed together with the
results for $N_f=2$ and 3. Only a rather weak dependence on $N_f$ is
found in this (greatly magnified)
plot where the entropy is normalized to the free value,
and $T$ to the respective ($N_f$-dependent) $T_c \propto
\Lambda_{\overline{\hbox{MS}}}$.

These results are in good agreement with recent lattice results\cite{Karsch}
for $N_f=2$ and their estimated extrapolation to the continuum limit
and to the limit of massless quarks as we have already noted in 
Ref.~\cite{PLB}. In Fig.~\ref{figSNf023}, a conversion of the lattice result 
to the entropy is included as a gray band, and, indeed, for $T/T_c\gtrsim
2.5$ our NLA estimate turns out to lie {close to} the center of the
estimated error band of the lattice result.

The further result that for $N_f=3$ our NLA estimate for
${\cal S}/{\cal S}_{SB}$ as a function of $T/T_c$ 
is {approximately} the same fits nicely
to the recent lattice data for $N_f=3$ \cite{KLP}, which are consistent with a
coincidence of the asymptotic values of $P/P_{SB}$ and also for
${\cal S}/{\cal S}_{SB}$. A more detailed comparison of our
results with the lattice data, in particular at smaller
temperatures, is hardly worthwhile in view of the large uncertainties
associated with the extrapolation to the massless continuum limit.\footnote{In 
a recent paper the authors of Ref.~\cite{HLRST}
have reported an extremely good fit of the entire lattice data using
only the perturbative
first-order correction to the pressure, a bag constant and
a numerically integrated 2-loop $\beta$-function. However, this
agreement has been achieved with the lattice results which still contain
finite-cut-off effects. In Ref.~\cite{Karsch,KLP}, the size of
the estimated correction for the continuum limit
is given as $+15\pm5$\%. These corrections are essential for
the good agreement with our results as shown in Fig.~\ref{figSNf023}.
Conversely, the results of Ref.~\cite{HLRST} remain even below
the plot area of Fig.~\ref{figSNf023}.}

{In our previous works \cite{PRL,PLB} we have been considering
a simple Pad\'e-improved inclusion
of the NLO asymptotic mass correction in place of the NLA 
form (\ref{barmascorr}). A comparison of the respective results
shows that our estimate of the effects of an approximately self-consistent
treatment of NLO corrections to the self-energies is fairly robust,
with the main uncertainties coming from the choice of the
renormalization scale.}

\subsubsection{Density}

For nonvanishing chemical potential, where lattice data are missing to
determine precisely the critical temperature or density in terms of
$\Lambda_{\overline{\mathrm MS}}$, we can nevertheless translate our
results into functions of $T/\Lambda_{\overline{\mathrm MS}}$ and
$\mu/\Lambda_{\overline{\mathrm MS}}$ provided we choose
the renormalization scale $\bar\mu$ as a suitable combination of $T$ and
$\mu$. If, as we have assumed, the spacing of the
Matsubara frequencies, $2\pi T$, gives a good choice for the
renormalization scale $\bar\mu$ at zero density, it seems plausible
to adopt the diameter of the Fermi sphere, $2\mu$, in the case of
zero temperature. This choice of a relative factor of $\pi$ is particularly
natural when considering the form of the leading-order result for the fermionic
thermal masses, Eq.~(\ref{MF}), where $T$ and $\mu/\pi$ appear on equal
footing.

In Fig.~\ref{figN3} we give the numerical
results for the quark density ${\cal N}_{HDL}$
at $T=0$ for $N=3$ and $N_f=3$ as a function of 
$\mu/\Lambda_{\overline{\mathrm MS}}$ for the
range $\bar\mu=\mu\ldots 4\mu$. In this case we do not attempt to
include NLO corrections, for they do not contribute terms of
order $g^3$. NLO corrections to the hard fermion self-energy
are in fact responsible for completing the
plasmon effect at order $g^4\log(g)$,
but a complete calculation of the former would be needed to determine that
part of the constant under the logarithm that comes from the
spectral properties of quasiparticles rather than explicit order-$g^4$
interactions, which are dropped in the approximation ${\cal N}'=0$.

The dashed line in Fig.~\ref{figN3} gives the strictly perturbative
result at order $g^2$. The result corresponding to a simple quasiparticle
model with mass $M=M_\infty$ is not included; from Fig.~\ref{figNN0}
it is clear that it is in between the HDL result and the order-$g^2$
one, and somewhat closer to the latter.

The perturbative result up to and including order $g^4$
has been calculated by Freedman and McLerran \cite{FM} {
and by Baluni \cite{Balu}. However, it has been
obtained in a particular momentum subtraction scheme.
In order to convert this to the gauge-independent
$\overline{\hbox{MS}}$ scheme, one should replace the
scale parameter $\mu_0$
in Ref.~\cite{FM} ($M$ in Refs.~\cite{Balu,Kapusta})
according to
\begin{equation}\label{MSbarconversion}
\mu_0=\bar\mu \exp\left\{ [(151+36\a+9\a^2)N-40N_f]/[24(11N-2N_f)] \right\},
\end{equation}
where $\alpha$ is the gauge parameter used in the momentum subtraction scheme
calculation.
In particular for $N=3$ and uniform chemical potentials one finds\footnote{The
numerical coefficients have been assembled from Eq.~(4.46) of Ref.~\cite{Balu}
using Eq.~(\ref{MSbarconversion}), thus avoiding some
unnecessary accumulated rounding errors that are present in
the final results (5.14) and (5.15) of Ref.~\cite{Balu}.
The actual error in the above
numerical coefficients is probably about 1 in the next-to-last digit.}
\begin{eqnarray}
P/P_0 &=& 1-2{\alpha_s(\bar\mu)\0\pi}- \Bigl[ 10.347 - 0.536 N_f + N_f \log{N_f
        \alpha_s(\bar\mu)\0\pi}  \nonumber\\
&&\qquad\qquad\qquad\qquad\qquad\qquad 
+ \;(11-{2\03}N_f)\log{\bar\mu\0\mu}\; \Bigr]
({\alpha_s(\bar\mu)\0\pi})^2+O(\alpha_s^3), \\
{\cal N}/{\cal N}_0 &=&1-2{\alpha_s(\bar\mu)\0\pi}- 
\Bigl[ 7.597 - 0.369 N_f + N_f \log{N_f
        \alpha_s(\bar\mu)\0\pi} \nonumber\\
&&\qquad\qquad\qquad\qquad\qquad\qquad 
+ \;(11-{2\03}N_f)\log{\bar\mu\0\mu}\; \Bigr]
({\alpha_s(\bar\mu)\0\pi})^2+O(\alpha_s^3).
\end{eqnarray}
With $N_f=3$, this is included in Fig.~\ref{figN3}
 by the dotted lines\footnote{In Fig.~4 of our previous
publication Ref.~\cite{PLB}, the perturbative
order-$g^4$ result was not correctly included because of
an incomplete scheme conversion.}. 
Although the perturbative order-$g^4$ result constitutes a
substantial correction of the order-$g^2$ result, perturbation
theory at zero temperature and high densities is clearly much
better behaved than at high temperatures---the interaction terms
are increased by less than 50\% for $\mu \gtrsim 2 
\Lambda_{\overline{\mathrm MS}}$ ($\mu \gtrsim 3 
\Lambda_{\overline{\mathrm MS}}$ in the case of $P/P_0$).

\begin{figure}
\epsfxsize=8.5cm
\centerline{\epsfbox[70 180 540 640]{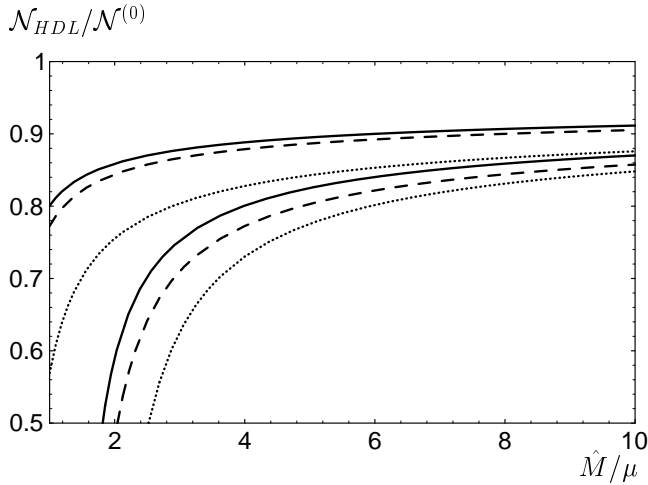}}
\vskip0.2cm
\caption{The result for the quark density for $N_f=3$ in the HDL
approximation for $\bar\mu=\mu\ldots 4\mu$ (full lines) compared with
the perturbative results at order $g^2$ (dashed lines) and order
$g^4$ (dotted lines).
\label{figN3}}
\end{figure}

On the other hand, the nonperturbative ${\cal N}_{HDL}$ result
is rather close to the perturbative order-$g^2$ result,
showing even a slight {\em decrease} of the interaction contribution compared
to the latter. The ${\cal N}_{HDL}$ result contains already a 
fraction of the coefficient of the $\a_s^2\log(\a_s)$ term, 
together with a subset of the true
higher-order contributions. It would be interesting to see
how an NLA calculation, which would complete the
$g^4\log(g)$ coefficient, compares with the perturbative order-$g^4$ result.
We intend to investigate that in future work.
}

\section{Conclusions and outlook}

We have shown that it is possible to perform
a resummation of HTL's which is free of overcounting and UV
problems through approximately self-consistent calculations of
the thermodynamical functions of QCD, without the
need for thermal counterterms. The two-loop skeleton approximation
for the free energy reduces to effectively one-loop expressions
for the entropy and the density but with dressed propagators.
With the latter approximated by the HTL/HDL propagators we
reproduce correctly the
leading-order interaction terms\footnote{In Ref.~\cite{Peshier99}
an attempt has been made to resum the HTL self-energies directly
on the level of the skeleton representation of the free energy.
However this relies on an arbitrary modification of the functional $\Phi$
which, although it yields the correct $g^2$ effects (by construction), 
does not respect the correct combinatorial factors and thus violates
the proper counting of the higher-order diagrams.}.
In fact, the latter can be
expressed entirely in terms of the asymptotic thermal mass.
The so-called plasmon-effect contributions of order $g^3$ on
the other hand are only partly accounted for
by HTL self-energies and propagators; the remaining contributions
arise, rather unconventionally, from NLO corrections to the self-energy of
hard particles at the light-cone as given by standard
HTL perturbation theory.

This is to be contrasted with a direct HTL resummation of the
one-loop pressure \cite{ABS}. There the plasmon effect is contained
completely in the soft contributions, whereas the leading-order interaction
terms are over-included and corrected only in a
(very complicated) two-loop calculation. 

We would like to recall that while the HTL and NLA approximations
that we  have considered are manifestly gauge independent, this is
not the case for our starting point, the self-consistent 
$\Phi$-derivable two-loop order approximation itself. {
The corresponding gap equation would involve unphysical
gauge dependent features as well as an incomplete lowest order
$\beta$-function both of which enter at order $g^4$, i.e.\ beyond
the perturbative accuracy of a two-loop approximation.
These are automatically dropped in our current approximations.}
Further improvements, beyond our HTL and NLA approximations
would require to also improve upon the self-consistent two-loop
approximation. In order to achieve gauge independence in
(approximately) self-consistent resummations one should
obviously turn to approximations which
include dressed vertices, using for instance the  formalisms which have been
developed  long ago by de Dominicis and Martin \cite{LW} and
also Freedman and McLerran \cite{FM}. The strategy, in principle,
would be to include vertex corrections,
together with increasingly better approximations
for the quasiparticle propagators. That is, with increasing number
of loops in $\Phi$, the building blocks in this scheme---the
self-consistent propagators and vertices---should be also improved.
However, a practical implementation of such a scheme in the case of
nonabelian gauge theories  seems to be hopelessly complicated. It is
therefore gratifying that  the approximate propagator
renormalization that we have  presented turns out to be already a good 
approximation. 

In the expressions that we use for the entropy and density
the main contribution comes from the vicinity of the light-cone
where hard thermal loops remain accurate also at hard momenta and 
provide the asymptotic masses. We have proposed a
procedure of including NLO corrections {through
approximately self-consistent}
corrections to the thermal masses of the hard excitations only. 
The  NLO corrections to the asymptotic masses can
be calculated more accurately by means of standard
HTL perturbation theory, the details of
which are postponed to a forthcoming publication. 

The numerical evaluation of our results 
combined with a two-loop
renormalization group improvement turn out to
compare remarkably well 
with available lattice data at zero quark chemical potential,
which supports the picture according to which much of the effects
of the interactions in the quark-gluon plasma can be
adequately described by means of
weakly interacting gluonic and fermionic (HTL) quasiparticles.

Extensions of the present work which are in progress concern
the evaluation for general $\mu>0$ and $T>0$, and the integration
of entropy and density to the thermodynamic pressure $P(\mu,T)$,
similarly to what has been done in simple quasiparticle models
in Ref.~\cite{PKSmu} (see also \cite{PLB}).
Maxwell's relations, which constitute the corresponding integrability
conditions, are satisfied up to and including order $g^3$ upon
inclusion of the NLO contributions; beyond that order they
give constraints on a possible renormalization-group improvement
and it seems interesting to further pursue the present
approach of combining the physical content of the
perturbatively derivable hard thermal/dense self-energies
with nonperturbative expressions for entropy and density,
which in self-consistent two-loop order approximations only depend
on the spectral properties of quasiparticle excitations.

\acknowledgments

The authors would 
like to thank R. Baier, D. B\"odeker, 
E. Braaten, U. Heinz, F. Karsch, S. Leupold,
and M. Strickland
for valuable discussions.

\appendix

\section{The plasmon effect in the QCD entropy}
\label{AppPlasmon}

In this Appendix, we shall explicitly verify that our approximation
for the entropy (cf. Eqs.~(\ref{SQCDD}) and (\ref{SQCDF}))
contains indeed the right perturbative correction of order $g^3$.
Recall that
${\cal S}_3$ involves two types of contributions:
the LO entropy of the soft gluons (longitudinal and transverse;
cf. Eqs.~(\ref{SL3}) and (\ref{ST31})), and the NLO entropy
of the hard particles (transverse gluons and fermions),
as determined by the corresponding NLO self-energies on
the light-cone (cf. Eqs.~(\ref{ST32}) and (\ref{deltaSQCD})).

Our strategy will be as follows: In Sect. A.1, we shall
rewrite the soft gluon entropy in a way which will be
convenient later. Then, in Sects. A.2 and A.3 we shall compute
the NLO self-energy $\delta \Pi_T$ of a hard transverse
gluon, and the corresponding contribution $\delta{\cal S}_T$
to the entropy. This will complete the derivation of the
plasmon effect for a purely gluonic plasma. The extension
to a plasma with fermions will be finally considered, in Sect. A.4.

\subsection{The entropy of soft gluons}

From Eqs.~(\ref{SL3}) and (\ref{ST31}), the order-$g^3$
contribution of the soft gluons reads
\bea\label{SHTLD1}
{\cal S}_3^{\mathrm soft}&=&
-\int\!\!{d^4k\0(2\pi)^4}\,\frac{1}{\omega}\left\{\Im
\Bigl[\log(1+ D_0\hat\Pi) - \hat\Pi D_0\Bigr]\,-\,
\Im\hat \Pi\Re(\hat D-D_0)\right\}\nonumber\\
&=&{\cal S}_3^{(a)}+\Delta {\cal S}_3,\eeq
where ${\cal S}_3^{(a)}=({\del} P_3/\del T)|_{\5m_D}$
(cf. Eq.~(\ref{SDERPa})), and $\Delta {\cal S}_3$ is defined
in Eq.~(\ref{DELTASS}). In Sect. \ref{secSHTL}, we have mentioned
that $\Delta {\cal S}_3$ has been numerically found to vanish,
because of a compensation between the electric and the magnetic 
contributions to Eq.~(\ref{DELTASS}). In what follows, however,
we shall not use this information, but rather consider separately
these electric and magnetic contributions, and show how they
combine with the corresponding contributions to the NLO entropy
of the hard particles $\delta{\cal S}$. Specifically,
we shall verify that the identity in Eq.~(\ref{S-H}) holds separately
in the electric and the magnetic sector.

To this aim, it is convenient to rewrite Eq.~(\ref{DELTASS})
in a slightly different form by
using $\Im\5\Pi \Re\5 D = 
\Im (\5\Pi \5 D) - \Re \5\Pi\Im\5 D$,
and then integrating the first term:
\beq
\int\!\!{d^3k\0(2\pi)^3}
\int\frac{d\o}{2\pi\o}\,\Im \Bigl(\5\Pi_L(\5 D_L - D_L^{(0)})\Bigr)
\,=\,{\5 m_D^2\02} \int\!\!{d^3k\0(2\pi)^3}\left({1\0k^2} - {1\0{k^2 +
\5 m_D^2}}\right)\,=\,{\5 m_D^3\0{8\pi}}\,.\eeq
This yields (with $\5\rho_{L,T}=2\Im\5 D_{L,T}$, 
cf. Eq.~(\ref{rhos})):
\bea\label{SHTLL}
\Delta {\cal S}_3=
N_g\int\!\!{d^4k\0(2\pi)^4}\,{1\02\o}\left\{
\5\rho_L \Bigl(\Re \5\Pi_L-\5m_D^2\Bigr)-
2\Bigl(\5\rho_T - \rho_T^{(0)}\Bigr)\Re \5\Pi_T
\right\}\,\equiv\,
\Delta {\cal S}_L^{(3)}+\Delta {\cal S}_T^{(3)}\,,\eea
where we have also used
the following ``sum-rule'' (cf. Eq.~(\ref{Dspec})):
\beq\label{sumruleL}
\int\frac{d\o}{2\pi}\,\frac{\5\rho_L(\o,k)}{\o}\,=\,
\,\frac{1}{k^2}\,-\,\frac{1}{k^2+\5m_D^2}\,.\eeq
Given the complicated structure of the HTL self-energies
and spectral functions, the integrals in Eq.~(\ref{SHTLL})
cannot be further evaluated in closed form.
But this is actually not needed: indeed, the cumbersome terms
in these expressions will be shortly shown to cancel against similar 
terms in ${\cal S}_3^{\mathrm hard}$, the order-$g^3$ contribution
of the hard particles, to be computed below.

\subsection{The NLO gluon self-energy}

We shall now compute the NLO self-energy contribution
$\delta \Pi_T$ of a hard transverse gluon.
This is determined by the effective
one-loop diagrams in Fig.~\ref{figdPit} where one of the internal
lines is a soft gluon ($L$ or $T$)
with the HTL-dressed propagator $\5 D-D_0$ (the subtraction 
of the free propagator $D_0$ is ensures that
the loop integral is saturated by soft momenta).
The other line in each of these diagrams is hard and transverse, and
therefore undressed.

We are interested only in the transverse projection of $\delta \Pi_{\mu\nu}$:
\beq
\delta \Pi(p) \equiv \delta \Pi_T (p)
\equiv \,{1\02}\,(\delta^{ij}-\hat p^i \hat p^j)\,\delta \Pi_{ij}(p).
\eeq
We write $\delta\Pi_T=\delta \Pi^l+\delta\Pi^t$,
where the upper indices refer to the soft internal lines in
these diagrams, and compute only the 
longitudinal contribution $\delta \Pi^l$ in more detail.
(The calculation of the transverse contribution is completely
analogous.)
This involves two of the diagrams in Figs.~\ref{figdPit}: 
the tadpole $\delta \Pi^l_a$ and the non-local
diagram $\delta \Pi^l_b\,$.
The tadpole gives
\beq\label{tad}
\delta \Pi^l_a&=&-g^2 N\int [{\rm d}k]
\Bigl(\5 D_L(k)-D_L^{(0)}(k)\Bigr)=-g^2 N\int\!\!{d^4k\0(2\pi)^4}
\,\5\rho_L(k_0,k)n(k_0)\nn&\simeq&
-g^2 NT\int\!\!{d^4k\0(2\pi)^4}\,{1\0k_0}\,\5\rho_L(k_0,k)
\,=\,-\,\frac{g^2NT\5m_D}{4\pi}\,,\eeq
where the Matsubara sum in the first line has been performed 
by using the spectral representation (\ref{Dspec}), and
in the second line we have replaced $n(k_0)\simeq T/k_0$ (as 
appropriate at soft energies), and then performed the energy
integral with the help of the sum-rule (\ref{sumruleL}).
The final result  in Eq.~(\ref{tad}) is indeed of order
$g^2T\5m_D \sim g^3 T^2$, as expected.

The non-local diagram in Fig.~\ref{figdPit}.b yields:
\beq\label{b1}
\Bigl(\delta \Pi_{b}^l\Bigr)_{ij}
(p)\,=\,-2\,{g^2N\02}\int [{\rm d}k]\,
(2p_0+k_0)^2\,D^{(0)}_{ij}(p+k)\Bigl(\5 D_L(k)-D_L^{(0)}(k)\Bigr),\eeq
where $D^{(0)}_{ij}(q)$ is the free magnetic propagator in the
Coulomb gauge,
\beq
D^{(0)}_{ij}(q)\,=\,(\delta_{ij}-\hat q_i \hat q_j)\,{-1\0
q_0^2-q^2}\,,\eeq
and the factor 2 in front of the integral reflects the
two possible ways to choose the soft longitudinal line
among the two internal lines in the original one-loop diagram.

The transverse projection of Eq.~(\ref{b1}) involves
${1\02}\,
(\delta^{ij}-\hat p^i \hat p^j)(\delta_{ij}-\hat q_i \hat q_j)$,
where ${\bf q}={\bf k+p}$. Since $p\sim T$, we have
$ \hat q_i ={p_i+k_i\0|{\bf p+k}|}\simeq\hat p_i$,
while the integral in Eq.~(\ref{b1}) will be
eventually dominated by soft $k$ momenta. In what follows,
we shall often perform such kinematical simplifications
relying on the fact that $k\ll p$. With this simplification,
the product of the transverse projectors above reduces to the
identity, so that
\beq\label{b2}
\delta \Pi_{b}^l(p)\,=\,-g^2N\int [{\rm d}k]\,
(2p_0+k_0)^2\,D_0(p+k)\Bigl(\5 D_L(k)-D_L^{(0)}(k)\Bigr).\eeq
To perform the Matsubara sum over $k_0$ we need
the following sums (with ${\bf q}\equiv {\bf k+p}$):
\beq
 T\sum_{k_0} D_0(p+k)\Bigl(\5 D_L(k)-D_L^{(0)}(k)\Bigr)&=&
\int\!\!{dk_0\0 2\pi}\int\!\!{dq_0\0 2\pi}\,\5\rho_L(k)
\rho_0(q)\,{n(q_0)-n(k_0)\0k_0-q_0+p_0},\\
T\sum_{k_0}k_0 D_0(p+k)\Bigl(\5 D_L(k)-D_L^{(0)}(k)\Bigr)
&=&\int\!\!{dk_0\0 2\pi}\int\!\!{dq_0\0 2\pi}\,k_0\,\5\rho_L(k)
\rho_0(q)\,{n(q_0)-n(k_0)\0k_0-q_0+p_0},\nn 
T\sum_{k_0}k_0(k_0+p_0) D_0(p+k)\Bigl(\5 D_L(k)-D_L^{(0)}(k)\Bigr)
&=&\int\!\!{dk_0\0 2\pi}\int\!\!{dq_0\0 2\pi}\,k_0\,q_0\,\5\rho_L(k)
\rho_0(q)\,{n(q_0)-n(k_0)\0k_0-q_0+p_0}\,.\nonumber\eeq
This finally yields, for the retarded self-energy,
\beq\label{b3}
\delta \Pi_{b}^l(p)\,=\,-g^2N\int\!\!{d^4k\0(2\pi)^4}
\int\!\!{dq_0\0 2\pi}\,\5\rho_L(k)
\rho_0(q)\,[4p_0^2+3p_0k_0+k_0q_0]\,
{n(q_0)-n(k_0)\0k_0-q_0+p_0+i\epsilon}\,.\eeq

To compute the entropy (\ref{ST32}) we need
the light-cone projection of the
real part of this self-energy, $\Re\delta \Pi_{b}^l(p_0=p)$.
Note that in previous calculations of the damping rate,
it was rather the {\it imaginary} part of this same self-energy
which was required \cite{P3,damping}.
The calculation of the imaginary part is easier since the
LO contribution $\sim g^2 T^2$ can be immediately extracted
from Eq.~(\ref{b3})
by neglecting $n(q_0) \sim 1$ against $n(k_0) \simeq T/k_0 \gg 1$,
and keeping only the large external momentum $4p_0^2$ in the numerator.
This, together with
\beq
\Im\,{1\0k_0-q_0+p_0+i\epsilon}\,=\,-\pi\delta(k_0-q_0+p_0)
\,\simeq\,-\pi\delta(k_0-k\cos\theta+p_0-p),\eeq
leads to the following, standard, result for the longitudinal part of
the damping rate \cite{damping} :
\beq
\gamma_l\,\equiv\,-\,{\Im\delta \Pi_{b}^l(p_0=p)\02p}
\,=\,{g^2NT\02}
\int\!\!{d^3k\0(2\pi)^3}\int\!\!{d\o\0\o}\,\5\rho_L(\o,k)\,
\delta(\o-k\cos\theta).\eeq
If we perform, however, the same simplifications on the
real part, then the would-be LO result turns out to vanish,
by parity (with ${\bf P}$ denoting the principal value):
\beq\label{naive}
\Bigl(\Re\delta \Pi_{b}^l(p_0=p)\Bigr)_{naive}\,=\,
g^2NT\int\!\!{d^3k\0(2\pi)^3}\int\!\!{d\o\0\pi\o}\,\5\rho_L(\o,k)\,\,
{\bf P}\,{1\0\o-k\cos\theta}\,=\,0\,.\eeq
In fact, this is only to be expected: the terms in Eq.~(\ref{naive})
are formally of order $g^2T^2$, while we know that
$\Re\delta \Pi$ should be rather of order $g^3T^2$.
Thus, in order to extract the leading contribution to $\Re\delta \Pi$
from Eq.~(\ref{b3}) one has to push the kinematical approximations
one step further as compared to the damping rate.
In particular, we need the expansion of the statistical factors
in Eq.~(\ref{b3}) to LO and NLO order:
\beq
n(k_0)-n(q_0)\simeq \,{T\0k_0}\,-\,{2n(q_0)+1\02}\,.\eeq
We shall denote by $\Re\delta \Pi_{b1}^l$ the contribution 
coming from $T/k_0\,$, and by $\Re\delta \Pi_{b2}^l$
the remaining one due to $(2n(q_0)+1)/2$. These quantities will be
evaluated at $p_0=p$, so they
are functions of the three-momentum $p$ alone. We have:
\beq\label{b11}
\Re\delta \Pi_{b1}^l(p)&=&g^2NT\int\!\!{d^3k\0(2\pi)^3}
\int\!\!{dk_0\0 2\pi k_0} \,\5\rho_L(k)
\int\!\!{dq_0\0 2\pi}\,[4p^2+3pk_0+k_0q_0]\,{\bf P}\,
{\rho_0(q)\0k_0-q_0+p}\,\nn &=&
g^2NT\int\!\!{d^3k\0(2\pi)^3}
\int\!\!{dk_0\0 2\pi k_0} \,\5\rho_L(k)(2p+k_0)^2\,\Re D_0(p+k)\,,\eeq
where in the second line we have identified the (retarded) free
propagator via its spectral representation.
In $\Re\delta \Pi_{b2}$, we can restrict ourselves to the LO
term $4p^2$ in the denominator, and to the positive-energy pole
$q_0=|{\bf p+k}|\simeq p+k\cos\theta$ in the spectral function
$\rho_0(q_0, |{\bf p+k}|)$. This yields:
\beq\label{b21}
\Re\delta \Pi_{b2}^l(p)&\simeq &-g^2Np(2n(p)+1)
\int\!\!{d^4k\0(2\pi)^4}\,
{\5\rho_L(k_0,k)\0k_0-k\cos\theta}\,.\eeq

\subsection{The NLO entropy of hard gluons}
\label{appdPit}

Let us consider first a purely gluonic plasma,
in which case the hard gluon self-energy 
$\delta\Pi_T=\delta \Pi^l+\delta\Pi^t$,
is all what we need to compute the NLO entropy
${\cal S}_3^{\mathrm hard}\equiv 
\delta{\cal S}^l+\delta{\cal S}^t$.
As before, we focus on the longitudinal contribution
$\delta{\cal S}^l$; by inserting Eqs.~(\ref{tad}), (\ref{b11}) 
and (\ref{b21}) in Eq.~(\ref{ST32}), we obtain
$\delta{\cal S}^l=\delta{\cal S}_{1}^l+
\delta{\cal S}_{2}^l$, where:
\beq\label{DS31}
\delta{\cal S}_{1}^l&\equiv&
- N_g\int\!\!{d^4p\0(2\pi)^4}\,\rho_0(p)\,{\6n(p_0)\0\6T}\,
\Bigl[\delta \Pi^l_a+\Re\delta \Pi_{b1}^l\Bigr]\nn
&=&-g^2NN_gT\int\!\!{d^4k\0(2\pi)^4}
\,{\5\rho_L(k)\0k_0}
\int\!\!{d^4p\0(2\pi)^4}\,{\6n(p_0)\0\6T}\,\rho_0(p)\,
\Bigl[(2p+k_0)^2\,\Re D_0(p+k)-1\Bigr]\nn&\simeq&
N_gT\int\!\!{d^4k\0(2\pi)^4}
\,{\5\rho_L(k)\02k_0}\,{\del\0\del T}
\Re\5\Pi_L(k_0,k)\,.\eeq
In writing the last line above, we have identified the
one-loop contribution to the self-energy of the
soft longitudinal gluon due to hard transverse gluons. 
To the order of interest,
this is precisely the HTL $\5\Pi_L$.
The second piece $\delta{\cal S}_{2}^l$ of the entropy reads:
\beq\label{DS32}
\delta{\cal S}_{2}^l&\equiv&
- N_g\int\!\!{d^3p\0(2\pi)^3}\,{1\0p}\,{\6n(p)\0\6T}\,
\Re\delta \Pi_{b2}^l(p)\nn &=&g^2NN_g\int\!\!{d^3p\0(2\pi)^3}\,
(2n(p)+1){\6n(p)\0\6T}\int\!\!{d^4k\0(2\pi)^4}
{\5\rho_L(k_0,k)\0k_0-k\cos\theta}\,\nn&=&
{N_g\02}{\6\0\6T} \Bigl(T {\hat m}_D^2\Bigr) \int\!\!{d^4k\0(2\pi)^4}
{\5\rho_L(k_0,k)\0k_0-k\cos\theta}\nn&=&
-{N_g}\int\!\!{d^4k\0(2\pi)^4}\,{\5\rho_L(k)\02k_0}\,{\6\0\6T}
\Bigl\{T\Bigl(\Re \5\Pi_L-\5m_D^2\Bigr)\Bigr\}\,.\eeq
In going from the second to the third line above, the
following chain of identities has been used (see also Eq.~(\ref{MD0})):
\beq\label{dermd}
2g^2 N \int\!\!{d^3p\0(2\pi)^3}\,[1 + 2n(p)]\,{\6n(p)\0\6T}\,=\,
2 g^2 N
{\6\0\6T}\int\!\!{d^3p\0(2\pi)^3}\,n(p)[1 + n(p)]\,=\,\nn
{}\qquad=\,-\,2g^2 N {\6\0\6T}\int\!\!{d^3p\0(2\pi)^3} \,T\,{\6 n\0\6 p}
\,=\,{\6\0\6T} \Bigl(T {\hat m}_D^2\Bigr) \,.\eeq
Furthermore, in writing the last line in Eq.~(\ref{DS32}),
 we have identified
the longitudinal HTL $\Re \5\Pi_L$ as follows (compare to
Eq.~(\ref{PiL})):
\beq\label{PILL}
\Re\5\Pi_L(\o,k)&=&-\5m_D^2\int{d\Omega\04\pi}\frac{k\cos\theta}
{\o-k\cos\theta}\,.\eeq

By adding Eqs.~(\ref{DS31}) and (\ref{DS32}), we finally deduce
the following expression for the longitudinal piece of the
NLO entropy:
\beq\label{DS3}
\delta{\cal S}_{}^l&=&N_gT\,{\6\5m_D^2\0\6T}
\int\!\!{d^4k\0(2\pi)^4}\,{\5\rho_L(k)\02k_0}\,-\,
{N_g}\int\!\!{d^4k\0(2\pi)^4}\,{\5\rho_L(k)\02k_0}\,
\Bigl(\Re \5\Pi_L-\5m_D^2\Bigr)\nn&=&
T\,{\6\5m_D^2\0\6T}\,\frac{N_g\5m_D}{8\pi}
\,-\,\Delta {\cal S}_L^{(3)},\eeq
with $\Delta {\cal S}_L^{(3)}$ as defined in Eq.~(\ref{SHTLL}).
An entirely similar calculation shows that
the remaining, transverse, piece $\delta{\cal S}^t$
cancels against the transverse contribution
$\Delta {\cal S}_T^{(3)}$ to $\Delta {\cal S}_3$, Eq.~(\ref{SHTLL}):
\be\label{S3T0}
\delta{\cal S}^t\,+\,\Delta {\cal S}_T^{(3)}\,=\,0.\ee
That is, the total contribution of the soft {\it transverse} gluons
to the plasmon effect cancels away, as it should.

All together, Eqs.~(\ref{SHTLD1}), (\ref{SHTLL}),
(\ref{DS3}) and (\ref{S3T0})
provide the expected result for the order-$g^3$ effect in the
entropy of the purely gluonic plasma where
$T(\6_T\5m_D^2) = 2\5m_D^2$:
\beq
{\cal S}_3^{\mathrm soft}+{\cal S}_3^{\mathrm hard} \,=\,
\frac{N_g\hat m_D^3}{12\pi}\,+\,
T\,{\6\5m_D^2\0\6T}\,\frac{N_g\5m_D}{8\pi}\,=\,
\frac{N_g\hat m_D^3}{3\pi}\,.\eeq
Moreover, it can be easily recognized that Eqs.~(\ref{DS3}) 
and (\ref{S3T0}) are equivalent to the longitudinal, and, 
respectively, transverse components of Eq.~(\ref{S-H}), as they should.

\subsection{Adding the fermions}

The previous results are easily extended to a QCD plasma with 
fermions. The entropy ${\cal S}_3^{\mathrm hard}$ in this case
involves also the NLO self-energies of the hard fermions, 
$\delta \Sigma_\pm\,$:
\bea\label{barS}
{\cal S}_3^{\mathrm hard}=-\int\!\!{d^4p\0(2\pi)^4}\left\{
N_g{\6n(p_0)\0\6T}\,\rho_0(p)\Re\delta \Pi_T\,
+\,2NN_f {\6f\0\6T} \sum_{s=\pm}\rho_s(p)
\Re\delta\Sigma_s\right\}.\eea
Once again, we focus on the contribution $\delta{\cal S}^l$
of the soft {\it longitudinal} gluons, and
use the integral over the hard momentum $p$ 
in Eq.~(\ref{barS}) to reconstruct the HTL $\5\Pi_L$.
Here, this involves both a hard gluon loop and a hard fermion loop, 
which enters via the self-energies $\delta \Sigma_\pm^l$.

For instance, the fermionic analog of
$\Re\delta \Pi_{b2}^l$, Eq.~(\ref{b21}), reads:
\beq
\Re\delta \Sigma_{\pm\,2}^l(p)\,=\, -{g^2C_f\02}[1 - 2f_\pm(p)]
\int\frac{d^4k}{(2\pi)^4}\,\frac{\5\rho_L(k_0,k)}{k_0-k\cos\theta}
\,,\eeq 
which, when inserted into Eq.~(\ref{barS}), determines the
following contribution to the NLO entropy (compare to Eq.~(\ref{DS32})):
\beq\label{SLAF}
\delta{\cal S}^l_2
&=&g^2 N_g\int\!\!{d^3p\0(2\pi)^3}\biggl\{
N (1 + 2n){\6n\0\6T}+{N_f\02}\sum_{s=\pm}
(1 - 2f_s){\6f_s\0\6T}\biggr\}
\int\frac{d^4k}{(2\pi)^4}\,\frac{\5\rho_L(k_0,k)}{k_0-k\cos\theta}
\nn&=&
{N_g\02}{\6\0\6T} \Bigl(T {\hat m}_D^2\Bigr) \int\!\!{d^4k\0(2\pi)^4}
{\5\rho_L(k_0,k)\0k_0-k\cos\theta}\,.\eeq
(We have used here $(1 - 2f)\6_Tf=
\6_T[f(1-f)] = -\6_T(T\6_kf)$, 
together with Eq.~(\ref{MD}) for the Debye mass.)
This is formally the same result as for pure glue, Eq.~(\ref{DS32}),
except that, here, $\5 m_D$ is the {\it full} HTL
Debye mass, which includes contributions from fermions.

Similarly, the other contribution
$\delta{\cal S}_1^l$ preserves the form in
Eq.~(\ref{DS31}) where, however, $\5\Pi_L$ is now the full HTL
in a theory with fermions. Thus the final result in Eq.~(\ref{DS3})
is formally unchanged, but it now applies to a QCD plasma
with fermions, for which $T(\6_T\5m_D^2) = 2m_T^2$
(cf. Eq.~(\ref{MTM})). 

Consider finally the order $g^3$ effect in the quark density:
as explained in the main text, this comes entirely from
the NLO corrections $\delta\Sigma_\pm$ to the hard fermion 
self-energies, and, more precisely, from the longitudinal sector
alone (the soft transverse effects eventually cancel,
as in the case of the entropy). Thus, ${\cal N}_3=
\delta{\cal N}^l$, with $\delta{\cal N}^l$
given by the same equations as above,
except for the replacement of the temperature derivatives
by derivatives with respect to $\mu$. Thus
$\delta{\cal N}^l\equiv
\delta{\cal N}^l_1+\delta{\cal N}^l_2$,
where (cf. Eqs.~(\ref{DS31}) and ~(\ref{DS32})):
\beq\label{N3A}
\delta{\cal N}^l_1\,=\,
N_gT\int\!\!{d^4k\0(2\pi)^4}
\,{\5\rho_L\02k_0}\,{\del\0\del \mu}
\Re\5\Pi_L(k_0,k)\,,\eeq
and 
\be\label{N3B}
\delta{\cal N}^l_2
=
{N_gT\02}{\6{\hat m}_D^2\0\6\mu} \int\!\!{d^4k\0(2\pi)^4}
{\5\rho_L(k_0,k)\0k_0-k\cos\theta}=
-{N_g T}\int\!\!{d^4k\0(2\pi)^4}\,{\5\rho_L\02k_0}\,{\6\0\6\mu}
\Bigl(\Re \5\Pi_L-\5m_D^2\Bigr)\,.\ee
As in the entropy, the non-local terms involving
$\Re \5\Pi_L(k)$ cancel in the sum
of the two contributions above, and we are left with the
following simple expression:
\beq
{\cal N}_3\,=\,
N_gT\,{\6\5m_D^2\0\6\mu}
\int\!\!{d^4k\0(2\pi)^4}\,{\5\rho_L(k)\02k_0}\,=\,
\frac{N_gT  m_\mu^2\5m_D }{4\pi \mu}\,.\eeq

\end{document}